\documentclass[twocolumn,epjc3]{svjour3}        
\journalname{Eur. Phys. J. C}

\usepackage{lmodern}

\usepackage{cuted}
\usepackage{comment}
\usepackage[titles]{tocloft}
\usepackage{bm}
\usepackage{amssymb}
\usepackage{amsmath}
\usepackage{mathpazo}
\DeclareSymbolFont{cmroman}{OT1}{cmr}{m}{n}
\DeclareMathSymbol{\cmUpsilon}{\mathalpha}{cmroman}{"07}
\renewcommand{\Upsilon}{\cmUpsilon}

\usepackage{booktabs}
\usepackage{graphicx}
\usepackage{adjustbox}
\usepackage{booktabs}

\usepackage{graphicx}
\usepackage{float}
\usepackage{multirow}
\usepackage{lineno}
\usepackage{subcaption}
\usepackage{amsmath,bm}
\usepackage{slashed}

\newcommand{\beq}[1]{\begin{equation}}
\newcommand{\eeq}[1]{\end{equation}}
\newcommand{\nn}{\nonumber}

\usepackage{color}

\newcommand{\C}{\mathcal{C}}

\DeclareUnicodeCharacter{2212}{-}
\usepackage{xcolor}
\usepackage{soul}
\usepackage[normalem]{ulem}
\definecolor{note_image}{HTML}{4682B4}

\definecolor{note_problem}{HTML}{DC143C}

\newcommand{\new}[1]{{\color[rgb]{0,0,1}{#1}}} % OLD strikeout normal font

\def\GeV{{\rm GeV}}
\def\GeV2{{\rm GeV}^2}

\newcommand{\Q}{{\cal Q}}

\newcommand{\bT}{b_\sT}

\newcommand{\sT}{{\scriptscriptstyle T}}
\newcommand{\CS}{{\rm CS}}

\newcommand{\mc}[1]{\mathcal{#1}}

\newcommand{\koneT}{\bm{k}_{1\sT}}
\newcommand{\ktwoT}{\bm{k}_{2\sT}}
\newcommand{\qT}{\bm{q}_{\sT}}
\renewcommand{\bT}{\bm{b}_{\sT}}
\newcommand{\bme}[1]{\bm{\scriptstyle #1}}

\newcommand{\bt}{b_\sT}

\newcommand{\qt}{q_\sT}

\newcommand{\ce}[1]{Eq.~(\ref{#1})}
\newcommand{\cf}[1]{{Fig.~\ref{#1}}}

\makeatletter
\renewcommand\appendix{%
  \par
  \setcounter{section}{0}%
  \setcounter{subsection}{0}%
  \gdef\thesection{\@Alph\c@section}%
}
\makeatother
\usepackage[numbers,sort&compress]{natbib}
\usepackage[colorlinks,citecolor=blue,linktoc=all,linkcolor=blue]{hyperref}
\usepackage{scalerel}
\usepackage{tikz}
\usetikzlibrary{svg.path}

\definecolor{orcidlogocol}{HTML}{A6CE39}
\tikzset{
  orcidlogo/.pic={
    \fill[orcidlogocol] svg{M256,128c0,70.7-57.3,128-128,128C57.3,256,0,198.7,0,128C0,57.3,57.3,0,128,0C198.7,0,256,57.3,256,128z};
    \fill[white] svg{M86.3,186.2H70.9V79.1h15.4v48.4V186.2z}
                 svg{M108.9,79.1h41.6c39.6,0,57,28.3,57,53.6c0,27.5-21.5,53.6-56.8,53.6h-41.8V79.1z M124.3,172.4h24.5c34.9,0,42.9-26.5,42.9-39.7c0-21.5-13.7-39.7-43.7-39.7h-23.7V172.4z}
                 svg{M88.7,56.8c0,5.5-4.5,10.1-10.1,10.1c-5.6,0-10.1-4.6-10.1-10.1c0-5.6,4.5-10.1,10.1-10.1C84.2,46.7,88.7,51.3,88.7,56.8z};
  }
}

\usepackage{hhline}

\newcommand\orcidicon[1]{\href{https://orcid.org/#1}{\mbox{\scalerel*{
\begin{tikzpicture}[yscale=-1,transform shape]
\pic{orcidlogo};
\end{tikzpicture}
}{|}}}}

\title{NNLL fit of the transverse-momentum-dependent distribution of unpolarised gluons to LHCb data on $J/\psi$-pair production}

\author{
Dani\"el Boer\thanksref{addr1}\protect\orcidicon{}
\and
Jelle Bor\thanksref{addr1,addr2}\protect\orcidicon{}
\and
Alice Colpani Serri\thanksref{addr3}\protect\orcidicon{}
\and
Miguel G. Echevarria\thanksref{addr4}\protect\orcidicon{0000-0003-2888-8526}
\and
Jean-Philippe~Lansberg\thanksref{addr2}\protect\orcidicon{0000-0003-2746-5986}
\and 
Samuel F. Romera\thanksref{addr4}\protect\orcidicon{0009-0007-0853-6212}
\and 
Pieter Taels\thanksref{addr2,addr5}\protect\orcidicon{}
}

\institute{
Van Swinderen Institute for Particle Physics and Gravity, University of Groningen, Nijenborgh 4, 9747 AG Groningen, The Netherlands\label{addr1}
\and
Universit\'e Paris-Saclay, CNRS, IJCLab, 91405 Orsay, France\label{addr2}
\and
Faculty of Physics, Warsaw University of Technology, plac Politechniki 1,00-661, Warszawa, Poland\label{addr3}
\and
Department of Physics and EHU Quantum Center, University of the Basque Country EHU, 48940, Leioa, Spain\label{addr4}
\and
Department of Physics, University of Antwerp, 2020 Antwerpen, Belgium\label{addr5}
\\~ \\
\email{d.boer@rug.nl, j.bor@rug.nl, alice.colpani\_serri.dokt@pw.edu.pl, jean-philippe.lansberg@in2p3.fr, samuel.fernandez@ehu.eus, pieter.taels@uantwerpen.be}
}

\date{\today}
\begin{document}

\date{\today}

\maketitle

\begin{abstract}
    We perform the first fit of the unpolarised gluon transverse-momentum-dependent distribution $f_1^g$ up to next-to-next-to-leading-logarithmic accuracy to the double-differential distribution in the invariant mass and transverse momentum of the latest $J/\psi$-pair production LHCb data. We review previous attempts to describe the older $J/\psi$-pair LHCb data using $f_1^g$, and discuss two novel methodologies which allow us to account in the fit for both the perturbative and PDF uncertainties to construct $f_1^g$ in the perturbative region. It turns out that the consideration of these uncertainties is crucial for a meaningful fit to such low-scale gluon-sensitive data. 
\end{abstract}

\tableofcontents

\section{Introduction}\label{short_introduction}
Quantum Chromodynamics (QCD) has achieved remarkable phenomenological success through factorisation frameworks that separate perturbative from non-perturbative dynamics. One such framework is Transverse Momentum Dependent (TMD) factorisation~\cite{Collins:2011zzd,Angeles-Martinez:2015sea}, applicable when a small transverse momentum $q_{\sT}$ is probed that is much smaller than the hard scale $Q$ of the process. In TMD factorisation, the non-perturbative three-dimensional momentum structure of hadrons in terms of quarks and gluons is parameterised by TMD parton distributions functions (TMD PDFs, TMDs for short), which need to be extracted from experimental (or lattice) data. While quark TMDs have been extensively studied, see e.g.~\cite{Barry:2023qqh,Bacchetta:2024qre,Bacchetta:2024yzl,Bacchetta:2025ara,Moos:2025sal}, using Drell–Yan and semi-inclusive deep-inelastic scattering data, gluon TMDs remain much less constrained because they do not contribute to these processes at leading twist and evolve independently from their quark counterparts.

Gluon TMDs can be extracted from TMD-factorisable processes dominated by gluon fusion. A particularly promising channel is the hadroproduction of quarkonium pairs, where the invariant mass of the pair sets the hard scale $Q$ and the vector sum of the transverse momenta of the bound states provides the small transverse momentum scale $q_{\sT}$. For $S$-wave vector mesons, such as $J/\psi$ or $\psi(2S)$, TMD factorisation holds to a very good approximation, since the quarkonia are predominantly produced through the hadronisation of charm-anticharm pairs in a colour-singlet $^3S_1$ state~\cite{Lansberg:2014swa,Lansberg:2019adr,Lansberg:2019fgm}, strongly suppressing final-state radiation. 

A first Gaussian leading-order fit~\cite{Lansberg:2017dzg} (without the effects of TMD evolution) of the unpolarised gluon TMD  to single-differential LHCb data~\cite{LHCb:2016wuo} of $J/\psi$ pairs with average invariant mass $\langle M_{\psi\psi}\rangle \!=\! 8\;\mathrm{GeV}$ resulted in a value $\langle k_{\sT}^2\rangle \!=\! 3.3\pm 0.8\;\mathrm{GeV}^2$ for the average transverse momentum dependence of $f_1^g$. This was a clear indication of the importance of TMD evolution, since the size of purely intrinsic transverse momentum should be of the order of the proton mass. A subsequent study~\cite{Scarpa:2019fol,Scarpa:2020sdy} including TMD evolution up to next-to-leading log (NLL) confirmed the effect of momentum broadening and reaffirmed the feasibility to reliably extract gluon TMDs from this process. 

In this work, we present the first fit with evolution up to next-to-next-to-leading logarithmic accuracy (NNLL) of the unpolarised gluon TMD $f_1^g$ to LHCb data of inclusive (prompt) $J/\psi$-pair production~\cite{LHCb:2023ybt}, measured double-differentially in the invariant mass $Q$ and the transverse momentum $q_{\sT}$ of the pair. 
The relatively low values of $Q$ make this process sensitive to non-perturbative effects, and the fact that different bins in $Q$ are provided makes it possible to better study the sensitivity to TMD evolution.
Our results nicely complement those in the recent work~\cite{Anedda:2026cox} providing an NNLL gluon TMD fit to data on Higgs production, probing TMD evolution at a much higher (fixed) scale.

The paper is organised as follows. In section~\ref{context}, we sketch the broad context of our research. Technical details on quarkonium production, TMD factorisation, and TMD evolution are provided in section~\ref{methodology}. The fit is performed using three distinct approaches, presented in sections~\ref{sec:classic_fit},~\ref{bT_fit}, and~\ref{qT_fit}. After an extensive discussion in section~\ref{discussion} on the resulting extractions of the gluon TMD, conclusions are drawn in section~\ref{conclusions}.

\section{Context}\label{context}
\subsection{TMD factorisation and gluon TMDs}
TMD factorisation is rigorously proven (at leading twist) for the Drell-Yan process~\cite{Collins:2011zzd,Echevarria:2011epo} and for semi-inclusive deeply inelastic scattering (SIDIS)~\cite{Ji:2004wu}. In the last years, considerable effort has been devoted to the extraction of quark TMDs from Drell-Yan and SIDIS data; see, e.g.,~\cite{Barry:2023qqh,Bacchetta:2024qre,Bacchetta:2024yzl,Bacchetta:2025ara,Moos:2025sal}. Despite their importance, much less is known on gluon TMDs. This is due to the fact that, at leading twist, they do not contribute to the Drell-Yan or SIDIS cross sections, nor do they mix with quark TMDs through evolution.\footnote{In collinear factorisation, the situation is drastically different. Both quark and gluon PDFs can be extracted from deeply-inelastic scattering measurements, since gluon PDFs appear at higher perturbative orders in the cross section and mix with their quark counterparts through DGLAP evolution.} Apart from the works~\cite{Lansberg:2017dzg,Scarpa:2019fol,Anedda:2026cox} mentioned in the previous section~\ref{short_introduction}, a numerical analysis of the linearly polarised gluon TMD $h_1^{\perp g}$ in the context of Higgs production is presented in~\cite{Gutierrez-Reyes:2019rug}. An extraction of the unintegrated gluon distribution, closely related to the gluon TMD, in the context of low-$x$ QCD can be found in~\cite{Kutak:2012rf}.

In our work we make use of quarkonium-pair production which, similarly to the case of the Higgs boson, is dominated by gluon fusion~\cite{Boer:2012bt,denDunnen:2014kjo}.
As we will argue below, the hadroproduction of a pair of vector quarkonia of the same flavour can indeed be described within TMD factorisation, up to relativistic corrections. The fact that the hard scale $Q$ of this process is comparatively low is an advantage for probing non-perturbative contributions, and performing measurements at different invariant masses provide a handle on TMD evolution. 

\subsection{Quarkonium-production mechanisms}
Before going into details on the fit, a few words on the production mechanism of quarkonium states are in order. Early descriptions of (single) quarkonium production were either based on the Colour Evaporation Model (CEM)~\cite{Fritzsch:1977ay,Halzen:1977rs}, or on the Colour Singlet Model (CSM)~\cite{Chang:1979nn, Berger:1980ni, Baier:1983va}. In the former, the production of the heavy-quark pair is taken to be largely decorrelated from its hadronisation, in line with the principle of parton-hadron duality. In the latter, the heavy-quark pair is produced at short distances with the same quantum numbers as the final quarkonium state. The hadronisation of the pair into the bound state is encoded in the non-relativistic wave function at the origin, obtained from potential models by solving the Schr\"odinger equation or from decays. 

Nowadays, most theory descriptions of quarkonium are based on non-relativistic QCD (NRQCD)~\cite{Bodwin:1994jh}, an effective field theory in which the average velocity $v$ of the heavy quarks in the quarkonium rest frame provides a small expansion parameter, in addition to the strong coupling $\alpha_s$. For $S$ waves, NRQCD reduces to the CSM at the lowest order in $v$ but, when including higher-order relativistic corrections, allows for the heavy-quark pair to be produced in all quantum states permitted by the conservation laws, including colour-octet (CO) states. After the hard scattering, the pair hadronises into the observed quarkonium by emitting soft gluons. The non-perturbative hadronisation is parametrised by Long-Distance Matrix Elements (LDMEs) which, for colour-singlet (CS) states, reduce up to $v^2$ corrections to the wave function at the origin. The CO LDMEs, on the other hand, are essentially unconstrained and need to be extracted from experimental data.

For the production of single $S$-wave quarkonia such as $J/\psi$ and $\Upsilon$, CO-state contributions are suppressed by a power $v^4$ with respect to that of the CS state~\cite{Bodwin:1994jh,Cho:1995vh,Cho:1995ce}. As demonstrated in~\cite{Lansberg:2019fgm}, CO contributions to the differential di-$J/\psi$ cross section
lie below the percent level, except at very large rapidity separations irrelevant for our TMD analysis (see~\cite{Scarpa:2019fol}). In our study we can, therefore, safely describe both quarkonia by their CS states only. 
This is also one of the main reasons why, at leading order in $v^2$, TMD factorisation holds, as CO states connect with each other and with the incoming protons through gluon radiation, which would break (or, at least, complicate) the factorisation like in the hadroproduction of back-to-back hadrons~\cite{Collins:2007nk} or back-to-back photon-hadron pair~\cite{Rogers:2013zha}.

\subsection{Single versus double parton scattering}

For TMD factorisation to apply to the hadroproduction of a $J/\psi$ pair, the reaction must take place via single parton scattering (SPS). However, the cross section receives contributions from double parton scattering (DPS), where each proton provides two incoming partons that independently participate in two separate hard scattering processes~\cite{Lansberg:2014swa}. In the experimental data from the LHCb collaboration that we use for our fit~\cite{LHCb:2023ybt}, these DPS contributions are removed. The subtraction method relies on the different behaviours of the SPS and DPS cross sections as a function of the rapidity difference $\Delta y$ between both mesons~\cite{Kom:2011bd}. 
Indeed, in DPS, the produced $J/\psi$'s are uncorrelated, thus more likely to be produced with a large rapidity separation than in SPS, where they are correlated. 
In~\cite{LHCb:2023ybt}, the DPS distribution is obtained by combining two measured single-inclusive differential cross sections from Ref.~\cite{LHCb:2015foc}, assuming that the two $J/\psi$'s are uncorrelated. The normalisation of the DPS cross section is then determined from the data in the range $1.8<\Delta y<2.5$, where the SPS contribution is negligible according to NRQCD calculations~\cite{Baranov:2011zz,Sun:2014gca,Likhoded:2016zmk,Lansberg:2019fgm} which is then tested by looking at the azimuthal dependence in the transverse plane.

\subsection{TMD shape functions in quarkonium production}
There are some important theoretical subtleties regarding NRQCD in TMD kinematics, i.e., when the quarkonium transverse momentum $q_{\sT}$ is much smaller than the hard scale $Q$ of the process. In TMD factorisation, the evolution equations resum large Sudakov logarithms in the ratio $q_{\sT}/Q$. These Sudakov logarithms stem from the radiation of soft-collinear gluons. But soft gluons with momenta $\!\sim\! m_Qv$ participate in the binding mechanism of the heavy-quark pair. Therefore, when $q_{\sT}\!\sim\! m_Qv$, both momentum modes overlap, and much care is needed to avoid double counting. The usual NRQCD approach does not suffice, since it does not distinguish between modes below the hard scale $2m_Q$. Two generalisations of NRQCD that distinguish soft ($\!\sim\! m_Qv)$ from ultrasoft ($\!\sim \!m_Q v^2$) gluons are pNRQCD~\cite{Brambilla:1999xf} and vNRQCD~\cite{Luke:1999kz,Manohar:1999xd}. In particular, in Refs.~\cite{Echevarria:2019ynx,Fleming:2019pzj,Echevarria:2024idp}, quarkonium was studied within TMD factorisation in an effective field theory approach, combining vNRQCD with Soft-Collinear Effective Theory (SCET)~\cite{Bauer:2000ew,Bauer:2000yr,Bauer:2001ct,Bauer:2001yt}. One remarkable outcome of these studies is the introduction of novel non-perturbative objects called TMD Shape Functions (TMD ShFs).\footnote{Named after the shape functions to resum soft radiation at the kinematic endpoint, originally in leptonic B-meson decay and later applied in NRQCD, see, e.g.,~\cite{Neubert:1993ch,Bigi:1993ex,Mannel:1994pm,Rothstein:1997ac,Beneke:1997qw,Beneke:1999gq,Fleming:2003gt}.} TMD ShFs parametrise the non-perturbative spatial distribution of the heavy-quark pair inside the bound state, and are the generalisation of the LDMEs to the TMD regime.\footnote{In the recent works \cite{Copeland:2025vop,copeland2026}, additional non-perturbative structures called TMD soft transition functions were identified.} They have been introduced earlier as phenomenological ‘smearing functions' in~\cite{Bacchetta:2018ivt}. In Refs.~\cite{Boer:2020bbd,Boer:2023zit}, they were studied from the perturbative matching of TMDs with collinear factorisation. Studies of the impact of the CO TMD ShFs on the electroproduction of quarkonium can be found in Refs.~\cite{Maxia:2025zee,Echevarria:2025oab}.

As mentioned above, we describe the $J/\psi$ mesons in the process under consideration by their CS states, as CO states are part of higher-order corrections in the $v^2$ expansion. 
As we will argue in Sec.~\ref{subsec:factorisation} and discussed in detail in a future work~\cite{dijpsifactorization}, at leading order in $v^2$ the cross section can be shown to factorise in terms of the standard LDMEs, without the need for a CS TMD ShF. To this accuracy, the $\qT$-smearing of the di-$J/\psi$ pair is, therefore, entirely due to the incoming gluon TMD PDFs.

\section{Methodology}\label{methodology}
\subsection{Factorisation}\label{subsec:factorisation}
We consider the process $p(P_{1})\!+\!p(P_{2})\!\to\!\mathcal{Q}(P_{\Q,1})\!+\!\mathcal{Q}(P_{\Q,2})\!+\!X$, where the $\,^{3}S_{1}^{[1]}$ states are denoted by $\mathcal{Q}$ and the colliding protons by $p$.\footnote{In this work we only consider proton-proton reactions. It is, however, trivial to extend the theory and methodology to other light hadrons, or to asymmetric collisions such as proton-pion scattering.} The phase space of the outgoing quarkonia is parametrised in terms of the vector sum $\qT\!\equiv\!\mathbf{P}_{\Q,1}\!+\!\mathbf{P}_{\Q,2}$ of the transverse momenta, the invariant mass $Q^{2}\!=\!(P_{\mathcal{Q},1}+P_{\mathcal{Q},2})^{2}$, the rapidity $y\equiv y_{\psi \psi}\!=\!(1/2)\ln\frac{2(P_{\mathcal{Q},1}^{+}\!+\!P_{\mathcal{Q},2}^{+})^{2}}{Q^{2}\!-\!\qT^{2}}$ of the pair,\footnote{The rapidity of the pair, $y_{\psi \psi}$, is also denoted $y$, for simplicity.} and their relative solid angle $\mathrm{d}\Omega$. The incoming gluons carry fractions $x_1\!=\!\frac{P_{\mathcal{Q},1}^{+}+P_{\mathcal{Q},2}^{+}}{P_1^+}\!=\!\sqrt{\frac{Q^2-\qT^2}{s}}e^{+y}$ and $x_2\!=\!\sqrt{\frac{Q^2-\qT^2}{s}}e^{-y}$ of their parent proton longitudinal momentum, where $\sqrt{s}$ is the center-of-mass energy of the colliding protons. The differential cross section reads: \footnote{The notation used is: $v^{+}\equiv v\cdot\bar{n},v^{-}\equiv v\cdot n$, where $n$ and $\bar{n}$ are light-like vectors defined as $n^{\mu}\!\equiv\!\frac{1}{\sqrt{2}}(1,0,0,1)\!=\!(1^{+},0^{-},\boldsymbol{0})$ and $\bar{n}^{\mu}\!\equiv\!\frac{1}{\sqrt{2}}(1,0,0,-1)\!=\!(0^{+},1^{-},\boldsymbol{0})$. Moreover, we always write transverse momenta in bold and directly in Euclidean space. Purely spatial momenta in three-dimensional Euclidean space are written with an arrow: $\vec{v}\!=\!(v_{1},v_{2},v_{3})$.} 
\begin{equation}
\begin{aligned}  &\frac{\mathrm{d}\sigma}{\mathrm{d}Q\mathrm{d}y\mathrm{d}^{2}{\qT}\mathrm{d}\Omega}\,
=H_{\mu\nu\rho\sigma}(y,Q,m_c,\Omega;\mu)\\
&\times\int\mathrm{d}^{2}\bT\,e^{i\bT\cdot\mathbf{q}_{\sT}}
 f_{g/p}^{\mu\nu}(x_{1},\bT;\zeta_1,\mu)\,
 f_{g/p}^{\rho\sigma}(x_{2},\bT;\zeta_2,\mu)\\
 & \times
\left[\left< \mathcal{O}^{J/\psi}\big({}^3S_1^{[1]}\big)\right>(\mu)\right]^2
 \;.
 \label{eq:factorisationtheorem}
\end{aligned}
\end{equation}
Eq.~\eqref{eq:factorisationtheorem} describes the TMD factorisation of the cross section in terms of hard part $H$, which only depends on the large momentum scales, while all the long-distance physics is encoded in four distinct hadronic correlation functions: two gluon TMD PDFs $f_{g/p}^{\mu\nu}$ corresponding to each of the colliding protons, and two CS LDMEs pertaining to the formation of each quarkonium.

A complete derivation of the factorisation theorem for double-$J/\psi$ production in Eq.~\eqref{eq:factorisationtheorem}, which holds at leading twist in $q_T/Q$ and at leading order in $v^2$, will be presented elsewhere~\cite{dijpsifactorization}.
Schematically, it relies on: i) the decoupling of soft initial-state radiation from the CS quarkonium states~\cite{Copeland:2026yqa}; ii) the fact that soft radiation between well-separated colour singlets cancels~\cite{Bodwin:2010fi}, such that both $J/\psi$'s can be described by separate CS LDMEs; iii) TMD factorisation of the initial state into two gluon TMD PDFs, similar to Higgs production~\cite{Echevarria:2015uaa}; iv) the universal soft function, describing initial-state radiation and decoupling from the final state according to i), being absorbed into the gluon TMD PDFs in order to regularise rapidity divergences.
We note that, for this case of CS quarkonium production, there is no need for TMD ShFs.

The gluonic structure of an unpolarised proton can be parametrised in terms of an unpolarised ($f_{1}^{g}$) and linearly polarised ($h_{1}^{\perp g}$) gluon TMD PDFs~\cite{Mulders:2000sh,Meissner:2007rx,Boer:2016xqr,Echevarria:2015uaa} 
\begin{equation}
\begin{aligned}
f_{g/p}^{\mu\nu}(x_{i},k_{\sT i};\zeta_{i},\mu) & \!=\!\frac{1}{2x_{i}}\!\Big[\!-\!g_{\sT}^{\mu\nu}f_{1}^{g}(x_{i},\mathbf{k}_{\sT i}^{2};\zeta_{i},\mu)\\
 & \!+\!\frac{k_{\sT}^{\mu}k_{\sT}^{\nu}-\frac{1}{2}k_{\sT}^{2}g_{\sT}^{\mu\nu}}{M_{p}^{2}}h_{1}^{\perp g}(x_{i},\mathbf{k}_{\sT i}^{2};\zeta_{i},\mu)\!\Big]
\end{aligned}
\end{equation}
where $i\!=\!1,2$, and where $\zeta_i$ and $\mu_i$ are the rapidity- and renormalisation scales on which we will elaborate in section~\ref{SecEvo}. 
\subsection{Cross section}
The cross section can be written explicitly as~\cite{Lansberg:2017dzg}:
\begin{align}
&\frac{\mathrm{d}\sigma}{\mathrm{d}Q \mathrm{d}y \mathrm{d}^2\qT \mathrm{d}\Omega} 
=\mathcal{K}(Q,y;\mu) 
\nonumber \\
&\times\Bigg\{F_1(Q,\theta_{\CS};\alpha_s(\mu))\ \mc{C} \Big[f_1^gf_1^g\Big](x_{1,2},\qT;\mu) \nonumber \\
  &\;\;+ F_2(Q,\theta_{\CS};\alpha_s(\mu))\ \mc{C} \Big[w_2h_1^{\perp g}h_1^{\perp g}\Big](x_{1,2},\qT;\mu)
   \nonumber\\
  &\;\;+ \cos2\phi_{\CS} \label{eq:TMDcrosssection}
  \\
&\;\;\quad\times\!\!\Bigg(F_3(Q,\theta_{\CS};\alpha_s(\mu)) \ \mc{C} \Big[w_3 f_1^g h_1^{\perp g}\Big](x_{1,2},\qT;\mu)  \nonumber \\
   &\;\; \quad\quad+\! F'_3(Q,\theta_{\CS};\alpha_s(\mu))\mc{C} \Big[w'_3 h_1^{\perp g} f_1^g\Big](x_{1,2},\qT;\mu)\!\!\Bigg)\!  
   \nonumber\\
  &\;\;+  \cos 4\phi_{\CS}\nonumber\\
  &\;\;\quad\times
   F_4(Q,\theta_{\CS};\alpha_s(\mu)) \mc{C}\! \Big[w_4 h_1^{\perp g}h_1^{\perp g}\Big]\!(x_{1,2},\qT;\mu)\! 
  \!\Bigg \} \nonumber
  \\
 & \times
\left[
\left< \mathcal{O}^{J/\psi} ({}^3S_1^{[1]}) \right>\big(\mu\big) \right]^2
  \,.\nonumber
\end{align}
In the above expression, an overall dependence on short-distance kinematics is contained in the prefactor $\mathcal{K}$. 
The solid angle between the outgoing quarkonium states and the interaction plane is expressed in the Collins-Soper frame: $\mathrm{d}\Omega\!=\!\mathrm{d}\!\cos\theta_{\CS}\mathrm{d}\phi_{\CS}$~\cite{Collins:1977iv}. 
The hard-scattering coefficients $F_i$ only depend on $\theta_{\CS}$, the invariant mass $Q$, and the running coupling $\alpha_s$. Their tree-level expressions for quarkonium-pair production can be found in~\cite{Lansberg:2017dzg} following from the uncontracted amplitudes from~\cite{Qiao:2009kg}.  
Finally, the TMD convolutions appearing in \ce{eq:TMDcrosssection} are defined as follows:
\begin{equation}
\begin{aligned}\label{eq:Cwfg}
&\mathcal{C}[w\, f\, g](x_{1,2},\qT;\mu)\\
&\!\equiv\! \int\!\! \mathrm{d}^{2}\koneT\!\! \int\!\! \mathrm{d}^{2}\ktwoT\,
  \delta^{(2)}(\koneT\!+\!\ktwoT\!-\!{\qT}) \\
  &\quad\times w(\koneT,\ktwoT)  f(x_1,\koneT^{2};\zeta_1,\mu)\, g(x_2,\ktwoT^{2};\zeta_2,\mu) \,,
\end{aligned}  
\end{equation}
where $w(\koneT,\ktwoT)$ denotes a TMD weight. 
The weights in Eq.~\eqref{eq:TMDcrosssection} are common to all gluon-fusion processes originating from unpolarised proton collisions and
read (see Ref.~\cite{Lansberg:2017tlc,Bor:2025ztq}):
\begin{equation}
\begin{aligned}
w_{2} & = \frac{1}{4M_{p}^{4}}\bigg[ 2(\koneT\cdot\ktwoT)^{2} - \koneT^{2}\ktwoT^{2}\bigg] \hspace{1mm}, \\
w_{3} & = \frac{1}{2M_{p}^{2}\bm{q}_{\sT}^{2}}[\ktwoT^{2}  \bm{q}_{\sT}^2-2(\ktwoT \cdot  \bm{q}_{\sT})^{2} ] \hspace{1mm}, \\ w_{3}' &= \frac{1}{2M_{p}^{2}\bm{q}_{\sT}^{2}}[\koneT^{2}  \bm{q}_{\sT}^2-2(\koneT \cdot  \bm{q}_{\sT})^{2} ] \hspace{1mm}, \\
w_{4} & \!=\! 2 \bigg(\!\frac{\koneT\cdot\ktwoT}{2M_{p}^{2}} \!-\! \frac{(\koneT\cdot \bm{q}_{\sT}) (\ktwoT\cdot \bm{q}_{\sT})}{M_{p}^{2}\bm{q}_{\sT}^{2}}\bigg)^{\!2} - \frac{\koneT^{2}\ktwoT^{2}}{4M_{p}^{4}} \;.
\label{Eqweights}
\end{aligned}
\end{equation}

To disentangle the different contributions in the cross section~\ce{eq:TMDcrosssection}, one can study specific angular-momentum observables.
The simplest one is the cross section integrated over the azimuthal angle $\phi_{\CS}$, where the terms with a $\cos(2\phi_{\CS},4\phi_{\CS})$-dependence drop out:
\begin{equation}
\begin{aligned}
&\int\frac{\mathrm{d}\phi_{\CS}}{2\pi}  \frac{\mathrm{d}\sigma}{\mathrm{d} Q \mathrm{d} y \mathrm{d}^2 \qT \mathrm{d} \Omega} \\
&=
\mathcal{K}\bigg\{F_1\, \mc{C} \Big[f_1^{\,g}f_1^{\,g} \Big]
+F_2\, \mc{C} \Big[w_2h_1^{\perp\, g}h_1^{\perp\, g}\Big]\bigg\}\;.
\label{eq:phi_av_xsection}
\end{aligned}
\end{equation}
This is what we will focus on in this analysis.

Second, one can consider $\cos(n\phi_{\CS})$-weighted differential cross sections:
\begin{equation}
\begin{aligned}
\langle  \cos(n\phi_{\CS}) \rangle \equiv\frac{\displaystyle \int \!\!\mathrm{d}\phi_{\CS} \cos(n\phi_{\CS})\,  \frac{\displaystyle \mathrm{d}\sigma}{\mathrm{d} Q \mathrm{d} y \mathrm{d}^2 \qT \mathrm{d} \Omega}}{\displaystyle\!\!\int \!\!\mathrm{d}\phi_{\CS} \frac{\mathrm{d}\sigma}{\mathrm{d} Q \mathrm{d} y \mathrm{d}^2 \qT \mathrm{d} \Omega}}\; .
\end{aligned}
\end{equation}
These quantities correspond to (half of) the relative size of the $\cos(2\phi_{\CS},4\phi_{\CS})$-modulations present in the TMD cross section \ce{eq:TMDcrosssection} divided by the  $\phi_{\CS}$-independent component:
\begin{equation}
\begin{aligned}\label{asym_exp}
\langle\cos 2\phi_{\CS}\rangle  \! &= \!\frac{1}{2}\frac{F_3 \mc{C} \Big[w_3 f_1^{\,g} h_1^{\perp\, g} \Big] \!+\! F'_3 \mc{C} \Big[w'_3 h_1^{\perp\, g} f_1^{\,g} \Big]}{F_1\, \mc{C} \Big[f_1^{\,g}f_1^{\,g} \Big]+F_2\, \mc{C} \Big[w_2h_1^{\perp\, g}h_1^{\perp\, g}\Big]} \, ,\\
\langle\cos 4\phi_{\CS}\rangle \!  &=\! \frac{1}{2}\frac{F_4 \mc{C}\! \left[w_4 h_1^{\perp\, g}h_1^{\perp\, g}\right]}{F_1\, \mc{C} \Big[f_1^{\,g}f_1^{\,g} \Big]+F_2\, \mc{C} \Big[w_2h_1^{\perp\, g}h_1^{\perp\, g} \Big]}\; .
\end{aligned}
\end{equation}

In Ref.~\cite{LHCb:2023ybt}, LHCb performed the very first measurement of these modulations. They found a non-zero (negative) value for $\langle\cos 4\phi_{\CS}\rangle$ with a significance of about 1.6~$\sigma$ which provides a very first hint of a non-zero $h_1^{\perp g}$. Yet, as we will discuss later, its impact on the $\phi_{\CS}$-averaged $J/\psi$-pair cross section Eq.~\eqref{eq:phi_av_xsection} via the term  $F_2\, \mc{C} \Big[w_2h_1^{\perp\, g}h_1^{\perp\, g}\Big]$ is negligible in view of the other uncertainties in our analysis. This follows from the smaller size of $h_1^{\perp g}$  with respect to $f_1^g$ and the suppression of $F_2$ in the kinematical domain that we consider~\cite{Scarpa:2019fol}.

\subsection{Normalisation}
The normalisation of the current theoretical evaluation~\cite{Sun:2014gca,Lansberg:2014swa,Lansberg:2019fgm} of $J/\psi$-pair-production cross sections suffers from very large uncertainties from the factorisation and renormalisation scales, the value of the charm mass, the NRQCD LDMEs, and the collinear gluon PDFs. To mitigate the effect of these uncertainties, while nearly fully preserving the sensitivity on the shape of the gluon TMDs, we perform our fit on a self-normalised $\qt$-differential cross section. The LHCb data are given in three bins of the invariant mass $Q\equiv M_{\Q\Q}$ of the pair. For each bin in $Q$, we only consider those data points with $\qt<Q/2$, i.e., where TMD factorisation can reasonably be applied. Therefore, to normalise the data, we divide each bin in $Q$ by the discrete integral over $\qt$ from $0$ up to $Q/2$.

In practice, this normalisation allows us to disregard in the cross section both the hard factors and the LDMEs, since they are global factors which do not depend on $b_T$ and cancel in the ratio.

\subsection{Conventional approaches to TMD evolution \label{SecEvo}}

\subsubsection{Generalities about TMD evolution}
The most common way to implement the scale evolution (see Refs.~\cite{Collins:1984kg,Collins:1989gx,Collins:2011zzd,Echevarria:2011epo,Echevarria:2014rua,Echevarria:2015uaa}) of the TMD PDFs is by working in transverse-coordinate or  $b_{\sT}$ space, with $\bT$ being the conjugate variable to $\qT$. Like collinear PDFs, TMDs depend on the renormalisation scale $\mu$. However, they contain an additional dependence on the so-called rapidity scale $\zeta$ which arises from the regularisation of gluons with infinite rapidity in the definition of the hadron correlator~\cite{Collins:1981uk}. 
Consistency of the TMD factorisation framework requires that the rapidity scales only appear in the cross section in the combination $\zeta_1 \zeta_2 \!=\! Q^4$. We enforce this relation by setting the rapidity scales of the two TMD PDFs in the convolution~\ce{eq:Cwfg} as:
\begin{align}
\zeta_{1}  = M^{2}_{p}x_{1}^{2}e^{2(y_{1}-y_{s})}\,, \quad \zeta_{2}  = M^{2}_{p}x_{2}^{2}e^{2(y_{s}-y_{2})}\;,
\label{eq:zetadef}
\end{align}
where $y_{1,2}$ denote the rapidities of the colliding protons, and $y_{s}$ is an arbitrary cut-off that cancels in the cross section.\footnote{Of course, for the scattering of other types of hadrons, the masses in Eq.~\eqref{eq:zetadef} must be changed accordingly.} 
As usual, the renormalisation scale $\mu$ in the hard-scattering coefficient $\mathcal{K}$ should be set to $\mu\!\sim\! Q$ to avoid large logarithms.

To implement the TMD evolution, we define the TMD PDFs in  $b_{\sT}$ space:
\begin{equation}
\label{eq:TMDhatdef}
\begin{aligned}
\tilde{f}_1^{\,g}(x,\bm b_{\sT}^2;\zeta,\mu) & = 
\int\!\mathrm{d}^2\bm k_{\sT}\, e^{-i\bme b_{\sT} \cdot \,\bme k_{\sT}}f_1^{\,g}(x,\bm k_{\sT}^2;\zeta,\mu)
\;,
\\
\tilde{h}_1^{\perp\, g}(x,\bm b_{\sT}^2;\zeta,\mu) & =  
\int\!\mathrm{d}^2\bm k_{\sT}\, \frac{(\bm b_{\sT}\cdot\,\bm k_{\sT})^2-\frac{1}{2}\bm b_{\sT}^2 \bm k_{\sT}^2}{\bm b_{\sT}^2 M_p^2} \\ 
&\hspace{1cm}\times e^{-i\bme b_{\sT}\cdot\,\bme k_{\sT} }h_1^{\perp\, g}(x,\bm k_{\sT}^2;\zeta,\mu) 
\;.
\end{aligned}
\end{equation}
Note that $\tilde{h}_1^{\perp\, g}$ is not exactly given by a simple Fourier transform; rather, it is proportional to the second $\bm b_{\sT}^2$ derivative which systematically appears in the convolutions~\eqref{eq:Cwfg} and, hence, in the cross section~\eqref{eq:TMDcrosssection}.

Indeed, the convolutions $\C[w f g](x_1,x_2,\qT)$ can be written in terms of the TMDs~\eqref{eq:TMDhatdef} in  $b_{\sT}$ space as follows (suppressing for clarity the scale dependences): 
\begin{equation}
\begin{aligned}
&\C[f_{1}^{g}f_{1}^{g}] 
\!=\!
\int_{0}^{\infty}  \frac{\mathrm{d}b_{\sT}}{2\pi} \!  b_{\sT}J_{0}(b_{\sT}q_{\sT})\tilde{f}_{1}^{g}(x_{1},\bT^{2})  \tilde{f}_{1}^{g}(x_{2},\bT^{2})\,, 
\\
&\C[w_{2}h_{1}^{\perp g}h_{1}^{\perp g}]
\!= \!
\int_{0}^{\infty} \!\!\frac{\mathrm{d}b_{\sT}}{2 \pi} \!  b_{\sT}  J_{0}(b_{\sT}q_{\sT})  \tilde{h}_{1}^{\perp g}(x_{1},\bT^{2})  \tilde{h}_{1}^{\perp g}(x_{2},\bT^{2})\,, 
\\
&\C[w_{3}f_{1}^{g}h_{1}^{\perp g}] 
\! =\!
\int_{0}^{\infty} \! \! \frac{\mathrm{d}b_{\sT}}{2 \pi}  \!   b_{\sT}  J_{2}(b_{\sT}q_{\sT})  \tilde{f}_{1}^{g}(x_{1},\bT^{2})  \tilde{h}_{1}^{\perp g}(x_{2},\bT^{2})\,, 
\\
&\C[w_{3}'h_{1}^{\perp g}f_{1}^{g}] 
\! =\!
\int_{0}^{\infty} \! \! \frac{\mathrm{d}b_{\sT}}{2 \pi}  \!  b_{\sT}  J_{2}(b_{\sT}q_{\sT})  \tilde{h}_{1}^{\perp g}(x_{1},\bT^{2}) \tilde{f}_{1}^{g}(x_{2},\bT^{2}) \,, 
\\
&\C[w_{4}h_{1}^{\perp g}h_{1}^{\perp g}] 
\! =\!
\int_{0}^{\infty} \! \! \frac{\mathrm{d}b_{\sT}}{2 \pi}  \! b_{\sT}  J_{4}(b_{\sT}q_{\sT})  \tilde{h}_{1}^{\perp g}(x_{1},\bT^{2})  \tilde{h}_{1}^{\perp g}(x_{2},\bT^{2}) \,,
\end{aligned}\label{eq:convbT}
\end{equation}
where $q_{\sT}\!\equiv\!|\qT|$ and $b_{\sT}\!\equiv\!|\bT|$. 
The Bessel functions of the first kind come from the angular integrals of the different weights since the TMDs do not depend on the direction of $\bT$. Their order depends on the weight, and is higher for more complicated angular dependences.  We refer to Appendix A of~\cite{Scarpa:2020sdy} for their derivation.

Now, the advantage of working in  $b_{\sT}$ space becomes clear as, in this space, the TMD evolution is multiplicative:
\begin{equation}
\begin{aligned}\label{Cf1f1_mub}
\tilde{f}_1^{\,g}(x,\bT^2;\zeta_f,\mu_f) & =  
e^{-\frac{1}{2}S_\text{P}(b_{\sT};\zeta_f,\mu_f,\zeta_i,\mu_i)}\tilde{f}_1^{\,g}(x,\bT^2;\zeta_i,\mu_i)
,
\\
\tilde{h}_1^{\perp\, g}(x,\bT^2;\zeta_f,\mu_f) & = 
e^{-\frac{1}{2}S_\text{P}(b_{\sT};\zeta_f,\mu_f,\zeta_i,\mu_i)}\tilde{h}_1^{\perp\, g}(x,\bT^2;\zeta_i,\mu_i)\,,
\end{aligned}
\end{equation}
where the evolution takes the TMD PDFs from the initial scales $(\zeta_i,\mu_i)$ to the final scales $(\zeta_f,\mu_f)$.
In the above formula, $S_\text{P}$ stands for the \emph{perturbative Sudakov factor} of the TMD PDFs:
\begin{equation}
\begin{aligned}
&S_\text{P}(b_{\sT};\zeta_f,\mu_f,\zeta_i,\mu_i)=
2 \mathcal{D}_\text{P}(\bT,\mu_i)\ln\frac{\zeta_f}{\zeta_i} \\ &\quad 
+2\!\int_{\mu_i}^{\mu_f}\!\!\frac{\mathrm{d}\bar\mu}{\bar\mu}\!
\Bigg[ \!\Gamma(\alpha_s(\bar\mu^2)\!)\!\ln\frac{\zeta_f}{\bar\mu^2} 
\!+\! \gamma(\alpha_s(\bar\mu^2)\!)\!\Bigg]
\,.
\end{aligned}
\label{eq:Sudakovdef}
\end{equation}
The physical meaning of the Sudakov factor can be understood by considering the fact that we are studying the cross section in a restricted region of phase space (in this case, the regime where $\qT^2\!<\!Q^2$). This restriction influences the cancellation of soft and collinear divergences between virtual and real higher-order perturbative corrections, leading to large so-called Sudakov logarithms. The perturbative Sudakov factor resums these logarithms, which would otherwise violate unitarity, and can be interpreted as the probability for no resolvable radiation to occur between the initial and final scales. Eq.~\eqref{eq:Sudakovdef} depends, amongst others, on the (perturbative) \emph{Collins-Soper} kernel $\mathcal{D}_\text{P}(\bT,\mu_b)$, which determines the $\zeta$-dependence of the radiation.
Moreover, the \emph{cusp anomalous dimension} $\Gamma$ controls the large Sudakov double logarithms stemming from soft-collinear gluon emissions. 
Likewise, $\gamma$ stands for the non-cusp anomalous dimension of the gluon TMD PDFs, which controls the large Sudakov single logarithms caused by radiation that is either soft or collinear.
We note that the Sudakov factor $S_\text{P}$ is spin-independent, and is thus the same for all (un)polarised TMDs~\cite{Echevarria:2014rua,Echevarria:2015uaa,Echevarria:2012pw}.

The TMD PDFs must be evolved up to the hard scale of the process, i.e., $\mu_f\!\sim \!Q$. The final rapidity scales of the two gluon TMDs must satisfy the requirement $\zeta_{1f} \zeta_{2f} \!= \!Q^4$, hence we set them to $\zeta_{1f}\!=\zeta_{2f}\!=\! \mu_f^2$. We choose the initial scales to be $\mu_i^2 \!=\!\zeta_{1i}\!=\!\zeta_{2i}\!=\!\mu_b^2$, with $\mu_b\!\equiv\! b_0/b_{\sT}$ ($b_{\sT}\!\equiv\!|\bT|$ and $b_0\!\equiv\! 2e^{-\gamma_E}\!\simeq\! 1.123$). With this choice, large logarithms in the Wilson coefficients $C_{g/j}$ (see Eq.~\eqref{eq:OPE f1g}) are minimised.

We stress that the evolution equations are only applicable in the perturbative region, thus at  $b_{\sT}$ sufficiently small compared to $\Lambda_{\rm QCD}$, not in the whole integration domain of the convolutions Eq.~\eqref{eq:convbT}. This is a well-known issue of working in $b_{\sT}$ space and we will shortly come back to it.

For $b_{\sT} \ll \Lambda_{\text{QCD}}^{-1}$, the \emph{perturbative} part of the TMDs can be obtained by matching on the $\qT$-integrated PDFs~\cite{Collins:2011zzd}, order by order in $\alpha_s$:
\begin{equation}
\begin{aligned}
\label{eq:OPE f1g}
    & \tilde{f}_1^{\,g}(x,\bT^{2};\zeta,\mu)  \\
    & = \sum_{j =q, \bar{q}, g} C_{g/j} (x,\bT^2;\zeta,\mu) \otimes f_1^{j/p}(x;\mu) +\mathcal{O}(b_{\sT}\Lambda_{\rm QCD}) \; ,
\end{aligned}
\end{equation}
where the convolution $\otimes$ refers to the longitudinal momentum fraction $x$.
In this work, the Wilson matching coefficients $C_{g/j}$ are considered up to NLO~\cite{Echevarria:2015uaa}, consistent with NNLL resummation accuracy.
As $h_1^{\perp\, g}$ describes the correlation between the gluon polarisation and its transverse momentum inside the unpolarised proton, it requires a helicity flip and, therefore, an additional gluon exchange. This is why its LO perturbative expression appears at $\mathcal{O}(\alpha_s)$~\cite{Sun:2011iw}:
\begin{equation}
    \begin{aligned}
    \label{h1pert}
    &\tilde{h}_1^{\,\perp g}(x,\bT^{2};\zeta,\mu)  = -\frac{\alpha_s(\mu)}{\pi}\int_x^1\!\!\frac{d\tilde{x}}{\tilde{x}}\Big(\frac{\tilde{x}}{x}-1\Big)\\
    &\qquad\qquad\times  \Big(C_A f_1^{g/p}(\tilde{x};\mu)+C_F \sum_{i=q,\bar q}f_1^{i/p}(\tilde{x};\mu)\Big)\\
    &\qquad\qquad+\mathcal{O}(\alpha_s^2)+\mathcal{O}(b_{\sT}\Lambda_{\rm QCD})\;.
    \end{aligned}
\end{equation}

The perturbative order at which the anomalous dimensions (cusp $\Gamma$, non-cusp $\gamma$ and Collins-Soper kernel $\mathcal{D}$) and matching coefficients (hard coefficient $H$ and TMD matching coefficients $C$) are needed for a given resummation accuracy is given in Table~\ref{tab:perturbative-orders}.

\begin{table}[t]
    \centering
    \renewcommand{\arraystretch}{1.25}
    \setlength{\doublerulesep}{1pt}
    \begin{tabular}{c||c|c|c|c|c}
        & $\Gamma$ & $\gamma$
        & $\mathcal{D}$ & $H$ & $C$ \\
        \hhline{=||=|=|=|=|=}
        NLL
        & $\alpha_s^2$
        & $\alpha_s^1$
        & $\alpha_s^1$
        & $\alpha_s^0$
        & $\alpha_s^0$
        \\
        \hhline{-||-|-|-|-|-}
        NNLL
        & $\alpha_s^3$
        & $\alpha_s^2$
        & $\alpha_s^2$
        & $\alpha_s^1$
        & $\alpha_s^1$
    \end{tabular}
    \caption{Perturbative orders required at NLL and NNLL resummation accuracies.}
    \label{tab:perturbative-orders}
\end{table}

The explicit expressions for $\Gamma$, $\gamma$, $\mathcal{D}$ and $C$ are conveniently summarised, e.g., in Ref.~\cite{Echevarria:2015uaa} (section 3 and appendix E).

\subsubsection{The conventional non-perturbative approaches}
Summarising the results discussed above, each TMD convolution in $b_{\sT}$ space is of the following form:
\begin{equation}
\label{eq:convolution}
\mathcal{C}[w\, f\, g] = \int_0^\infty\!\! \frac{\mathrm{d}b_{\sT}}{2\pi}\, b_{\sT} J_m(b_{\sT} q_{\sT})\,\tilde{W}(b_{\sT})\,,
\end{equation}
for integers $m$ and where $\tilde W$ is a simple product of (evolved) Fourier-transformed TMDs, $\tilde{f}(x,\bT^{2};\zeta,\mu)$ or $\tilde{h}_1^{\,\perp g}(x,\bT^{2};\zeta,\mu)$ with $\mu \sim \sqrt{\zeta} \sim Q$. Higher powers of $b_{\sT}$ can appear when the proton polarisation is considered~\cite{Kato:2024vzt, Boer:2001he}.

As mentioned above, the scale at which the TMDs are evaluated in $b_{\sT}$ space, which is initially $\mu_b \!\equiv \!2e^{-\gamma_E}/b_{\sT}$, must remain in the perturbative regime, where the matching relations Eqs.~(\ref{eq:OPE f1g}) and (\ref{h1pert}) are valid and the Sudakov factor~\eqref{eq:Sudakovdef} can be calculated perturbatively. However, the $b_{\sT}$-integration in the convolution~\eqref{eq:convolution} extends to infinity. Therefore, a method is needed to separate the small-$b_{\sT}$ perturbative from the large-$b_{\sT}$ non-perturbative regions.

A common way to do so is the $b_{\sT}^*$ prescription~\cite{Collins:1981va}, which consists in replacing $\tilde{W}(b_{\sT})$ in the convolution~\eqref{eq:convolution} by $\tilde{W}\big(b^*_{\sT}\big(b_{\sT}))$. Here, $b_{\sT}^*$ is an arbitrary function that tends to a fixed value $b_{\sT,\text{max}}$ when $b_{\sT} \to \infty$.
The conventional implementation is
\begin{equation}
\label{eq:conventional bstar}
    b_{\sT}^*(b_{\sT}) = \frac{b_{\sT}}{\sqrt{1 + b_{\sT}^2/b_{\sT,\text{max}}^2}} \; ,
\end{equation}
and $b_{\sT,\text{max}}$ is interpreted as the boundary between the perturbative and non-perturbative regions.
With this replacement, the TMD PDFs Eqs.~(\ref{eq:OPE f1g}) and (\ref{h1pert}) are never evaluated beyond the perturbative regime.
It should be noticed that one may choose $b_{T, \text{max}}$ to be small enough for the corrections scaling like $(b_{\sT,\text{max}} \Lambda_{\text{QCD}})^n$ to be negligible.
Indeed, it is common to take $b_{\sT,\text{max}}$ between $ 0.5\!-\!1.5\, \text{GeV}^{-1}$ in phenomenological analyses, equivalent to $0.1\!-\!0.3$~fm,  which correspond to $\mu_{b}(b_{\sT,\text{max}})\!=\!b_{0}/{b_{\sT,\text{max}}}\equiv{2e^{-\gamma_{E}}}/{b_{\sT,\text{max}}}$ from $2.25$ GeV to $0.75$ GeV, respectively.

On the other hand, as noted in Refs.~\cite{Boer:2014tka, Parisi:1979se}, for very small values of $b_{\sT}$ the scale $\mu_{b}$ can become arbitrarily large, which causes trouble in matching TMD and collinear factorisations. Moreover, it is natural to assume that TMD expressions should not be sensitive to QCD at scales beyond the scale $Q$ probed by the process.
This is the purpose of another prescription meant to prevent $\mu_{b}(b_{\sT})$ to exceed $Q$ when $b_{\sT}$ becomes small.\footnote{In Refs.~\cite{Aslan:2024nqg, Gonzalez-Hernandez:2022ifv}, a large transverse-momentum cut-off in transverse-momentum space is considered rather than a small $b_{\sT}$ cut-off.}
It consists in replacing $b_{\sT}$ by
\begin{align}
b_{\sT}'(b_{\sT}) = \sqrt{b_{\sT}^{2}+\frac{b_{0}^{2}}{Q^{2}}}\,.
\label{eq:bprime}
\end{align}

Both the $b_{\sT}^*$ and $b_{\sT}'$ prescriptions are implemented in the evaluations of Eq.~(\ref{eq:convolution}).
The order in which they are applied differs among works in the literature. However, we observe that the two replacements should be applied in the following order if one wishes that $S_\text{P}(b_{\sT}\to 0)\!=\!1$,\footnote{This condition stems from the interpretation of the exponential Sudakov factor as a no-emission probability, tending to unity for radiation with arbitrary large transverse momentum.}
\begin{equation}
\begin{aligned}
b_{\sT} & \to b_{\sT}' = \sqrt{b_{\sT}^2 + \frac{b_0^2}{Q^2}} \to b_{\sT}^{'*} = \sqrt{b_{\sT}^{*2} + \frac{b_0^2}{Q^2}} \; , \\
\mu_{b} & \to \mu_{b}'=\frac{b_{0}}{\sqrt{b_{\sT}^{2}+\frac{b_{0}^{2}}{Q^{2}}}} \to \mu_{b^*}' = \frac{b_{0}}{\sqrt{b_{\sT}^{*2}+\frac{b_{0}^{2}}{Q^{2}}}}\,.
\label{EqbTmethods}
\end{aligned}
\end{equation}
We also remark that:
\begin{equation}
\begin{aligned}
    \lim_{b_{\sT} \to 0} b_{\sT}^*(b_{\sT}') & = \frac{b_0/Q}{\sqrt{1+(b_0/Q)^2/b_{\sT,\text{max}}^2}} \; ,\\
    \lim_{b_{\sT} \to 0} b_{\sT}'(b_{\sT}^*) & = b_0/Q \; ,
\end{aligned}
\end{equation}
hence, for large values of $(b_0/Q)/b_{\sT,\text{max}}$ the prescription $b_{\sT}^*(b_{\sT}')$ deviates slightly from $b_0/Q$. In this work, we will always use the sequence of Eq.~\eqref{EqbTmethods}, i.e. $b_{\sT}^{'*}\!\equiv\!b_{\sT}^{'}\big(b_{\sT}^{*}\big)$ and $\mu_{b}^{'*}\!\equiv\!\mu_{b}^{'}\big(\mu_{b}^{*}\big)$.

In summary, the $b_{\sT}$ prescriptions in Eq.~(\ref{eq:conventional bstar}) and Eq.~(\ref{eq:bprime}) enforce
$Q\! \geq\! \mu_{b^*}' \!\geq\! b_0/b_{\sT,\text{max}}\! \gg\! \Lambda_{\text{QCD}}$,
yet allowing one to integrate over $b_{\sT}$ without restriction.
However, they have the unwanted side-effect of distorting the behaviour of $\mu_b(b_{\sT})$, and hence $\tilde{W}$, within the region where perturbative calculations are expected to be reliable.
We return to this issue below.

When $\mu_b$ approaches $\Lambda_{\mathrm{QCD}}$, one enters the region where non-perturbative physics becomes dominant.
To parametrise such effects, one introduces a so-called non-perturbative factor $S_{\text{NP}}$ through the two-step replacement:
\begin{equation}
\begin{aligned}
&\tilde{W}(b_{\sT},\{x_i\};Q)\\
&\to \tilde{W}(b_{\sT}^{*},\{x_i\};Q)e^{-\widehat{S}_\text{NP}(b_{\sT},\{x_i\};Q)} \\
&\to \tilde{W}^{\mathrm{pert}}(b_{\sT}^*,\{x_i\};Q)e^{-S_\text{NP}(b_{\sT},\{x_i\};Q)}\;,
\end{aligned}\label{EqNonPertS} 
\end{equation}
where, in the last line, $\tilde{W}^{\mathrm{pert}}$ stands for the purely perturbatively calculable part of $\tilde{W}$. 
$S_{\text{NP}}$ depends on the initial scale, the final scale $Q$, and indirectly on the prescription used to separate both perturbative and non-perturbative regions, in this case the $b_{\sT}^*$ prescription. We note that the object $\widehat{S}_\text{NP}$ appearing in the first replacement is, strictly speaking, not the same as the non-perturbative Sudakov factor $S_{\text{NP}}$.

Overall, with the above prescriptions and definitions, the simplest convolution becomes:
\begin{equation}\label{eq:Cff}
\begin{aligned}
&\C[f_{1}^{g}f_{1}^{g}](x_{1},x_{2}, q_{\sT};Q) = \int_{0}^{\infty} \frac{\mathrm{d}b_{\sT}}{2\pi} b_{\sT}\, J_{0}(b_{\sT}q_{\sT})
\\
&\times 
\tilde{f}_{1}^{g}(x_{1},b_{\sT}^{'*};\mu_{b^*}^{'2},\mu_{b^*}')  
\tilde{f}_{1}^{g}
(x_{2},b_{\sT}^{'*};\mu_{b^*}^{'2},\mu_{b^*}') 
\\
&\times
e^{- S_\text{P}(b_{\sT}^{'*};Q^2,Q,\mu_{b^*}^{'2},\mu_{b^{*}}')} 
e^{-S_\text{NP}(b_{\sT},x_1,x_2;Q)}
 \,,
\end{aligned}
\end{equation}
and similarly for any alternative $b_{\sT}$ prescriptions.
We note that similar expressions can be found for convolutions involving $\tilde{h}_{1}^{\perp g}$, where only the order of the Bessel function is different due to the weights \cite{Scarpa:2019fol}.

From Eq.~\eqref{eq:Cff}, the perturbatively calculable part of $W(b_{\sT})$ is given by:
\begin{equation}
\begin{aligned}
\label{eq:Wpert}
    &W^{\rm pert}(b_{\sT},x_1,x_2;Q) \\
    &\equiv\tilde{f}_{1}^{g}(x_{1},b_{\sT};\mu_{b}^2,\mu_{b})\,
    \tilde{f}_{1}^{g}(x_{2},b_{\sT};\mu_{b}^2,\mu_{b})\,\\
   & \times e^{- S_\text{P}(b_{\sT};Q^2,Q,\mu_{b}^2,\mu_{b})}  \,.
\end{aligned}
\end{equation}
It is important to note that all the functions entering the above expression are evaluated at $b_{\sT}$ (without the use of any $b_{\sT}^*$ prescription).

Regarding $S_\text{NP}$, the following form for two quark TMD PDFs was proposed long ago~\cite{Collins:1981va, Collins:1984kg}:
\begin{align}\label{SNPCollins}
S^{(2q)}_\text{NP}(b_{\sT},x_{1},x_{2};Q) = &\ln \bigg( \frac{Q}{Q_\text{NP}}\bigg) g_{K}(b_{\sT}) \\ +& g_{\text{TMD}_{a}}(x_{1},b_{\sT}) + g_{\text{TMD}_{b}} (x_{2},b_{\sT})  \,,\nn
\end{align}
where $g_{K}(b_{\sT})$, $g_{\text{TMD}_{a,b}}(x_{1,2},b_{\sT})$ are typically functions of $b_{\sT}^2$ where $Q_\text{NP}$ is a parameter with the dimension of mass that is chosen to be (close to) the smallest scale at which perturbation theory is expected to be valid. $S_\text{NP}$ satisfies the following general guiding principle: $\text{exp}[-S_\text{NP}]$ should be unity at $b_{\sT} \!=\! 0$, and it should vanish at large $b_{\sT}$ such that the TMD support in $b_{\sT}$ does not extend over distances much larger than typical confinement radius\footnote{Note however that $\bt$ is not an impact parameter but rather the conjugate variable of $\qt$.} and that the convergence of the (2D) Fourier transform is guaranteed.

To satisfy these conditions it is, however, sufficient to consider a simpler form for one single gluon TMD PDF:
\begin{equation}
\begin{aligned}
S^{(1g)}_\text{NP}(b_{\sT};x_i,Q)=\frac{1}{2}\left(A \,\ln\left(\frac{Q}{Q_\text{NP}}\right)+B\right)\,b_{\sT}^{2}\,, \label{eq:SNPsimple}
\end{aligned}
\end{equation}
where $A\,(\!=\!\! A^{(2g)})$ and $B\,(\!=\!\! B^{(2g)})$ are parameters to be determined. The $\ln Q$ dependence is kept as it naturally follows from the $\zeta_i$ dependence of the Sudakov factor.
From quark TMD fits, we  know that these parameters can depend on the numerical value of $b_{\sT,\text{max}}$~\cite{Aybat:2011zv}. Moreover, in the most general case, $B$ should be a function of the parton momentum fraction $x_i$ (see Subsec.~\ref{subsec:xdependence}). However, in our fit we will treat $A$ and $B$ as constants. Moreover, we directly fit the non-perturbative Sudakov factor in Eq.~\eqref{eq:Cff}, which we relate to the single-gluon one~\eqref{eq:SNPsimple} through the relation $S_{\mathrm{NP}}\!=\!2S_{\mathrm{NP}}^{(1g)}$.

Finally, the ingredients introduced above allow us to write an expression for the Collins-Soper kernel that is valid whether $\mu$ lies in the perturbative or non-perturbative domain. It is given by the sum of the perturbatively small-$b_{\sT}$ expression, calculable by matching; its (additive) renormalisation-group evolution driven by the cusp anomalous dimension $\Gamma$; supplemented by a genuinely non-perturbative contribution $\mathcal{D}_{\rm NP}(b_{\sT})$ that must be modelled or extracted from data. The $b_{\sT}^*$ prescription keeps the first two terms from entering the non-perturbative domain:
\begin{equation}
\begin{aligned}
\label{eq:CS kernel}
    &\mathcal{D}(b_{\sT}, \mu) \!=\!\mathcal{D}_{\rm P} (b_{\sT}^*,\mu_{b^*}\!) \!+\!\! \int_{\mu_{b^*}}^\mu \!\! \!\frac{\mathrm{d}\mu'}{\mu'}\! \Gamma (\mu')\! +\! \mathcal{D}_{\rm NP}(b_{\sT}) \, .
\end{aligned}
\end{equation}

\subsection{Limitations of existing approaches}
\label{Limitations of existing approaches}

Before elaborating on our fit methodologies, let us quickly review the latest results of the literature to date on di-$J/\psi$ hadroproduction as a probe of gluon TMDs.
As explained in the introduction, in recent years, di-$J/\psi$ production has been identified as one of the most promising channels to access the gluon transverse momentum structure of the proton.
In particular, the transverse momentum distribution of the di-$J/\psi$ cross section in terms of gluon TMDs was first investigated in Ref.~\cite{Lansberg:2017dzg}.

The effects of TMD evolution up to NLL accuracy and predictions for the self-normalised $\qt$-differential cross section were considered in Ref.~\cite{Scarpa:2019fol}. In that work, the non-perturbative factor $S_{\text{NP}}$ was modelled according to the functional form of \ce{eq:SNPsimple} with $B\! =\! 0$ and $Q_{\text{NP}}\!=\! 1\,\text{GeV}$, and by employing the $b_{\sT}^*(b_{\sT}')$ prescription (in that order) defined in Eqs.~(\ref{eq:conventional bstar}) and~(\ref{eq:bprime}).
The analysis explored a wide range of the non-perturbative parameter $A \!=\! [0.04,0.64]\,\text{GeV}^2$, together with the choice $b_{\sT,\max} \!=\! 1.5 \, \text{GeV}^{-1}$. 
However, the obtained theoretical $\qt$ distribution was not entirely satisfactory, exhibiting a peak at higher transverse momenta than the early LHCb data~\cite{LHCb:2016wuo}.\footnote{It should be noted, however, that those early LHCb data were not differential in the invariant mass.} This mismatch suggests that the perturbative evolution is ‘too fast', i.e., too much transverse momentum is generated when the hard scale increases, exacerbated by the fact that the cross section depends on two gluon TMDs. It turns out that, as explained in detail in Appendix~\ref{appendix:selection of bmax}, the problem is related to $b_{\sT,\max}$. With the large value chosen in Ref.~\cite{Scarpa:2019fol}, too much room is accorded to perturbative evolution, while the sensitivity to non-perturbative effects becomes negligible. Moreover, the reported uncertainty band was likely dominated by variations in the perturbative region driven by the $\bt^*$ prescription, rather than by genuine non-perturbative effects from the large-$b_{\sT}$ region.

In what follows, we revisit the procedures to implement TMD evolution, in particular the approaches to parameterise non-perturbative physics. The presented considerations are general and apply to processes involving two gluon TMDs in the initial state, such as single-$C\!\!=\!\!+1$-quarkonium or Higgs production.

\subsubsection{Systematic effects induced by the conventional modelling of non-perturbative physics}
\label{Systematic effects induced by the conventional modelling of non-perturbative effects}

As mentioned before, when $b_{\sT}$ is large, the product of TMDs in $b_{\sT}$ space, $\tilde{W}(b_{\sT})$, is sensitive to non-perturbative physics. This includes the intrinsic transverse momentum of the incoming partons, as well as long-range soft-gluon interactions between both partons and the two proton remnants. These effects are accounted for by making the replacement Eq.~\eqref{EqNonPertS}. Both the $b_{\sT}^*$ prescription and the parametrisation of $S_{\text{NP}}$ contain a degree of arbitrariness. In this section we will, before conducting the fit, explore the interplay between the different objects in the above expression.

First, there is freedom in the choice of $b_{\sT,\text{max}}$, above which a perturbative description becomes inadequate.
As discussed in detail in Appendices~\ref{appendix:selection of bmax} and \ref{appendix:Complementary discussion on NLL and bmax15}, we find that for the process and the kinematics under consideration, the region $b_{\sT}\! >\! 1 \, \text{GeV}^{-1}$ is irrelevant.\footnote{The precise choice of $b_{\sT,\text{max}}$ is not a prominent issue in quark-TMD phenomenology, such as in Drell-Yan or SIDIS processes, where the perturbative contribution to the observable happens to be distributed over a significantly broader range in $b_{\sT}$ than in the gluon case.} Throughout this work, we will work with the choice $b_{\sT,\text{max}} \!= \!0.5\,\text{GeV}^{-1}$.

Second, a common assumption is that the gluon $S_{\text{NP}}$ can be related to the quark one through Casimir scaling: $S_{\text{NP}}^{(1g)} \!= \!({C_A}/{C_F}) \!\times\! S_{\text{NP}}^{(1q)}$~\cite{Kulesza:2003wi,Bozzi:2005wk,Sun:2012vc,Bor:2022fga} due to the stronger colour charge of the gluon. 
We will also explore $S_{\text{NP}}^{(1g)}\!=\!\tfrac{1}{2}\!\times\!S_{\text{NP}}^{(1q)}$, in order to see what happens if the gluon $S_\text{NP}$ turns out to be smaller than the quark one.

For the non-perturbative Sudakov factor of a single quark, we consider the parametrisation in Eq.~\eqref{eq:SNPsimple} with the $B$ parameter replaced by the following $x_i$-dependent function:
\begin{equation}
\begin{aligned}
\label{eq:B(x)}
B^{(1q)} \longrightarrow
    B^{(1q)} (x_i) = b_1 \ln(10 x_i) + b_2\; , 
\end{aligned}
\end{equation}
and setting $Q_{\rm NP} \!=\! 1 \, \rm GeV$. This expression for $S_{\mathrm{NP}}^{(1q)}$ was used in Ref.~\cite{Landry:2002ix} to fit $Z$-boson production at the Tevatron. The obtained parameters from the fit were $A^{(2q)} \!=\! 0.68\,\text{GeV}^2$, $b_1 \!=\! -0.126\,\text{GeV}^2$ and $b_2\!=\!-0.291\,\text{GeV}^2$, for $b_{\sT,\max} \!=\! 0.5\,\text{GeV}^{-1}$.
In terms of the above parametrisation, the expression of the non-perturbative \emph{two-quark} Sudakov factor becomes
\begin{equation}
    \begin{aligned}
        S_{\mathrm{NP}}^{(2q)}&=\Big(A^{(2q)}\ln{\frac{Q}{Q_{\mathrm{NP}}}} + 2 b_2 \\
        &\qquad+ b_1 (\ln(10x_1)+\ln(10x_2))\Big)b_{\sT}^2\;.
    \end{aligned}\label{eq:twoquarkSudakov}
\end{equation}
From this, we construct the non-perturbative two-\emph{gluon} Sudakov factor as $S_{\mathrm{NP}}\!=\!c\!\times\! S_{\mathrm{NP}}^{(2q)}$ where $c$ is either $C_A/C_F$ or $1/2$.

\begin{figure*}[hbt]
    \centering
    \begin{comment}
    
    % first row
    \begin{subfigure}{0.49\textwidth}
        \centering
        \includegraphics[width=\linewidth]
        {Figures SF/SUDAKOVS_mu3p000_ipdf55_n2_bTmax0.5_merged.pdf}
        % \caption{$Q=3$ \text{GeV}, $b_{\sT,\max}=0.5~\mathrm{\text{GeV}}^{-1}$}
    \end{subfigure}
    \hfill
    \begin{subfigure}{0.49\textwidth}
        \centering
        \includegraphics[width=\linewidth]
        {Figures SF/SUDAKOVS_mu12p000_ipdf55_n2_bTmax0.5_merged.pdf}
        % \caption{$Q=12$ \text{GeV}, $b_{\sT,\max}=0.5~\mathrm{\text{GeV}}^{-1}$}
    \end{subfigure}

    \vspace{0.5cm}

    % second row
    \begin{subfigure}{0.49\textwidth}
        \centering
        \includegraphics[width=\linewidth]
        {Figures SF/SUDAKOVS_mu3p000_ipdf55_n2_bTmax1.5_merged.pdf}
        %\caption{$Q=3$ \text{GeV}, $b_{\sT,\max}=1.5~\mathrm{\text{GeV}}^{-1}$}
    \end{subfigure}
    \hfill
    \begin{subfigure}{0.49\textwidth}
        \centering
        \includegraphics[width=\linewidth]
        {Figures SF/SUDAKOVS_mu12p000_ipdf55_n2_bTmax1.5_merged.pdf}
        % \caption{$Q=12$ \text{GeV}, $b_{\sT,\max}=1.5~\mathrm{\text{GeV}}^{-1}$}
    \end{subfigure}
\end{comment}

\includegraphics[width=0.95\textwidth]{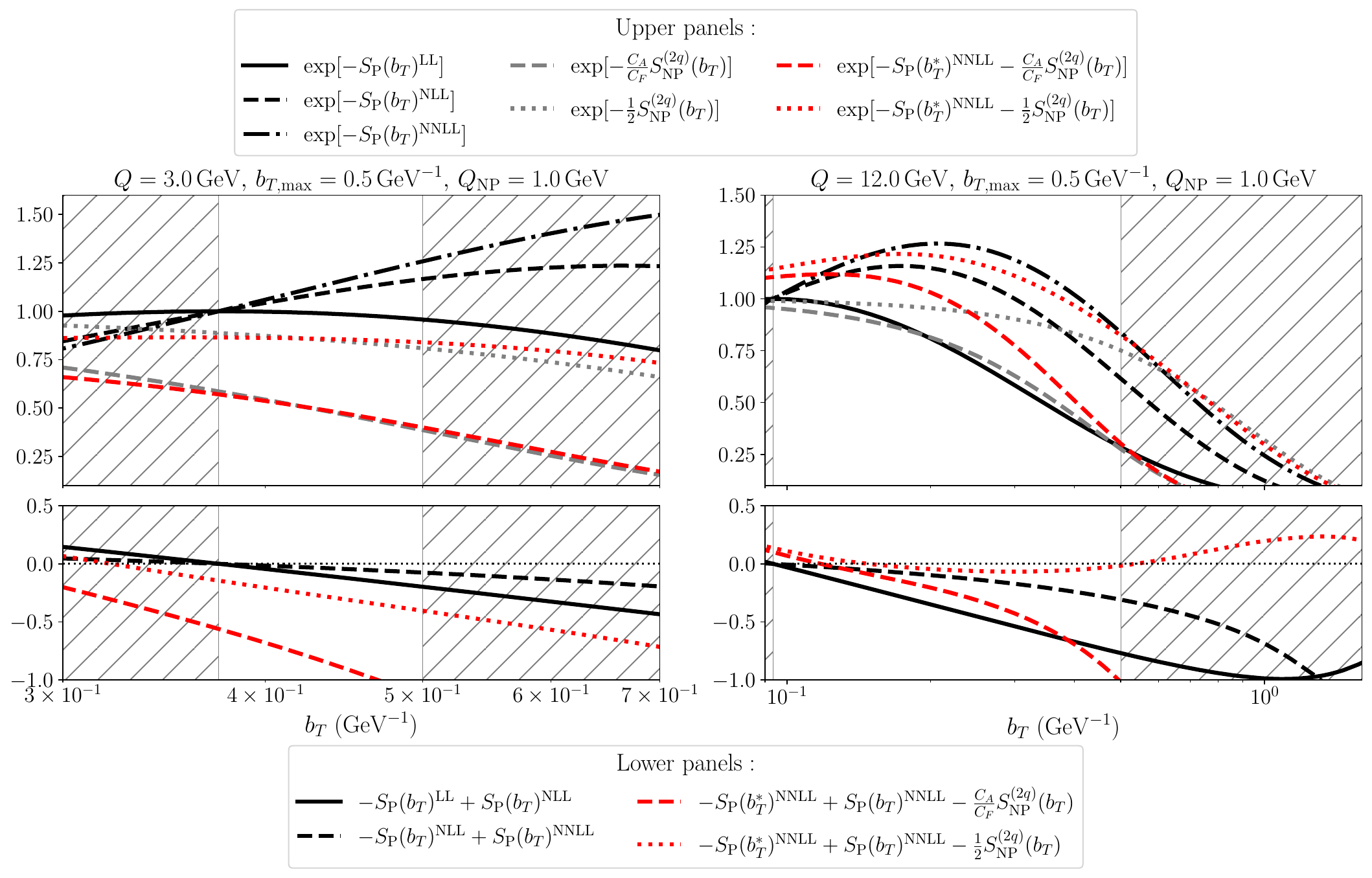}
    \caption{
Upper panel:    exponential perturbative LL, NLL and NNLL Sudakov factors $S_\text{P}$ (3 black curves), 
non-perturbative Sudakov factors using $S_{\text{NP}}^{(1g)} = (1/2, C_A/C_F)S_{\text{NP}}^{(1q)}$  for $b_{\sT,\text{max}}=0.5\text{ GeV}^{-1}$ (see Eq.~\eqref{eq:twoquarkSudakov}) (2 grey curves) and their combination at NNLL (2 red curves) for $Q=(3,12)~\text{ GeV}$ (left, right).
Lower panel: 
    differences between the perturbative Sudakov factors at successive logarithmic orders
    evaluated at $b_{\sT}$ (black), compared with the difference between the NNLL Sudakov evaluated at $b_{\sT}$ and the corresponding Sudakov at $b_{\sT}^*$ including the just defined $S_{\text{NP}}$ (red).
    The white areas represent the region $b_{\sT} \in [b_0/Q, b_{\sT,\text{max}}]$.}  
    
\label{fig:sudakov_combined}
\end{figure*}

Together with the conventional $b_{\sT}^*$ prescription, the above choices model the complete dependence of $\tilde{W}(b_{\sT})$ on non-perturbative phenomena. 
In Fig.~\ref{fig:sudakov_combined}, the thus obtained factors $\exp\!\big[\!-\!S_{\mathrm{NP}}\big]$ (grey curves) are shown. 
They are compared with the perturbative ones $\exp\!\big[\!-\!S_{\mathrm{P}}\big]$ (black curves) evaluated at leading-log (LL), NLL and NNLL accuracy, as well as with the total exponential factor (red curves) $\exp\!\big[\!-\!S_{\mathrm{P}}\!-\!S_{\mathrm{NP}}\big]$. The lower panels show the differences between the Sudakov factors themselves. The results are shown for two scales: $Q \!=\! 3\,\mathrm{GeV}$ (left) and $Q \!= \!12\,\mathrm{GeV}$ (right). The lower scale is typical for $\eta_{c}$ and $\chi_c$ production~\cite{Boer:2012bt}, while the higher scale is close to the invariant-mass bins of the di-$J/\psi$ production data from LHCb. The shaded regions correspond to $b_{\sT} \!<\! b_0/Q$ and $b_{\sT}\! >\! b_{\sT,\max}$, while the white band identifies the perturbative region defined by $b_{\sT} \!\in \![b_0/Q,b_{\sT,\max}]$.

For $Q\!=\!3\,\text{GeV}$, the black curves of the upper panel show that $\exp\!\big[\!-\!S_{\mathrm{P}}\big]$ at LL is approximately constant in the white area, and that the NLL and NNLL ones are monotonically increasing functions of $b_{\sT}$.
This is due to the low value of $Q$ leading to an extremely narrow perturbative region (white area), $b_{\sT}\! =\! [0.37, 0.50]\,\text{GeV}^{-1}$. In the case $Q\! =\! 12\,\text{GeV}$, the peak of each curve lies within the perturbative region which is much larger. 
In both cases, the non-perturbative exponentials (grey curves) exhibits a faster fall-off with increasing $b_{\sT}$ for the Casimir scaling than for the one where $S_{\mathrm{NP}}^{(1g)}\!=\!(1/2)S_{\mathrm{NP}}^{(1q)}$, as it should. The total exponential Sudakov factor (red curves) maintains this behaviour.
In the lower panels, the black lines show that the difference between NNLL and NLL calculations of $S_\text{P}$ is smaller than the difference between NLL and LL calculations. This indicates a satisfactory perturbative convergence. The exception is for $Q \!=\! 12\,\text{GeV}$ in the vicinity of $b_{\sT}\! \sim\! 1.5 \, \text{GeV}^{-1}$, which follows from the fact that $\alpha_s(\mu_{b})$ starts approaching non-perturbative large values.

The main observation is that, for both values of $Q$, the red curves differ from the black dash-dotted curve in the white area. This implies that the approach used to incorporate non-perturbative effects modifies the perturbative result for $S_\text{P}$ \emph{within the perturbative region.}
This modification is quantified in the lower panels. Ideally, from Eq.~\eqref{EqNonPertS}, the relation
\begin{equation}
\begin{aligned}
    S_{\rm P}(b_{\sT}^*)+S_{\rm NP}(b_{\sT})\simeq S_{\rm P}(b_{\sT})\;,
\end{aligned}\label{eq:SPdeviation}
\end{equation}
should hold in the perturbative $b_{\sT}$ region. In other words, $S_{\rm NP}$ should compensate for the change $b_{\sT}\to b_{\sT}^*$ inside $S_{\rm P}$.
This is clearly not what is observed. The deviation from the above identity, shown by the red curves, is sizeable throughout the entire white region.
As shown in the Appendix~\ref{appendix:Complementary discussion on NLL and bmax15}, this mismatch becomes even more pronounced for $b_{\sT,\max} \!=\! 1.5 \, \text{GeV}^{-1}$.

One might argue that the deviation from Eq.~\eqref{eq:SPdeviation} is due to the truncation of the perturbative calculation at NNLL. However, the lower panels show that the deviation from zero (red curves) is (almost) always larger than the perturbative uncertainty (black curves). Only in the specific case $Q\!=\!12\,$GeV and for the choice $S_{\rm NP}^{(1g)}\!=\!(1/2)S_{\rm NP}^{(1q)}$, one observes that Eq.~\eqref{eq:SPdeviation} approximately holds. 
To achieve a more systematic cancellation and preserve the (non-)perturbative behaviour in a controlled way, we propose to modify the conventional $b_{\sT}^*$ prescription and accordingly the $b_{\sT}$ dependence of $S_{\rm NP}$, as will be discussed in the next section.

In summary, the present section highlights the limitations of the conventional $b_{\sT}^*$ prescription introduced in Eq.~(\ref{eq:conventional bstar}) together with the conventional $S_{\text{NP}}$ model.
A lower value of $b_{\sT,\max}$ is motivated due to the ‘fast' evolution of gluons, but as $b_{\sT,\max}$ decreases, the extent of the perturbative region is significantly reduced. As a result, the freezing procedure encoded in the standard $b_T^*$ modifies a substantial portion of the perturbative calculation of the Sudakov factor, which is not systematically compensated by $S_{\text{NP}}$. Of course, the observable itself is controlled by $W(b_{\sT})$ in Eq.~\eqref{eq:Cff}, which thus also contains TMDs. 
A full analysis of the interplay of all the ingredients is presented in Appendix~\ref{appendix:Deviation from the perturbative part contribution}, where it is shown that the conclusions drawn here at the level of the Sudakov factor hold as well for the full $W(b_{\sT})$.

\subsubsection{Beyond the conventional $b_{\sT}^*$ prescription}
\label{sec:Beyond the conventional bstar prescription}

As demonstrated in the previous section, the $b_{\sT}^*$ prescription defined in Eq.~(\ref{eq:conventional bstar}) and the particular $b_{\sT}$ dependence adopted for $S_{\text{NP}}$ can lead to an excessively large impact on the perturbative part, diminishing one of the main advantages of the approach, namely the perturbative predictability below $b_{\sT,\max}$.

When $b_{\sT,\text{min}}$ and $b_{\sT,\text{max}}$ are relatively close, which is necessarily the case for low\new{-}scale processes, the perturbative expressions differ significantly, over a substantial portion of their domain of validity, when they are evaluated as functions of $\mu_{b}$ or as functions of $\mu_{b^*}'$.
As a consequence, in the Fourier transform back to $q_T$ space, the perturbative expressions are effectively evaluated at scales that do not correspond to their original $b_{\sT}$ values.
Several approaches have been proposed in the literature to avoid this issue~\cite{Aslan:2024nqg, Gonzalez-Hernandez:2023iso, Gonzalez-Hernandez:2022ifv, Bacchetta:2015ora, Laenen:2000de, Qiu:2000ga, Qiu:2000hf}. 
Here, however, we put forward an alternative approach. 

\paragraph{Prescription for $W^{\rm pert.}$}

There is a simple way to force the perturbative expressions as function of $\mu_{b^*}'$  to coincide with those as a function of $\mu_{b}$ within $b_{\sT,\text{min}}\! \leq \!b_{\sT}\! \leq \!b_{\sT,\text{max}}$ by slightly modifying the prescriptions as follows: 
\begin{equation}
\begin{aligned}
\label{eq:modified b*}
 b_{\sT}^{*} \to b_{\sT}^* (n) & = \frac{b_{\sT}}{\big(1+\left(b_{\sT}/b_{\sT,\text{max}}\right)^{n}\big)^{1/n}} \;  ,\\ 
b_{\sT}' \to b_{\sT}'(m) & = \big(b_{\sT}^{m}+\left( b_0/Q\right)^{m}\big)^{1/m} \; ,
\end{aligned}
\end{equation}
where $n$ and $m$ are large enough.
When $n\!=\!m\!=\!2$, the original prescriptions of Eqs.~(\ref{eq:conventional bstar}) and (\ref{eq:bprime}) are recovered. We remind the reader that we always employ the sequence of Eq.~\eqref{EqbTmethods}.

The larger the value of $n$ and $m$, the smaller the difference between the perturbative expressions as a function of $\mu_{b}$ or as a function of $\mu_{b^*}'$.
Increasing these parameters from $n \!=\! m \!=\! 2$ to $n \!=\! m \!=\! 10$ reduces the maximum relative deviation of $b_{\sT}^{'*}$ with respect to $b_{\sT}$ at the limiting points $b_{\sT,\text{min}}$ and $b_{\sT,\text{max}}$ from about $30\%$ to approximately $6\%$. Larger values further reduce this deviation, although the improvement becomes increasingly marginal, as is shown in Fig.~\ref{fig:n-scan of bTstar and bTdagger} in Appendix~\ref{appendix:Further discussion on bT*(n)}.

In what follows, we will perform the fit using three different approaches. In all three of them, we fix $m \!=\! 10$.
In two of our fits (see Sec.~\ref{sec:classic_fit} and~\ref{sec:Minimising chi2 with regularisation and n2}), we make the conventional choice $n \!=\! 2$. Our third fit, described in Sec.~\ref{sec:Minimising chi2 with regularisation and n10}, is conducted with the alternative prescription $n \!=\! 10$.

We notice that
\begin{equation}
\begin{aligned}
    \lim_{b_{\sT} \to 0} b_{\sT}'(b_{\sT}^*(b_{\sT})) &=b_0/Q \;,
\end{aligned}
\end{equation}
for any $m>0$ and $n>0$.
However, the limit for large $b_{\sT}$ is not exactly $b_{T,{\text{max}}}$,
\begin{equation}
    \lim_{b_{\sT} \to \infty} b_{\sT}'(b_{\sT}^*(b_{\sT})) =
    \left( b_{\sT,\text{max}}^m + \left( \frac{b_0}{Q} \right)^m \right)^{1/m} ,
\end{equation}
for any $m>0$ and $n>0$.
$b_{\sT,\text{max}}$ is only reached when $m$ is large,
\begin{equation}
    \lim_{m \to \infty} \left( b_{\sT,\text{max}}^m + \left( \frac{b_0}{Q} \right)^m \right)^{1/m} = b_{\sT,\text{max}} \; ,
\end{equation}
for $0<b_0/Q<b_{\sT,\text{max}}$.

\paragraph{Prescription for  $S_{\text{NP}}$.}

Correspondingly, $S_{\text{NP}}$ should contribute only in the region $b_{\sT} \!>\! b_{\sT,\text{max}}$ and should depend, in some way, on how the region $b_{\sT}\! \lesssim \!b_{\sT,\text{max}}$ is treated~\cite{Collins:1984kg,Collins:1981va}, i.e., it should be related to the $b_{\sT}^*$ prescription or at least to $b_{\sT, \text{max}}$ and $n$. 
Therefore, we replace the $b_{\sT}$ dependence of $S_{\text{NP}}$ by
\begin{equation}
\label{EqSnpGeneral}
    b_{\sT} \to b_{\sT}^\dagger(n, b_{\sT,\text{max}})  = \sqrt{ (b_{\sT}^n + b_{\sT,\text{max}}^n)^{2/n} - b_{\sT,\text{max}}^2} \; .
\end{equation}
The function $b_{\sT}^\dagger(n, b_{\sT,\text{max}})$ vanishes in the limit $b_{\sT}\! \to\! 0$, and for $n\!=\! 2$ it reduces to the standard $b_{\sT}^2$ behaviour. In particular, the larger the value the of $n$, the smaller $b_{\sT}^(n,b_{\sT,\text{max}})$ inside the perturbative domain, and hence the smaller the contribution from $S_{\text{NP}}\big(b_{\sT}^\dagger\big)$. This is illustrated in Fig.~\ref{fig:n-scan of bTstar and bTdagger} of Appendix~\ref{appendix:Further discussion on bT*(n)}. 
We use the prescription in Eq.~\eqref{EqSnpGeneral} in our third fit (Sec.~\ref{sec:Minimising chi2 with regularisation and n10}) where $n \!=\! 10$. For the other fits, $n\!=\!2$ such that $\exp\!\big[\!-\!S_{\mathrm{NP}}(b_{\sT})^\dagger\big]\overset{n\to2}{=}\exp\!\big[\!-\!S_{\mathrm{NP}}(b_{\sT})\big]$ has the conventional quadratic fall-off.

A full analysis of our approach is presented in Appendix~\ref{appendix:Deviation from the perturbative part contribution}, where we observe a systematic reduction of the deviation of $W^{\rm pert}(b_{\sT}^*) e^{- S_{\rm NP}(b_{\sT})}$ from $W^{\rm pert} (b_{\sT})$ in the perturbative region when $n \!=\! 10$.

\section{Conventional fit without novel theory constraints}\label{sec:classic_fit}
\subsection{Experimental data}

We analyse the self-normalised 
differential cross section of $J/\psi$-pair production via single-parton scattering in different bins of the average invariant mass of the $J/\psi$ pair, namely $\langle M_{QQ} \rangle \!=\! 6.6, \, 7.9$ and $11$~GeV:
\begin{align}
\frac{1}{\sigma}\frac{\mathrm{d}\sigma}{\mathrm{d}\qt} 
&\equiv  
\left(\int^{Q/2}_0 \mathrm{d}\qt \frac{\mathrm{d}\sigma}{\mathrm{d}\qt}\right)^{-1} 
\frac{\mathrm{d}\sigma}{\mathrm{d}\qt}
\,.
\end{align}
When plotted in bins of $\qt$, the expression above is integrated over the corresponding ranges in $\qt$ and divided by their width.
The analysis is based on the $\sqrt{s} \!=\! 13 \,$TeV data collected at LHCb \cite{LHCb:2023ybt} with the following cut on the rapidity of the pair: $2\!<\!y\!<\!4.5$.

Since TMD factorisation is reliable only in the regime where power corrections are negligible, we select the data bins that meet the following criterion:
\begin{equation} \label{eq:cut}
    \frac{ q_{\sT}}{\langle M_{\mathcal{Q} \mathcal{Q}} \rangle} \equiv \delta < 0.5
    \;.
\end{equation}
This choice is \textit{ad hoc} but we test its validity by analysing the theoretical predictions and their comparison with experimental data for values smaller than $\delta$.
We will come back to this later during in the discussion of the fit.
Note that this constraint applies to the average value of each bin.
In total, we fit 13 data points, with their statistical and systematic uncertainties.
We treat the statistical uncertainties as uncorrelated and the systematic ones as correlated.

The measured distributions are normalised to their integrated cross sections as follows: 
\begin{equation}
    \overline{\sigma}_i = \frac{\sigma_i}{\sigma} ,\quad  \sigma = \sum_k \sigma_k \Delta q_{\sT,k} \; ,
\end{equation}
where $\sigma_i$ denotes the measured differential cross section in bin $i$, $\Delta q_{\sT,i}$ is the corresponding bin width, and $\sum_i \overline{\sigma}_i \Delta q_{\sT,i} \!=\! 1$ by construction.
The covariance matrix of the normalised distribution is obtained through standard error propagation,
\begin{equation}
\begin{aligned}
    C_{\rm exp} \to J \, C_{\rm exp} \,J^T, \quad J_{ik}  = \frac{\delta_{ik} \sigma - \sigma_i \Delta q_{\sT,k}}{\sigma^2} \; ,
\end{aligned}
\end{equation}
where $J$ is the corresponding Jacobian matrix.

\subsection{Formalism}

We minimise the chi-squared function defined as
\begin{equation} \label{eq:chi2 exp}
    \chi^2 = \sum_Q \left( F_i(\theta) - D_i \right) \left( C_{\rm exp} \right)^{-1}_{ij} \left( F_j(\theta) - D_j \right) \; ,
\end{equation}
with $F_i$ being the theoretical model, $D_i$ representing the normalised experimental data, $\theta$ being the fit parameters $A$ and $B$ in Eq.~(\ref{eq:SNPsimple}), and $C_{\rm exp}$ the corresponding experimental normalised covariance matrix.
A separate covariance matrix is constructed for each $Q \!=\! \langle M_{QQ} \rangle$, implying that correlations between different $Q$ bins are neglected.
The $\chi^2$ is therefore obtained as the sum of the individual contributions from each $Q $ reported in the data, indicated by the summation over $Q$.
The Latin indices label the bins in $q_{\sT}$.

Although the fit successfully constrains the parameter $A$, no significant constraint can be placed on $B$.
The likelihood remains relatively flat along the $B$ direction, indicating that the available data do not provide sufficient sensitivity to determine its value.
Therefore, in the present fit, which we refer to as ``conventional", we consider $B\!=\!0$, and $A$ as the free parameter.
We anticipate that the introduction of theory constraints in the next sections will lift this degeneracy, enabling a simultaneous determination of both parameters $A$ and $B$.

The calculations are performed at NNLL accuracy, using the MSHT20 PDF set at NLO~\cite{Bailey:2020ooq},
while $\alpha_s$ is obtained from the values stored in the LHAPDF grid, which correspond to $\alpha_s(M_Z) \!=\! 0.118$ for MSHT20.
The number of flavours, $n_f$, changes according to the mass thresholds implemented by the used PDF set. 
The same is done for any other PDF sets.
Moreover, we use the $b_{\sT}^*$ prescription defined in Eq.~(\ref{eq:modified b*}) for $m\!=\!10$ (like everywhere) and  $n\!=\!2$ with $b_{\sT,\max} \!=\! 0.5 \, \rm GeV^{-1}$.
Finally, the minimisation procedure is performed using the \texttt{iMinuit} package \cite{iminuit}.

\subsection{Results}
\label{sec:Minimising chi2}

\begin{figure}[htbp]
    \centering
    \subfloat[]{
        \includegraphics[width=\linewidth]{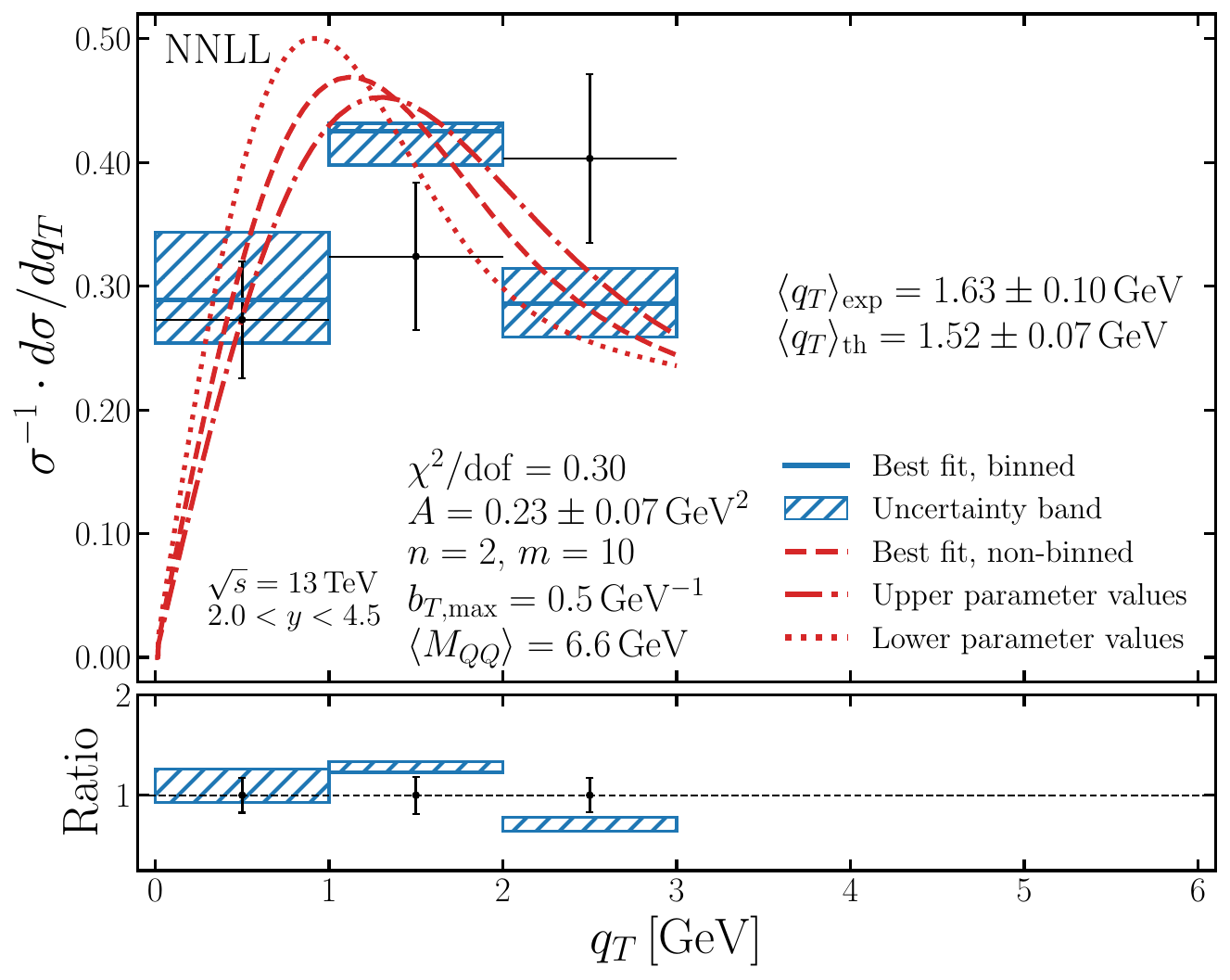}
   }\\
   \subfloat[]{
        \includegraphics[width=\linewidth]{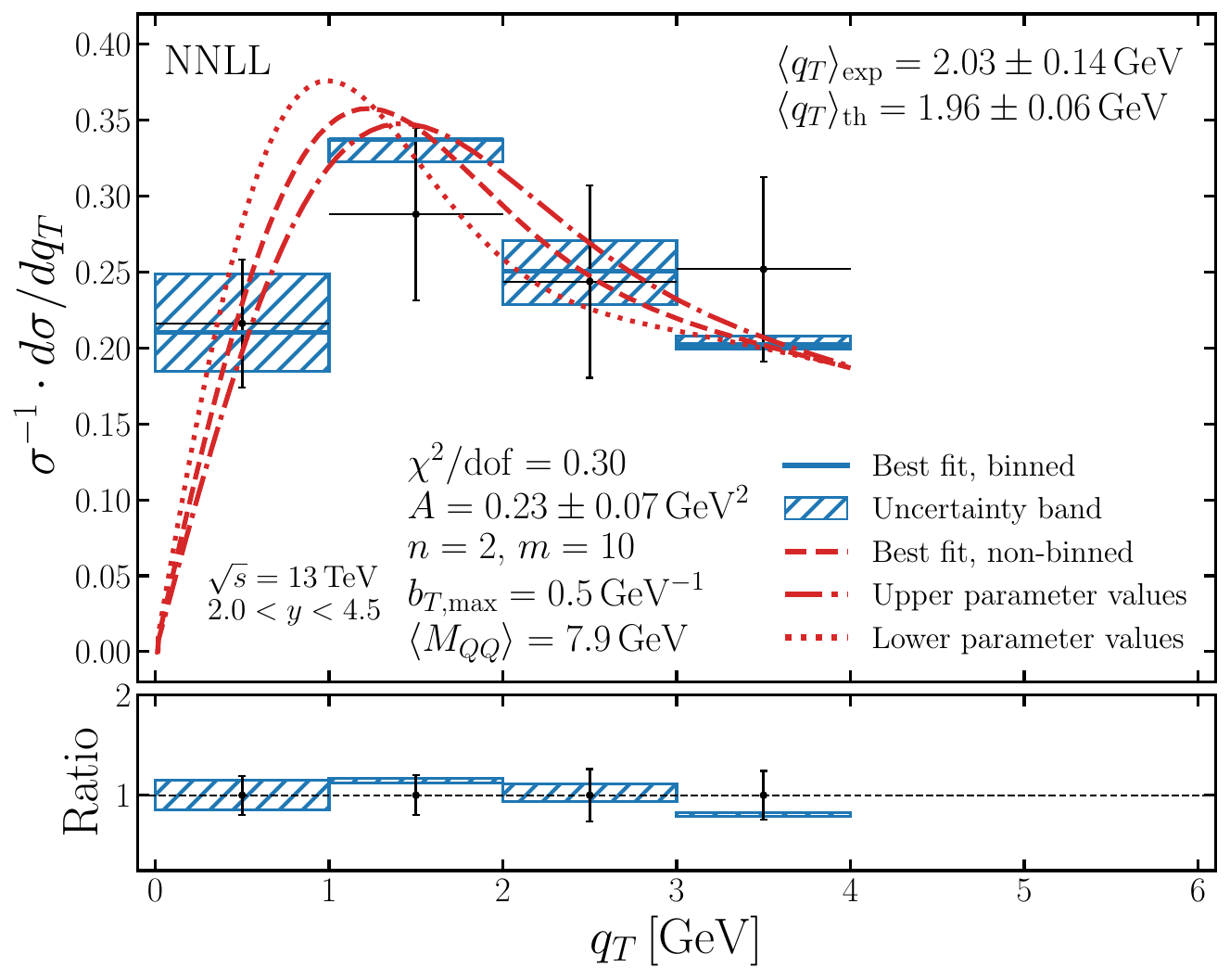}
   }\\\subfloat[]{
        \includegraphics[width=\linewidth]{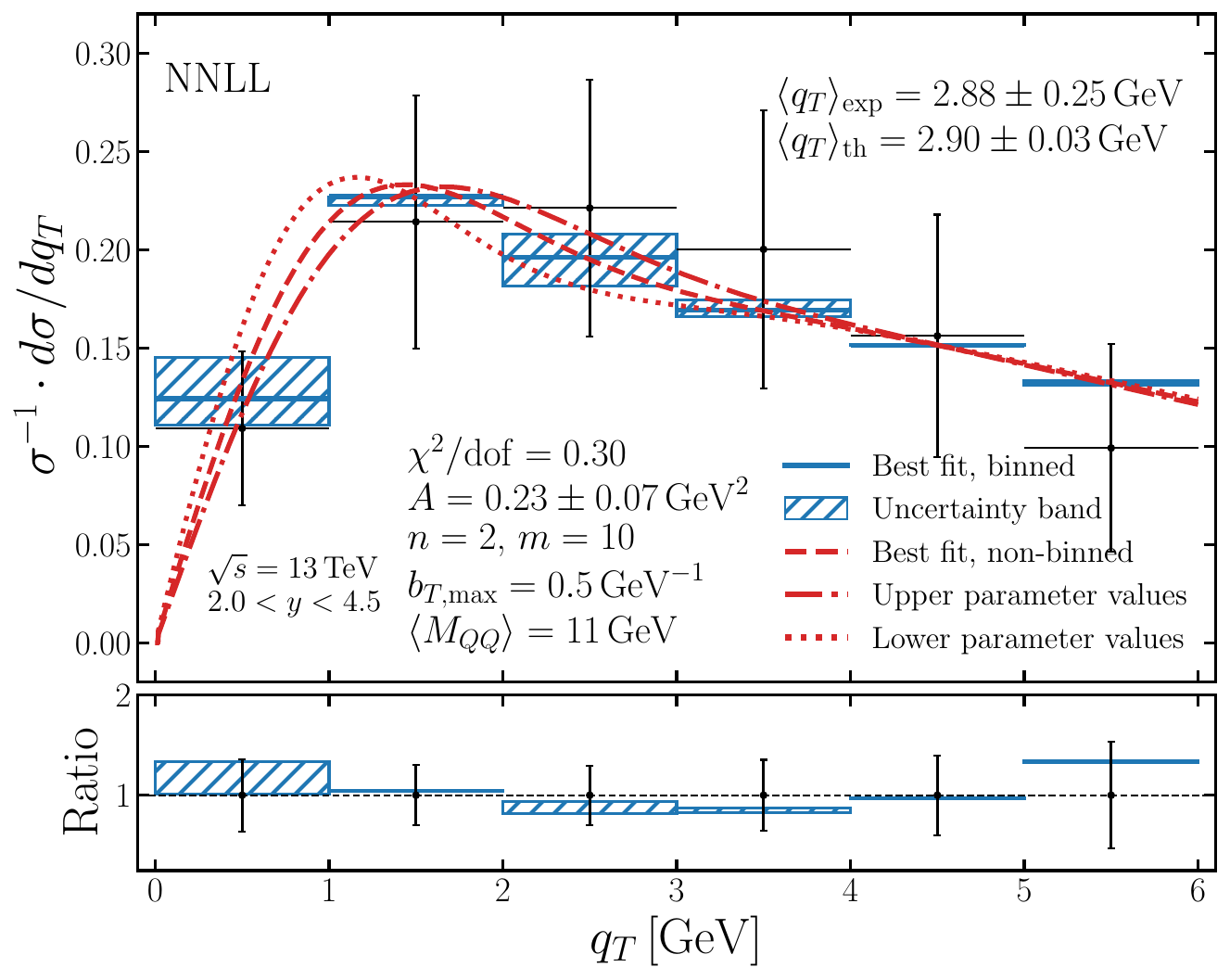}
}
    \caption{
    Conventional fit without theory constraint.
    Upper panels: normalised $q_{\sT}$ distribution of the theoretical and experimental differential cross section for the three bins of $\langle M_{QQ} \rangle$ (a-c).
    The binned and the unbinned theoretical cross sections are shown.
    The uncertainty bands correspond to the error propagation of the fit parameters.
    The average of $q_{\sT}$ is also displayed.
    Lower panels: ratio between experimental data and theoretical results.
    }
    \label{fig:Minimisation Chi2_exp}
\end{figure}

The results are
\begin{equation}
\label{eq:fit result M1 05}
    \chi^2/\text{dof} =  0.30\; , \quad A =0.23 \pm 0.07 \, \text{GeV}^2 \; .
\end{equation}
The resulting theoretical cross sections, together with the experimental data, are shown in the upper panels of Fig.~\ref{fig:Minimisation Chi2_exp}.
In the lower panels, the ratio theory/data is displayed.
The binned uncertainty bands to be compared to data are obtained by propagating the Hessian error of $A$. 
In addition we show three curves (in red) representing the envelope on our unbinned results.
Additionally, we calculate the average of $q_{\sT}$, i.e., $\langle q_{\sT} \rangle$, to be compared with the measured one.

Two observations concerning the numerical result of $A$ are worth being made.
First, in light of the discussion in Sec.~\ref{Systematic effects induced by the conventional modelling of non-perturbative effects}, according to $S_{\text{NP}}^{(1g)} \!\gtrsim\! (1/2)S_{\text{NP}}^{(1q)}$ and using the quark fit results reported in Eq.~(\ref{eq:twoquarkSudakov}) for $b_{\sT,\text{max}}\! =\! 0.5 \, \text{GeV}^{-1}$, one obtains the constraint $A\! \gtrsim\!  0.46, \, 0.51, \, 0.54 \, \text{GeV}^{2}$ for $Q \!=\!11, \, 7.9, \, 6.6 \, \rm GeV$, respectively, so the value reported in Eq.~(\ref{eq:fit result M1 05}) is approximately twice below these values which are already twice smaller than the quark ones, whereas one would expect a larger value. 
Second, following the Appendix~\ref{appendix:Deviation from the perturbative part contribution}, 
we find that the obtained value of $A$ gives rise to a substantial deviation between $W^{\rm pert}(b_{\sT}^*)e^{-S_{\rm NP}(b_{\sT})}$ and $W^{\rm pert}(b_{\sT})$ in the perturbative region, exceeding $100\%$.
These two findings imply that the present fit yields a value of $A$ for which the modification of the perturbative calculation induced by the prescription used to account for non-perturbative effects is excessively large.

It should be noticed that the fit favours this behaviour because it provides a good  description of the first $q_{\sT}$ bin for all $Q$ values, as can be seen from Fig.~\ref{fig:Minimisation Chi2_exp}.
As a consequence, 
the unbinned theoretical cross sections (red curves) display a behaviour that is not characteristic of a physical cross section, featuring a pronounced peak at small $q_{\sT}$. 
As explained in detail in Appendix~\ref{appendix:oscillations in qT}, the oscillations in $q_{\sT}$ arise from a change in curvature of the distribution in $b_{\sT}$ space, which generates a second peak with significant amplitude, giving rise to a tail that extends to large values of $b_{\sT}\!\sim\! 10 \, \text{GeV}^{-1}$.
Therefore, the resulting distribution is dominated by the peak of $S_{\rm NP}$.
This behaviour becomes more pronounced as the value of $A$ decreases.
It can also be seen that as $A$ increases (or equivalently, as the $S_{\text{NP}}$ gets steeper), the theoretical curve moves away from the first experimental $q_{\sT}$ bin.

We have explored larger values of $b_{\sT,\max}$.
At $b_{\sT,\max} \!=\! 0.75 \, \rm GeV^{-1}$, the extracted $A$ decreases ($A \!=\! 0.07 \!\pm\! 0.04\,\rm GeV^2$), and the unbinned distributions for the three Q values exhibit an even more pronounced behaviour with a peak around $q_{\sT} \!\simeq\! 0.5 \, \rm GeV$.
At $b_{\sT,\max} \!= \!1.5 \, \rm GeV^{-1}$,
we find that the observable is insensitive to the value of $A$ (or, more generally, to the non-perturbative parameters in $S_{\rm NP}$).
See Appendix~\ref{sec:Comparison with LHCb data at N$^2$LL} for a comprehensive analysis.
This behaviour is expected, as it was already anticipated by the analysis in $b_{\sT}$ space presented in Sec.~\ref{Systematic effects induced by the conventional modelling of non-perturbative effects} and further discussed in Appendix~\ref{appendix:selection of bmax}.

We have also investigated the dependence on the choice of the PDF set by repeating the minimisation using the NLO CT18~\cite{Yan:2022pzl}, NLO NNPDF40~\cite{NNPDF:2021njg} and NLO MSTW2008~\cite{Martin:2009iq} sets, obtaining qualitatively similar outcomes.
Furthermore, we have verified that the behaviour obtained in this first fit remains after a variation of the selection cut defined in Eq.~(\ref{eq:cut}) from $\delta \!<\! 0.5$ to $\delta\! <\! 0.25$.

Motivated by the above results, in the following sections we repeat the fit using two alternative novel minimisation procedures.
First, within the conventional approach ($n \!=\! 2$), we inject theory constraints in $\bt$ space derived from the perturbative uncertainties.
Second, using the approach proposed in this work with $ n\! =\! 10$, we impose theory constraints in $\qt$ space to avoid the presence of unphysical oscillations.

\section{Fit with theory constraints in \texorpdfstring{$\bt$}{bT} space}\label{bT_fit}
\label{sec:Minimising chi2 with regularisation and n2}

\subsection{Formalism}
In this method, we account for the uncertainty associated with truncating the perturbative calculations at a determined order in $\alpha_s$ 
in Eq.~(\ref{eq:Sudakovdef}) and in Eq.~(\ref{eq:OPE f1g}).
To estimate this uncertainty, we perform a $9$-point %\Remark{PT}{9 or 7?} \answer{SR}{9} 
scale variation by varying both the renormalisation and rapidity scales by a factor of two.

Following the notation of Eq.~(\ref{eq:chi2 exp}), we penalise values of the non-perturbative parameters $A$ and $B$ that produce an $F(\theta)$ in $b_{\sT}$ space lying outside the perturbative uncertainty band in the perturbative region. We remind the reader that we use $n \!= \!2$ in this fit.
In this way, we quantify the extent to which the deformation induced by the 
implementation of the $b_{\sT}^*$ prescription combined with the $S_{\rm NP}$ can be interpreted as a perturbative deformation.

Particularly, in this scenario, we now minimise the following quantity
\begin{equation}
\label{eq:minimisation chi2 + Dpert}
    \chi^2_{b_{\sT}} \equiv \chi^2 + D_{\rm pert} \; ,
\end{equation}
with
\begin{equation}
\label{eq:chi2 th}
    D_{\rm pert} = \sum_Q
    \left( \tilde{f}_i(\theta) - T_i^{0} \right)
    (C_{\rm th})^{-1}_{ij}
    \left( \tilde{f}_j(\theta) - T_j^{0} \right) \, .
\end{equation}
Here, $\tilde{f}(\theta)$ denotes the theoretical prediction in $b_{\sT}$ space corresponding to
$W^{\rm pert}(b^{'*}_{\sT})\exp\big[\!-\!S_{\mathrm{NP}}\big]$ and $T^{0}$ represents the theoretical prediction for $W^{\rm pert}(b_{\sT})$. 
The Latin indices label the points in $b_{\sT}$.
$C_{\rm th}$ is the theoretical covariance matrix obtained from scale variation in the perturbative region $b_{\sT} \!\in\! [b_0/Q, b_{\sT,\text{max}}]$.
Note that $D_{\rm pert}$ should not be interpreted as a statistical chi-squared distribution, but rather as a measure of the deformation of the perturbative prediction relative to the uncertainty estimated from scale variations.\footnote{
We remark that, in Eq.~(\ref{eq:chi2 th}), $\tilde{f}(\theta \!=\! 0)$ is not equal to $T^{0}$ due to the $b_{\sT}^*$ prescription, which changes the scale dependence of the former.}  The differences between $\tilde{f}(\theta \!=\! 0)$ and $T^{0}$ amount to $30\%$ on average.

The theoretical covariance matrix is given by
\begin{equation}
\begin{aligned}
\label{eq:covmat}
    (C_{\rm th})_{ij} & = \frac{1}{N_k - 1}
    \sum_{k = 1}^{N_k} 
    \left( T_i^{(k)} - T_i^{0} \right)
    \left( T_j^{(k)} - T_j^{0} \right) \\
    & + \frac{1}{4} \sum_{l = 1}^{(N_l-1)/2}
    \left( T^{(l,+)}_i - T^{(l,-)}_i \right)
    \left( T^{(l,+)}_j - T^{(l,-)}_j \right) \,.
\end{aligned}
\end{equation}
Here, $T^{(k)}$ denotes the $k^\text{th}$ scale variation of $W^{\rm pert}(b_T^{*'})e^{-S_{\rm NP}}$ and $N_k$ is the number of these scale variations.
Being consistent with $\chi^2$, we also assume no correlation among data sets for different $Q$ values, so there are three $C_{\rm th}$'s.
A discussion on the inversion of $C_{\rm th}$ can be found in Appendix~\ref{appendix: On the regularisation methods}.
In the second line we incorporate the PDF uncertainty.
Here, $T^{(l,+)}$ and $T^{(l,-)}$ denote $W^{\rm pert}(b_T^{*'})e^{-S_{\rm NP}}$ evaluated with the $l^\text{th}$ $+$ and $-$ members of the Hessian PDF set, respectively, with $N_l$ denoting the total number of PDF members.

Once again, we stress that $\tilde{F}(\theta)$ is computed using $b_{\sT}^*(n=2)$ and $S_{\text{NP}}(\theta)$ while $T^{(k)}$ and $T^{0}$ correspond to purely perturbative calculations, with neither a $b_{\sT}^*$ prescription nor any $S_{\text{NP}}$ influence. 
We recall that $m\!=\!10$ is kept fixed throughout this work, so the contribution of $b_{\sT}'$ remains unchanged in all cases.
In this way, the non-perturbative parameter space spanned by $A$ and $B$ is restricted by the perturbative uncertainty estimated in the perturbative domain.

The fit is performed at NNLL and the parameters are obtained using a Monte-Carlo-replica procedure. For each replica, pseudo-data sets are generated by sampling multivariate Gaussian distributions centred on the experimental measurements and theoretical predictions, using their respective covariance matrices. 
The $\chi^2_{b_{\sT}}$ function is then
minimised independently for each replica. 
The central values of the fit parameters are taken as the mean of the parameter distributions obtained from all convergent minimisations, while the associated uncertainties are determined from their standard deviations.

\begin{figure}[hbtp]
    \centering
   \subfloat[]{
        \includegraphics[width=0.98\linewidth]{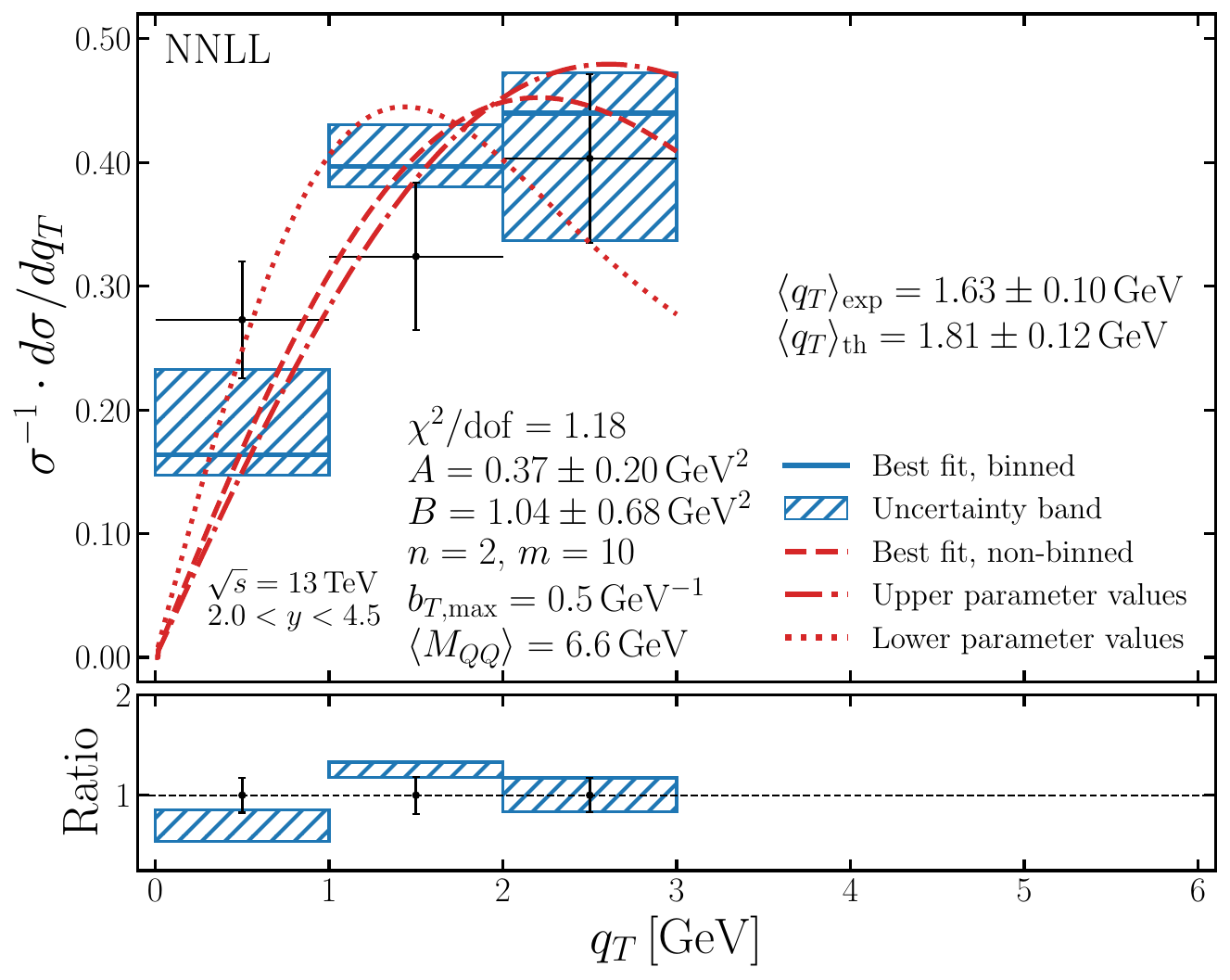}
    }\\ \subfloat[]{
        \includegraphics[width=0.98\linewidth]{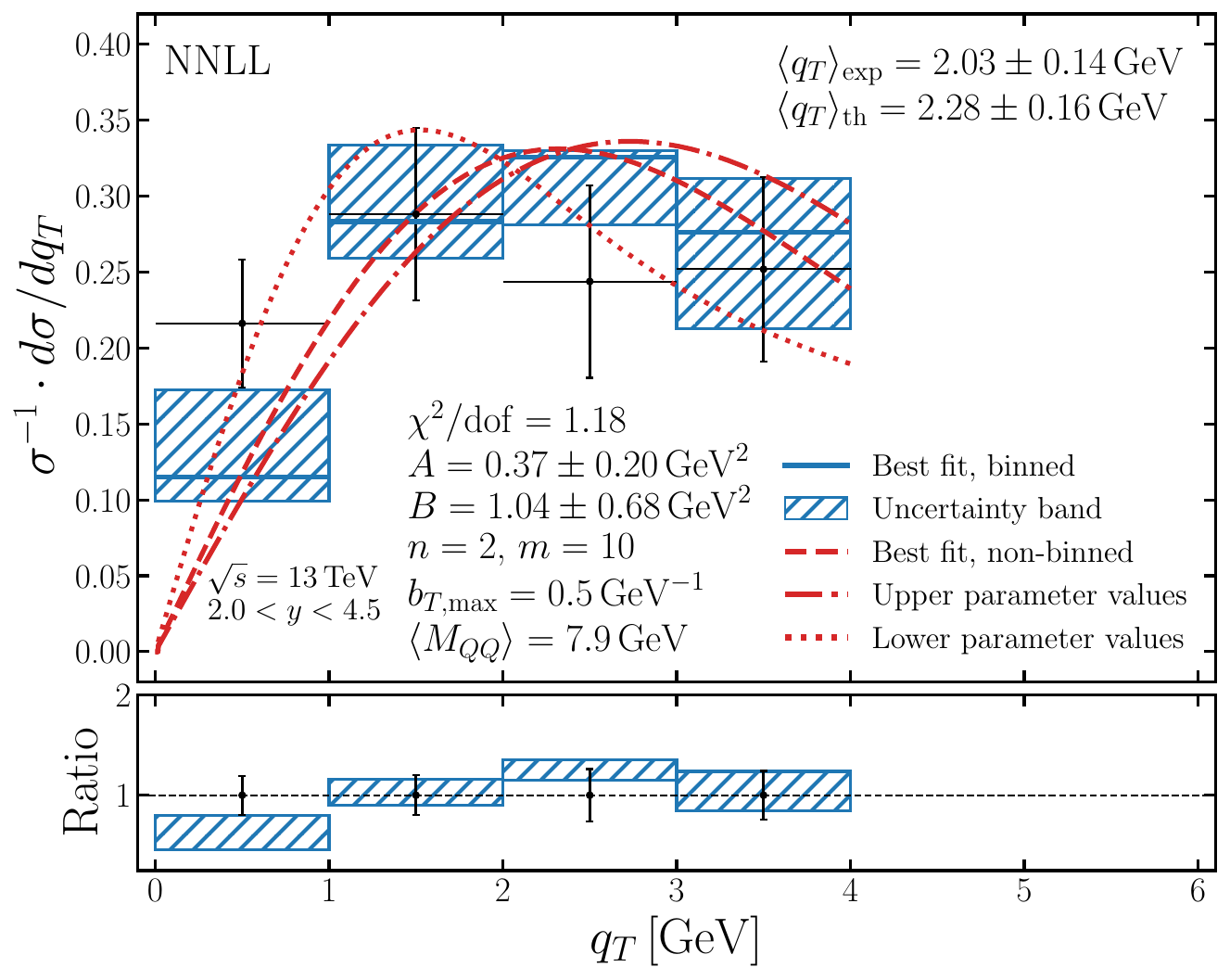}
    }\\ \subfloat[]{
        \includegraphics[width=0.98\linewidth]{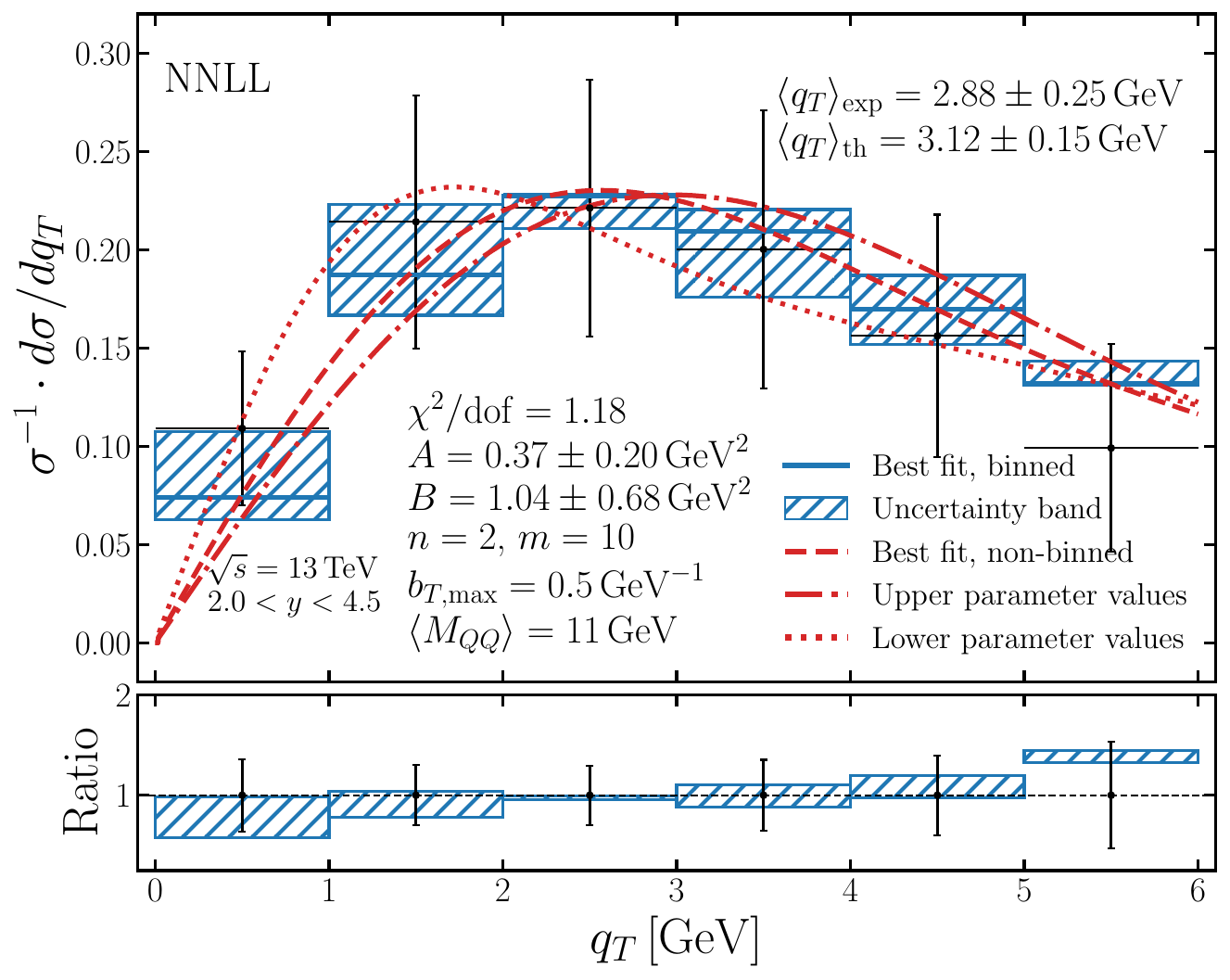}
   }
    \caption{
Same as \cf{fig:Minimisation Chi2_exp} for our fit with theory constraint in $b_{\sT}$ space.
    }
    \label{fig:Minimisation Chi2_exp + regularisation n = 2}
\end{figure}

\subsection{Results}

The results of the fit are
\begin{equation}
\begin{aligned}
\label{eq:MSHT20 LO 05}
    & \chi^2/\text{dof} = 1.18 \; ,\\
    & A =   0.37 \pm 0.20 \,\text{GeV}^2 \;, \\
    & B =  1.04 \pm 0.68 \, \text{GeV}^2 \; ,
\end{aligned}
\end{equation}
and the corresponding $q_{\sT}$ distributions are shown in Fig.~\ref{fig:Minimisation Chi2_exp + regularisation n = 2}.
{
Compared to Fig.~\ref{fig:Minimisation Chi2_exp}, the peaks of the theoretical cross sections are shifted towards larger $q_{\sT}$ values.
This behaviour is a consequence of imposing the theory-uncertainty constraints, which lead to an overall larger $S_\text{NP}$ and hence a faster decay with increasing $b_\sT$.
As a result, the distribution in $b_{\sT}$ is no longer dominated by the peak of $S_{\mathrm{NP}}$. Furthermore, the alteration of $W^{\rm pert}$ induced by $S_\text{NP}$ is found to be substantially reduced with respect to the result obtained using the conventional fit, and within the range of the perturbative uncertainties.

We find that penalising fit parameters yielding deviations which exceed the perturbative uncertainty in the perturbative region effectively overcomes the limitations discussed in Sec.~\ref{sec:Minimising chi2}.
This procedure yields an improved statistical quality of the fit, with a resulting value of the fit parameters that produces a cross section free from oscillatory behaviour. We further discuss the resulting TMD distributions in Sec.~\ref{discussion}.

We note that such a procedure can, and probably should, be applied to any TMD fit in general as it ensures that the $\bt^*$ prescription does not arbitrarily alter the perturbative input injected in the fit. 
Obviously, our method to include the theory uncertainties does not affect the outcome of any fitting procedure that is compliant with the perturbative uncertainties to begin with.

\section{Fit with theory constraints in \texorpdfstring{$q_{\sT}$}{qT} space}\label{qT_fit}
\label{sec:Minimising chi2 with regularisation and n10}

\subsection{Formalism}

\begin{figure}[hbtp]
    \centering
   \subfloat[]{
        \includegraphics[width=0.98\linewidth]{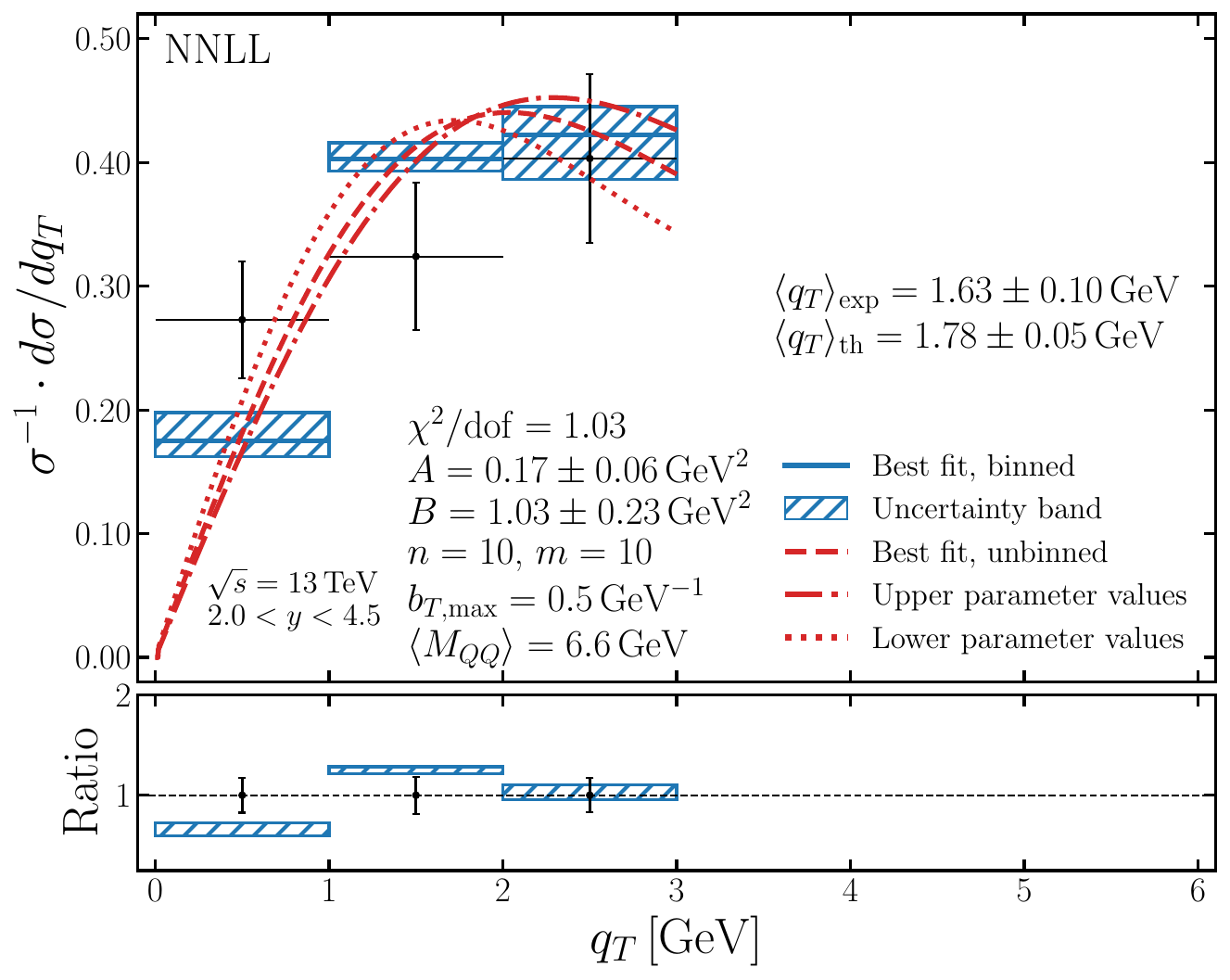}
    }\\
 \subfloat[]{
        \includegraphics[width=0.98\linewidth]{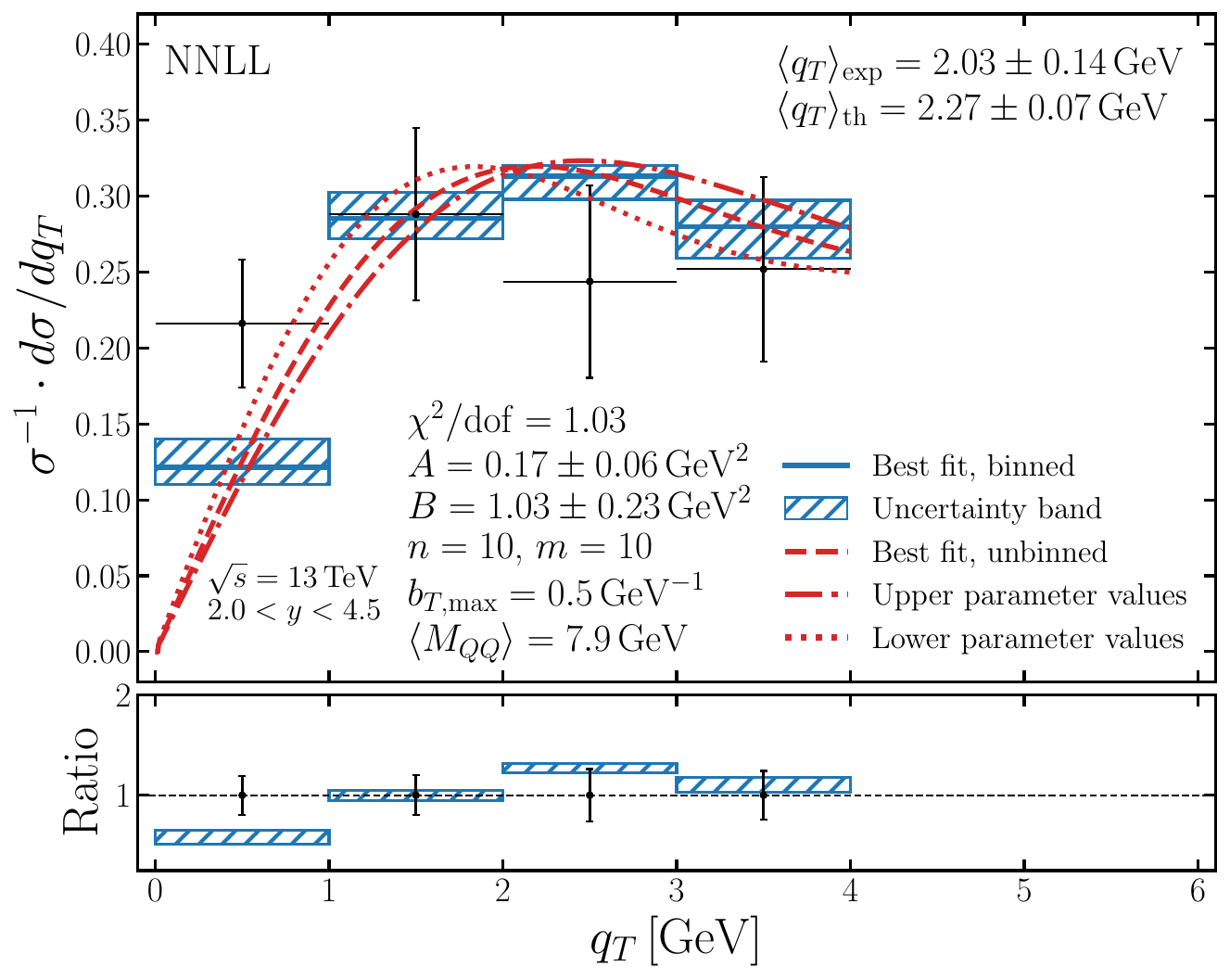}
   }\\
   \subfloat[]{
        \includegraphics[width=0.98\linewidth]{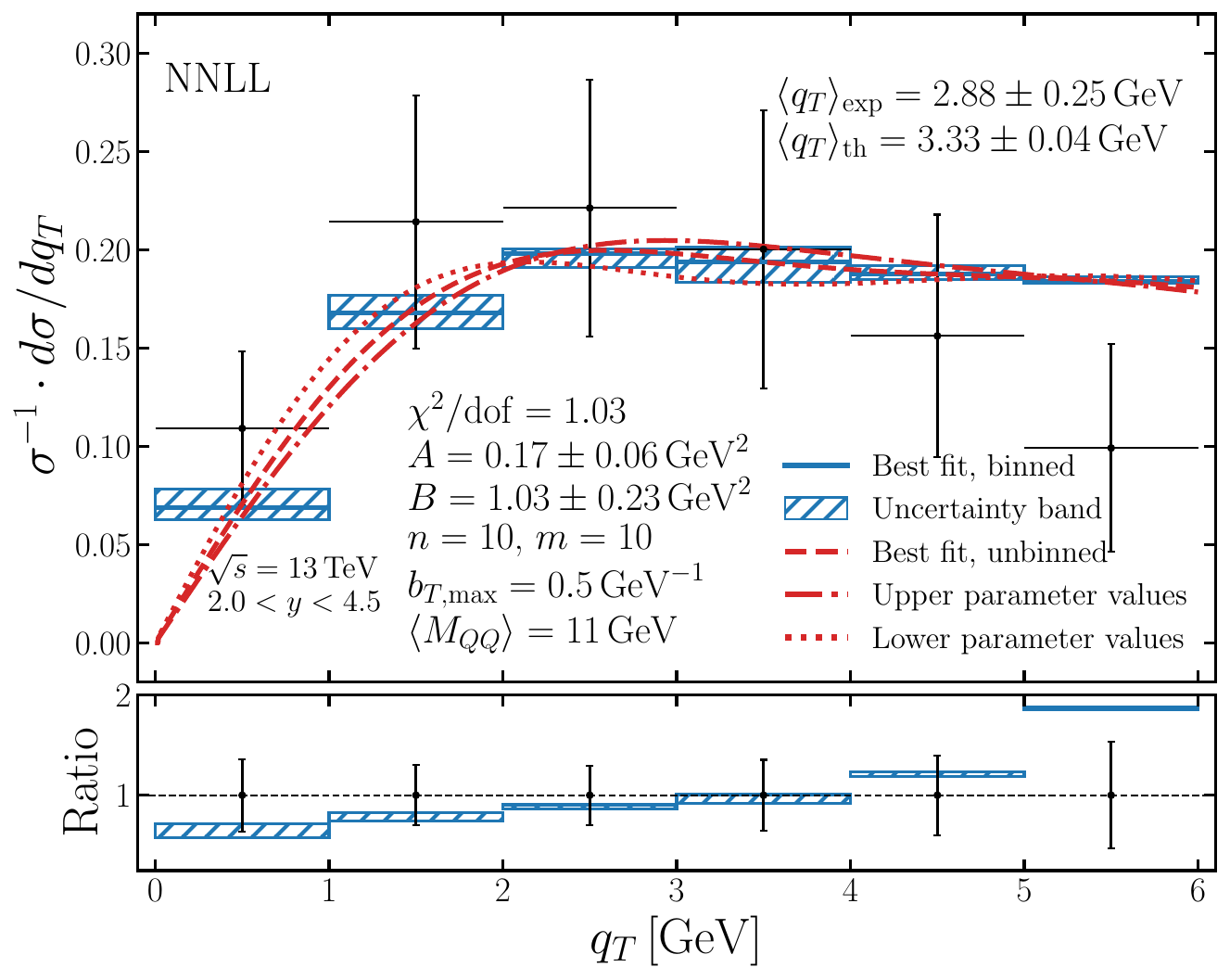}}
    \caption{
    Same as \cf{fig:Minimisation Chi2_exp} for our fit with theory constraint in $q_{\sT}$ space.
    }
    \label{fig:Minimisation Chi2_exp + regularisation, n=10}
\end{figure}

In this method, we perform the fit when $n \!=\! 10$ in Eq.~(\ref{eq:modified b*}) and in Eq.~(\ref{EqSnpGeneral}), i.e., we consider the scenario where the influence of the non-perturbative parameters into the perturbative part is minimised.
However, as already discussed, the oscillatory behaviour displayed in Fig.~\ref{fig:Minimisation Chi2_exp} and in Appendix~\ref{appendix:oscillations in qT} becomes more pronounced as $n$ increases.
To address this issue, we introduce a penalty on the non-perturbative parameters that give rise to oscillations in the physical $\qt$ space.

In particular, we constrain the fit by adding a penalty on the curvature of the $q_{\sT}$ distribution of the observable as follows
\begin{equation}
\label{eq:chi2_exp + regularisation}
    \chi^2_{q_{\sT}} \equiv \chi^2 + \lambda \int \mathrm{d}q_{\sT}
    \left[
    \max \left\{ 0,\left( \frac{1}{\sigma} \frac{\mathrm{d} \sigma}{\mathrm{d}q_{\sT}} \right)^{''} \right\}
    \right]^2 \; ,
\end{equation}
where $\chi^2$ is defined in Eq.~(\ref{eq:chi2 exp}), and where $\lambda$ is an arbitrarily large parameter enforcing that its accompanying factor is effectively vanishing when $\chi^2_{q_{\sT}}$ gets minimised.
The upper limit of the integral depends on the hard scale at which the penalty is applied.
The distribution becomes more oscillatory at lower hard scales. Hence, in order to ensure that the resulting parameter from the minimisation is as universal as possible, we decide to apply the penalty at $Q \!=\! 3$~GeV in the range $q_{\sT} \!=\! [0,3]$~GeV.\footnote{We note that corrections to the TMD cross section scaling like $\qt/Q$ appear outside ${\cal C}[ff]$ which is the object affected the curvature condition. This is why we find it justified to apply to condition up to $\qt\!=\!Q$ and not $\qt\!=\!Q/2$. In fact, nothing prevents to add additional universal theory constraints in $\qt$ space, at a different scale for instance, as long as no data taken in the corresponding phase space points are injected in the fit.} 
Note that this regularisation does not penalise the curvature itself; rather, it softly penalises negative curvature, which is not expected at low $\qt$.
To be realistic, the curvature has been computed from a cross section binned with a bin width of $0.125$~GeV, as if a measurement at $3$ GeV was performed.

In this scenario, we adopt a Monte-Carlo-replica approach, performing the calculation with 200 replicas.
Our replica generation follows Ref.~\cite{Kusina:2016fxy}, and is given by
\begin{equation}
\begin{aligned}
\label{eq:one replica}
    f_k & = f_0 \\
    & + \frac{f(2,1) - f(0.5,1)}{2} R^k_\mu + \frac{f(1,2) - f(1,0.5)}{2} R^k_\zeta \\
    & + \sum_{l = 1}^{(N_l-1)/2} \frac{f^{(l,+)} - f^{(l,-)}}{2} R_l^k
\end{aligned}
\end{equation}
where the superscript $k$ denotes the index of the replica. 
The factors $R$ are random numbers following the normal distribution $\mathcal{N}(0,1)$, which are different for each replica and for the three error sources.
The second line represents the perturbative uncertainty, while the third line represents the PDF uncertainty.
In particular, the function $f(c_\mu, c_\zeta)$ denotes $W^{\rm pert}(b_T^{*'})e^{-S_{\rm NP}}$, with $c_\mu$ and $c_\zeta$ representing the scale variation of the renormalisation and the rapidity scales, respectively.
On the other hand, the function $f^{(l,+/-)}$ corresponds to $W^{\rm pert}(b_T^{*'})e^{-S_{\rm NP}}$ calculated with the $l$-$(+/-)$ pair with $N_l$ the number of members of the corresponding PDF set.
In the first line, $f_0$ corresponds to the default $W^{\rm pert}(b_T^{*'})e^{-S_{\rm NP}}$ evaluated without scale variation and using the central member of the PDF set.
Moreover, we consider the same theoretical fluctuation for different values of $Q$, i.e., the $R$ factors do not depend on $Q$.

Once the fits for $N_{\rm rep}=200$ replicas on the experimental data have been performed, we calculate the weight for each one,
\begin{equation}
    w_k = \frac{e^{- \chi^2_k/2}}{\frac{1}{N_{\rm rep}} \sum_i^{N_{\rm rep}} e^{-\chi^2_i/2}} \; ,
\end{equation}
where $\chi^2_{k}$ is  the value of $\chi^2$ corresponding to the large-$\lambda$ asymptotic minimum of $\chi^2_{q_{\sT}}$  for  the $k^\text{th}$ replica. Doing so, the method effectively accounts for  (i) the perturbative and PDF uncertainties via the initial spread of the replicas, (ii) the absence of negative curvature via the $\lambda$ term when  $\chi^2_{q_{\sT}}$ is minimised and (iii) the experimental constraints from the $J/\psi$-pair cross section data via the re-weighting of the initial replicas using the value of $\chi^2$.

The effectiveness of the reweighting procedure can be estimated by computing the effective number of replicas, $N_{\rm eff}$, defined as,
\begin{equation}
    N_{\rm eff} = \exp \left[ \frac{1}{N_{\rm rep}} \sum^{N_{\rm rep}}_{k = 1}
    w_k \ln(N_{\rm rep}/w_k)\right]\;.
\end{equation}
This parameter indicates how many of the replica sets are effectively contributing to the reweighting procedure.
We obtain $N_{\rm eff}/N_{\rm rep}\!=\!0.47$.
This indicates a moderate reduction in the statistical ensemble, implying that the imposed constraint carries relevant information but remains broadly compatible with the prior distribution.

Finally, to compute the final reweighted results we use the following standard definition:
\begin{equation}
\begin{aligned}
    \langle \mathcal{O} \rangle & = \frac{1}{N_{\rm rep}} \sum^{N_{\rm rep}}_{k=1} w_k \mathcal{O}(f_k) \; ,\\
    \delta \langle \mathcal{O} \rangle & =
    \sqrt{
    \frac{1}{N_{\rm rep}} \sum_{k=1}^{N_{\rm rep}} \left(\mathcal{O}(f_k) - \langle \mathcal{O} \rangle \right)^2
    } \; .
\end{aligned}
\end{equation}
To determine the optimal asymptotic value of $\lambda$, we have evaluated $\langle \chi^2 \rangle$ over the range $\lambda \!=\! [10^{-2}, 10^7]$.
For $\lambda \!\lesssim\! 10$, the constraint is effectively inactive, and the fitted cross sections remain oscillatory.
As $\lambda$ increases, the constraint progressively suppresses the oscillatory behaviour, producing a rapid variation of $\langle \chi^2 \rangle$ until a plateau is reached for $\lambda \!\gtrsim\! 10^6$.
We therefore adopt $\lambda\! =\! 10^7$, where the fit is fully numerically stable.

\subsection{Results}

The results are
\begin{equation}
\begin{aligned}
    & \langle \chi^2 \rangle/\text{dof} =  1.03\, ,\\
    & A = \langle A \rangle \pm \delta \langle A \rangle =  0.17 \pm  0.06\, \text{GeV}^2 \; ,\\
    & B = \langle B \rangle \pm \delta \langle B \rangle =  1.03 \pm  0.23\, \text{GeV}^2 \; ,
\end{aligned}
\end{equation}
and the corresponding distributions in $q_{\sT}$ space are shown in Fig.~\ref{fig:Minimisation Chi2_exp + regularisation, n=10}.
These values are compatible with the results obtained in Eq.~(\ref{eq:MSHT20 LO 05}). 
We observe that the uncertainties of the parameters obtained by constraining in $b_{\sT}$ space are significantly larger than those obtained in the present section.
This is because the constraint imposed in $q_{\sT}$ space directly tackles the oscillations, and consequently is more stringent than the corresponding constraint defined in $b_{\sT}$ space.
The compatibility of the results obtained with the two methods demonstrates that both regularisation procedures are well founded and effectively address the problem identified in Sec.~\ref{sec:Minimising chi2}.

\section{Discussion}\label{discussion}
\subsection{Fit approaches}

In Sec.~\ref{sec:classic_fit} we found that the $\chi^2$ minimisation for the low-scale gluonic process considered here is deeply intertwined with the modification of $W^{\rm pert}$ beyond what can be sensibly tolerated from the perturbative uncertainties.
As a consequence, the resulting $q_{\sT}$ distributions do not exhibit a realistic behaviour.
We have proposed two complementary solutions to this issue, both based on additional theory constraints applied to the $\chi^2$ minimisation.
In the first method, described in Sec.~\ref{sec:Minimising chi2 with regularisation and n2}, we  keep the conventional smooth-$b_{\sT}^*$ prescription with $n\!=\!2$ with a modified $\chi^2$ minimisation accounting for the perturbative uncertainty below $b_{\sT,\max}$.
In the second method, Sec.~\ref{sec:Minimising chi2 with regularisation and n10}, we introduce a sharper $\bt^*$ prescription with $n\!=\!10$ for a cleaner separation between perturbative and non-perturbative inputs in $\bt$ space. This is complemented by a constraint in the physical $q_{\sT}$ space, penalising oscillations.
Interestingly, the results from both fits are compatible and, by construction, the modifications to $W^\text{pert}$ induced by the fit $S_\text{NP}$ are now systematically consistent with the perturbative uncertainties.

\subsection{PDF inputs}

Since the $J/\psi$-pair production data that we include in our fit probe a region in $x$ and $\mu$ where the differences among PDF sets are large, we have accounted for the PDF uncertainty directly in both fits as part of the theory uncertainties.
Indeed, we find that the dependence of the fitted parameters on the choice of the PDF set is not negligible.
We stress that the fit results obtained with different NLO PDF sets
(MSHT20~\cite{Bailey:2020ooq}, PDF4LHC15~\cite{Butterworth:2015oua} and CT18~\cite{Yan:2022pzl}) are mutually consistent \emph{only when their uncertainties are accounted for.} As a consequence, we believe that using only the  central PDF eigensets alone is to be avoided.
Note that we have performed our fits using Hessian PDF sets, but both our methodologies are equally applicable to Monte Carlo PDF sets (see e.g.~\cite{NNPDF:2021njg}), with the corresponding treatment of PDF uncertainties.

\subsection{$x$ dependence of $S_{\rm NP}$}\label{subsec:xdependence}

We did not attempt to fit the $x$ dependence of the gluon TMD PDF contribution to $S_{\rm NP}$, since the latest LHCb data~\cite{LHCb:2023ybt} are not  precise enough to show a possible pair-rapidity $y$ dependence within the LHCb acceptance, namely from 2 to 4.5.

We recall that the $x$ dependence of the gluon $S_{\rm NP}$ is expected to appear in $B\!\equiv\! B(x)$ as defined in Eq.~(\ref{eq:SNPsimple}).
As a numerical exercise, one can estimate the expected variation of $B(x_i)$ for a single gluon TMD induced by the variation of $x_i$ within the LHCb acceptance under the assumption of an $x_i$ dependence analogous to that observed for quarks in Eq.~(\ref{eq:B(x)}), thus parametrised as $B(x_i) \!=\! b_1 \ln (10x_i) \!+\! b_2$, and without applying any Casimir scaling for such a back-of-the-envelope calculation.
Since the difference $B(x_{i,\rm min})\! -\! B(x_{i,\rm max})\! =\! b_1 \ln(x_{i,\rm min} / x_{i,\rm max})$ depends only on the ratio of the $x_i$ values, the total variation over the LHCb range $y \!=\! [2,4.5]$ reduces to $\Delta B \!=\! b_1 \Delta y\! =\! 2.5 \, b_1$.
Using the slope obtained from the quark parametrisation from Ref.~\cite{Landry:2002ix}, namely  $b_1 \!=\! -0.126$ (for $Q_{\rm NP}\! =\! 1 \, \rm GeV\,$ and $b_{\sT,\rm max} \!=\! 0.5 \, \rm GeV^{-1}$),  this yields $\Delta B \!=\! -0.32 \, \rm GeV^2$.
Compared to the uncertainty on $B$ in the $b_{\sT}$-constrained fit from Eq.~\eqref{eq:MSHT20 LO 05}, this corresponds to $|\Delta B|\! =\! 0.44 \!\times\! \sigma_B$, where $\sigma_B\!=\!0.68$ is the standard deviation of the fitted value. Likewise, for the $q_{\sT}$-constrained fit, which yields a smaller uncertainty $\sigma_B \!=\! 0.10$, the same variation corresponds to $|\Delta B| \!=\! 1.4 \times \sigma_B$. One can conclude that the data could show some sensitivity to constrain $B(x_i)$.

However, the cross section depends on the convolution of {\it two} gluon TMDs, not one.
When considering ${\cal C} [ff]$, the $x_i$ dependences in $B(x_i)$ of each TMD collapse to a function of $x_1 x_2\!=\!Q^2/s$  with $\Delta B \!=\! b_1 \, \ln(Q_{\rm min}^2/Q_{\rm max}^2)$,  
which does not depend on the rapidity anymore, but rather on the hard scale.
In practice, for $Q_{\rm min}\! =\! 6.6 \, \rm GeV$ and $Q_{\rm max} \!=\! 11 \, \rm GeV$, we find $|\Delta B|\!=\! 0.43\!\times\! \sigma_B$ for the $b_{\sT}$-constrained fit, and $|\Delta B| \!=\! 1.3 \!\times\!\sigma_B$ for the $q_{\sT}$-constrained fit.

Above, we have reported on a joint fit of $A$ and $B$ with the assumption that $B$ does not depend on any $x_i$.
However, as we have just seen, if $B(x_i)$ indeed follows a form like $ b_1 \ln (10x_i) + b_2$, our extraction of $A$, namely of the coefficient of $\ln(Q^2/Q^2_0)$ in $S_{\text{NP}}$, will also contain the $b_1$ term which has a different physical origin. At this stage, we have no reason to believe that such a term is indicated to parametrise gluon TMDs as opposed to quark TMDs. 

To distinguish contributions scaling like $\ln Q^2$ from the evolution, such as $A$, from those like $b_1$ in which the hard-scale dependence comes from the combination of the momentum fractions, measurements at various collision energies should be performed.
In fact, $J/\psi$-pair production was studied at different collision energies: the first (out of three) LHCb analysis~\cite{LHCb:2011kri} as well as the only one by CMS~\cite{CMS:2014cmt} were carried out at $7$~TeV, while that of ATLAS~\cite{ATLAS:2016ydt} was at $8$ TeV rather than $13$ TeV. 
However, double-differential distributions in $\qt$ and $Q$ for these measurements were not reported. The same holds for the CMS $\Upsilon$-pair measurements at $8$ TeV~\cite{CMS:2016liw} and $13$ TeV~\cite{CMS:2020qwa}. Another very interesting option is to measure $\psi$-pair production with LHCb in the fixed-target mode at $\sqrt{s} \simeq 115$~GeV~\cite{Brodsky:2012vg} similarly to the single $J/\psi$ measurements~\cite{LHCb:2022sxs}. It has the additional advantage that in this energy range, the yield should be dominated by SPS~\cite{Lansberg:2015lva}. 

Of course, measurements at larger invariant mass will provide a more robust lever arm to disentangle the $A$ and $B$ terms. Clearly, CMS and ATLAS can do that by extending their  measurements mentioned above.
Lastly, LHCb in the fixed-target mode should be able to provide $\eta_c$ measurements at low enough $\qt$ ($<M_{\eta_c}/2$)~\cite{Hadjidakis:2018ifr}. 

Overall, we expect that gluon-sensitive data could be recorded for $Q^2$ ranging from $100$ to $1000$ $\mathrm{GeV}^2$ with some access to the $x$ dependence of $B$, through data taking at different energies and in different rapidity ranges.

\subsection{Evolution effect on the average of $q_{\sT}$.}

The averaged transverse momentum $\langle q_{\sT} \rangle$ is defined as the cross-section-weighted
average of $q_{\sT}$  over the measured range $[q_{\sT,\mathrm{min}},q_{\sT,\mathrm{max}}]$
and rapidity range $[y_{\mathrm{min}},y_{\mathrm{max}}]$.

Since the measured cross section is provided as discrete values
$\sigma_i \equiv (d\sigma/dq_{\sT})_i \, \Delta q_{\sT,i}$ in bins of width
$\Delta q_{\sT,i}$ centred at $q_{\sT,i}$, it becomes
\begin{equation}
    \langle q_{\sT} \rangle_{\rm exp} =
    \frac{\displaystyle\sum_{i} q_{\sT,i}\,\sigma_i}
    {\displaystyle\sum_{i} \sigma_i} \; ,
    \label{eq:meanpt_data}
\end{equation}
where the sum runs over all $q_{\sT}$ bins in the considered TMD region (i.e., $0\!<\! q_{\sT}\! <\! 0.5  \langle M_{\mathcal{Q} \mathcal{Q}} \rangle$).
The uncertainty on $\langle q_{\sT}\rangle_{\mathrm{exp}}$ is obtained by
propagating the statistical and systematic uncertainties on each $\sigma_i$, taking correlated systematic
uncertainties between bins into account.

The corresponding theoretical value is obtained by evaluating the same
weighted average using the (bin-integrated) theoretical cross section
$\sigma_i^{\mathrm{th}}$ predicted in each bin,
\begin{equation}
    \langle q_{\sT} \rangle_{\rm th} =
    \frac{\displaystyle\sum_{i} q_{\sT,i}\,\sigma_i^{\mathrm{th}}}
    {\displaystyle\sum_{i} \sigma_i^{\mathrm{th}}}
    =
    \frac{\displaystyle\int_{q_{\sT,\mathrm{min}}}^{q_{\sT,\mathrm{max}}}
    q_{\sT} \, \left(\frac{\mathrm{d}\sigma}{\mathrm{d}q_{\sT}}\right)_{\!\mathrm{th}} \mathrm{d}q_{\sT}}
    {\displaystyle\int_{q_{\sT,\mathrm{min}}}^{q_{\sT,\mathrm{max}}}
    \left(\frac{\mathrm{d}\sigma}{\mathrm{d}q_{\sT}}\right)_{\!\mathrm{th}} \mathrm{d}q_{\sT}} \; ,
    \label{eq:meanpt_theory}
\end{equation}
where $(\mathrm{d}\sigma/\mathrm{d}q_{\sT})_{\mathrm{th}}$ is obtained by integrating the
theoretical double-differential cross section over the full rapidity
range, $(\mathrm{d}\sigma/\mathrm{d}q_{\sT})_{\mathrm{th}} = \int_{y_{\mathrm{min}}}^{y_{\mathrm{max}}}
(\mathrm{d}^2\sigma/\mathrm{d}q_{\sT}\,\mathrm{d}y)_{\mathrm{th}}\, \mathrm{d}y$, before performing the $q_{\sT}$-bin
average.
The theoretical uncertainty on $\langle q_{\sT}\rangle_{\mathrm{th}}$
is estimated
by propagating the individual uncertainties on the fit parameters $A$ and $B$ in quadrature.

\begin{table}[htbp]
\centering
\begin{tabular}{lccc}
\toprule
 & \multicolumn{3}{c}{$Q(\equiv\langle M_{\mathcal{Q} \mathcal{Q}} \rangle$) (GeV)} \\
\cmidrule(lr){2-4}
{$\langle \qt \rangle$} & 6.6 & 7.9 & 11.0 \\
\midrule
Experiment & $1.63 \pm 0.10$ & $2.03 \pm 0.14$ & $2.88 \pm 0.25$ \\
\midrule
$b_{\sT}$-constrained fit    & $1.81 \pm 0.12$ & $2.28 \pm 0.16$ & $3.12 \pm 0.15$ \\
$q_{\sT}$-constrained fit    & $1.78 \pm 0.05$ & $2.27 \pm 0.07$ & $3.33 \pm 0.04$ \\
\midrule
Conventional fit    & $1.52 \pm 0.07$ & $1.96 \pm 0.06$ & $2.90 \pm 0.03$ \\
\bottomrule
\end{tabular}
\caption{Comparison of the theoretical predictions of $\langle  q_{\sT} \rangle$ in GeV from the three fits with the experimental results for 3 invariant masses.}
\label{tab:comparison qt_avg}
\end{table}

Table~\ref{tab:comparison qt_avg} collects $\langle q_{\sT} \rangle$ values  obtained from the experimental data and from the three fits presented in Secs.~\ref{sec:classic_fit}, \ref{sec:Minimising chi2 with regularisation and n2} and \ref{sec:Minimising chi2 with regularisation and n10}.
As expected, $\langle q_{\sT} \rangle$ increases with the hard scale, reflecting the progressive broadening of the transverse-momentum spectrum due to evolution.
The conventional fit reproduces the experimental values reasonably well over the entire $Q$ range, at the expense of an oscillating $\qt$ shape.
The introduction of the theoretical constraint systematically shifts the average transverse momentum towards larger values.
Furthermore, the $q_{\sT}$-constrained fit (Sec.~\ref{sec:Minimising chi2 with regularisation and n10}) consistently predicts a slightly larger average transverse momentum than the $b_{\sT}$-constrained fit (Sec.~\ref{sec:Minimising chi2 with regularisation and n2}).

\subsection{Data selection}

We have investigated the dependence of the fit on the TMD cut introduced in Eq.~(\ref{eq:cut}).
We considered the choices $\delta \!<\!0.3$ and $\delta \!<\! 0.4$.
As $\delta$ was decreased, the oscillatory behaviour of the cross section became more pronounced, while the extracted fit parameters systematically decreased.
At the same time, the $\chi^2/\rm{dof}$ was also reduced, reaching values significantly below unity.
These observations suggested that the fit became worse for cuts below $0.5$ with the current experimental data, and we did not pursue a more detailed analysis.

\subsection{Influence of the linearly-polarised gluon TMD $h_{1}^{\perp g}$}
\label{subsec:h1}

To quantify the relative importance of the different TMD contributions entering Eq.~(\ref{eq:TMDcrosssection}), we separately evaluated the terms proportional to $f_1^gf_1^g$ and $h_1^{\perp g}h_1^{\perp g}$.
We recall that the short-distance coefficient $F_2$ associated with the linearly-polarised-gluon contribution amounts~\cite{Scarpa:2019fol,Scimemi:2019cmh} in the acceptance of the LHCb measurement to only about $0.3\%-3\%$ of the unpolarised one $F_1$. In particular, away from threshold, 
for $M_{QQ} \gg M_Q$ and for $\cos \theta_{\rm CS} \!\to \!0$, corresponding to small $\Delta y$, $F_2 \to (81 M_Q^4 \cos^2\theta_{\rm CS}/2M_{QQ}^4) F_1$~\cite{Lansberg:2017dzg}. 
At threshold where $M_{QQ} \!\to\! 2 M_{Q}$,
$F_2\! \to\! 3 F_1/787$  and is even less important. This is reminiscient of the $\psi/\Upsilon+\gamma$ case where $F_2$ is~\cite{denDunnen:2014kjo} identically zero. 

In addition, the corresponding TMD convolution involving $h_1^{\perp g}h_1^{\perp g}$ is itself suppressed with respect to the $f_1^g f_1^g$ contribution by $\alpha_s^2$ in the perturbative region.
As a result, the linearly-polarised gluon TMDs contribute less than $0.5\%$ to the total cross section over the entire kinematic range considered, falling below $0.04\%$ for $Q\!=\!11~\mathrm{GeV}$.
The cross section is therefore overwhelmingly dominated by the unpolarised gluon TMD, and the contribution from linearly polarised gluons can be safely neglected in the present phenomenological analysis.

\subsection{Our fit gluon TMDs}

Figs.~\ref{fig:gluon TMDPDF extraction 1} and \ref{fig:gluon TMDPDF extraction 2} show the resulting unpolarised gluon TMD PDFs from the fits of Secs.~\ref{sec:Minimising chi2 with regularisation and n2} and \ref{sec:Minimising chi2 with regularisation and n10} in transverse-momentum space at NNLL.
 
\begin{figure}[htbp]
    \centering
    \subfloat[]{
        \includegraphics[width=\linewidth]{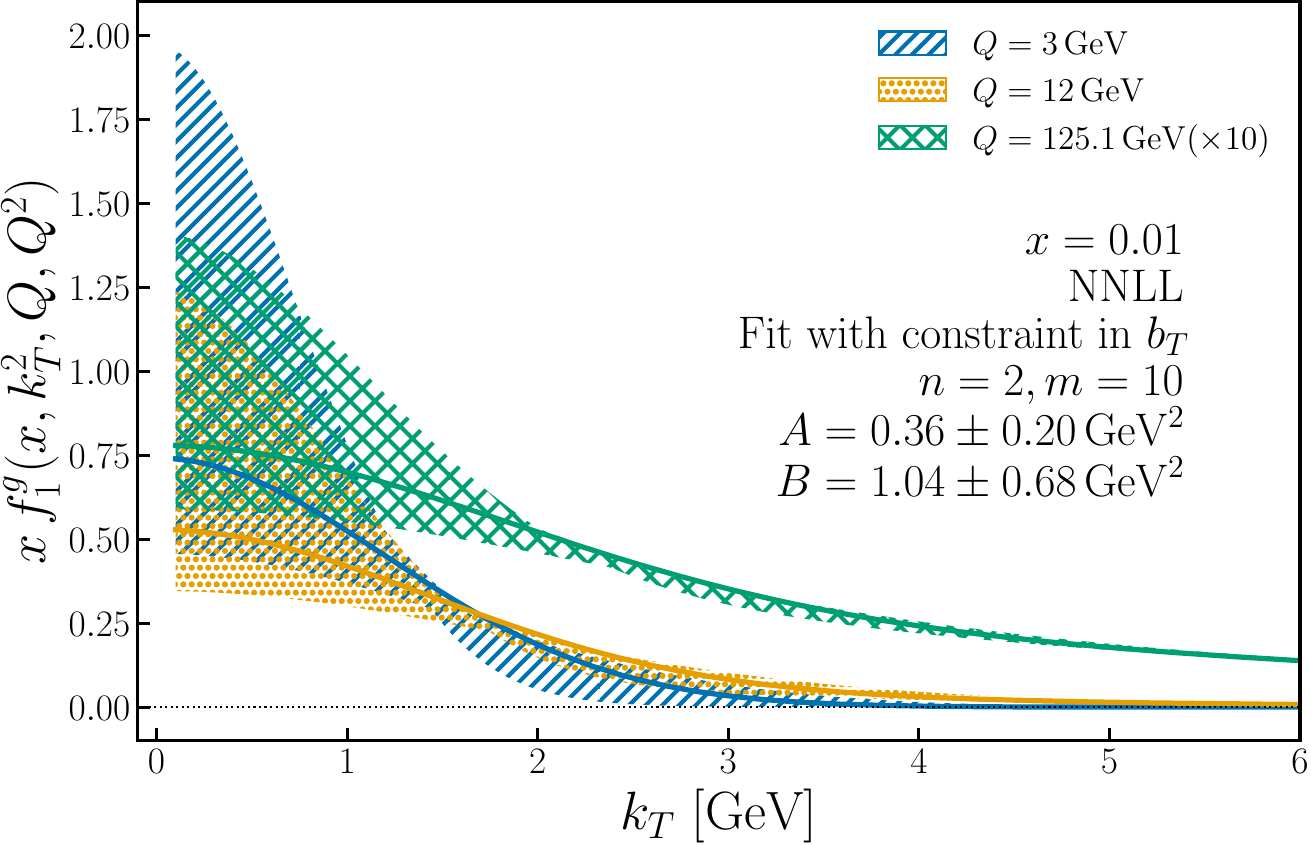}}\\
   \subfloat[]{
        \includegraphics[width=\linewidth]{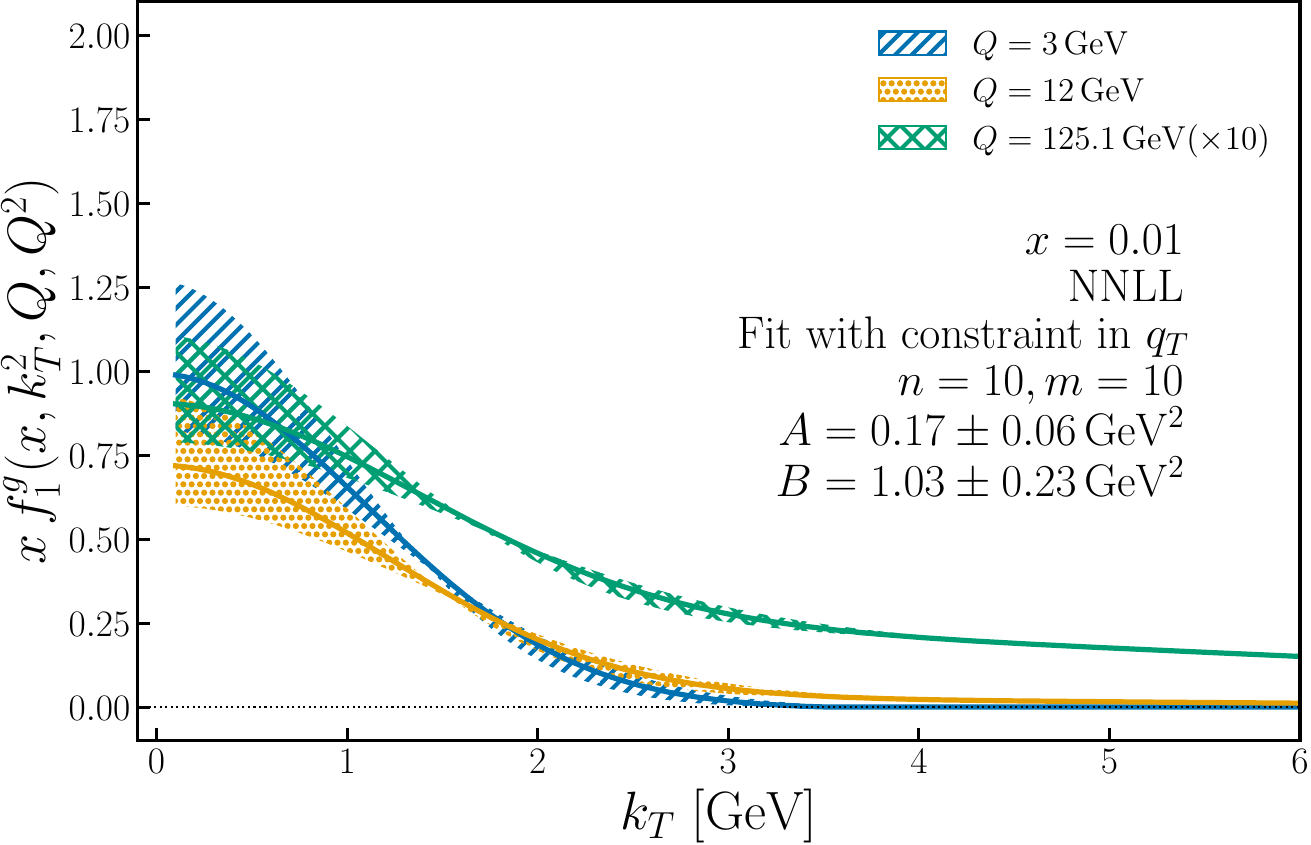}}
    \caption{
    Transverse-momentum distribution of our extracted gluon TMD PDF, $x f_1^g$, for $x\! = \!10^{-2}$ and $Q\!= \!3, \, 12$ and $125.1$~GeV.
    The calculation is performed at NNLL  with $b_{\sT,\max}\!=\! 0.5 \, \rm GeV^{-1}$ and using the fit results of Sec.~\ref{sec:Minimising chi2 with regularisation and n2} (a) and Sec.~\ref{sec:Minimising chi2 with regularisation and n10} (b).
    The bands are obtained by error propagation of $A$ and $B$.
    }
    \label{fig:gluon TMDPDF extraction 1}
\end{figure}
\begin{figure}[htbp]
    \centering
    \subfloat[]{
        \includegraphics[width=\linewidth]{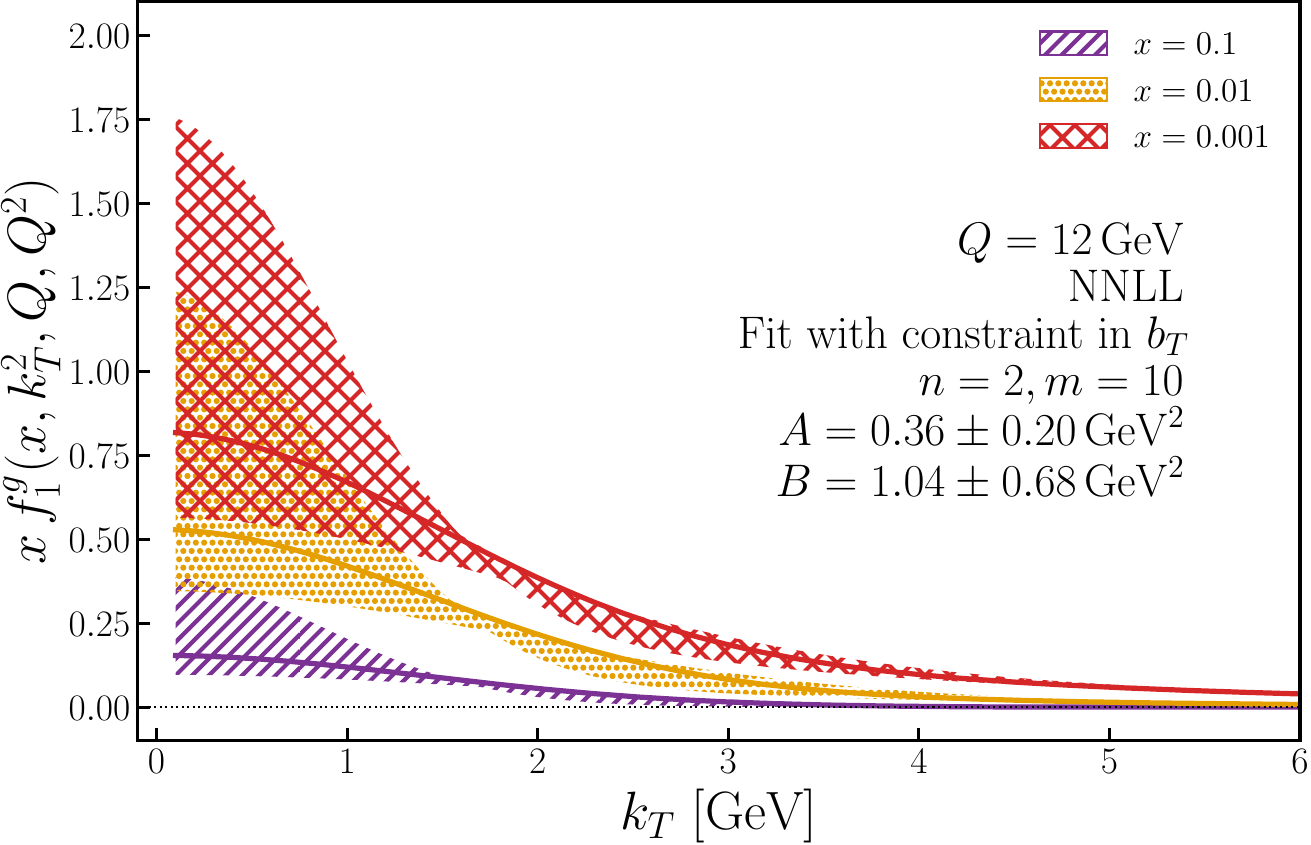}}\\
    \subfloat[]{
        \includegraphics[width=\linewidth]{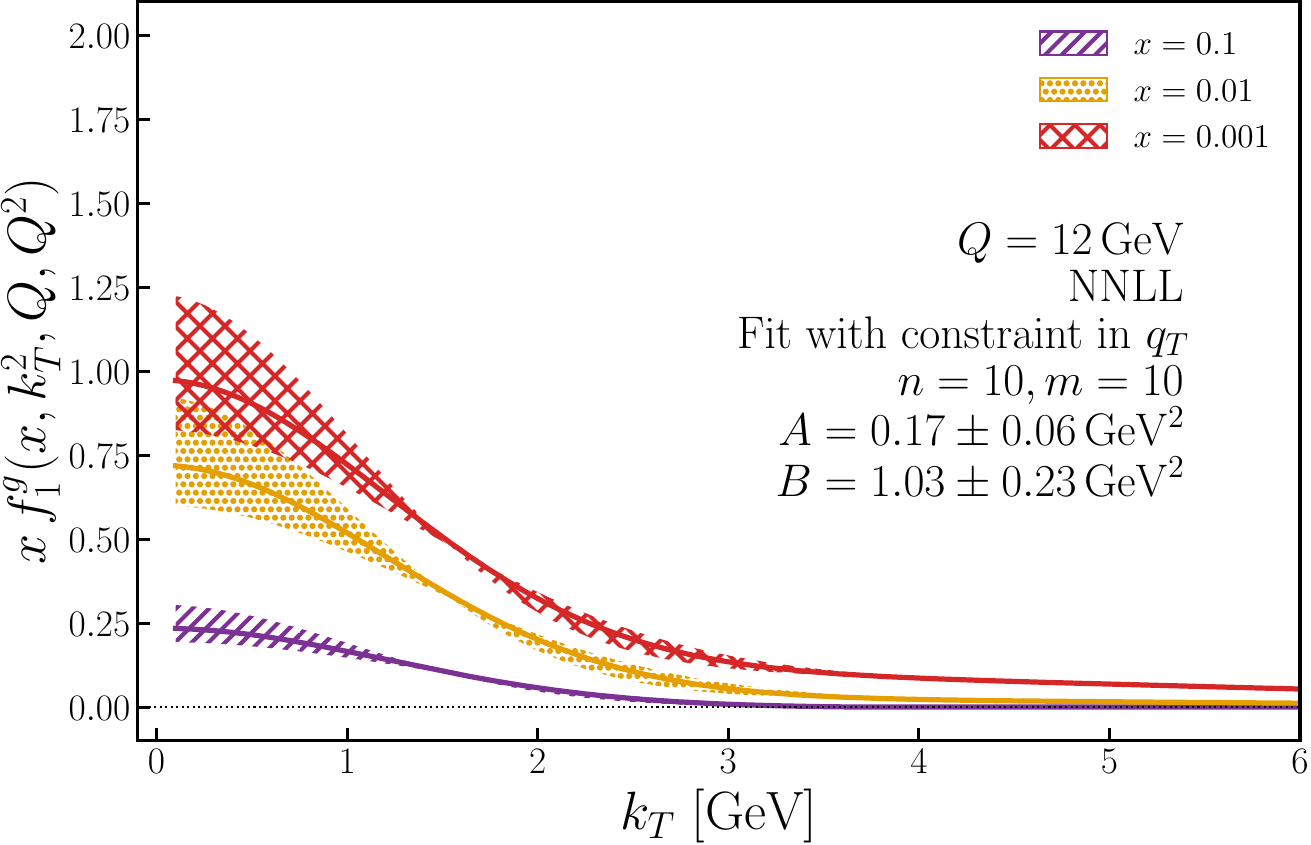} }
    \caption{
    Same as \cf{fig:gluon TMDPDF extraction 1} for $Q \!=\! 12$~GeV and $x\!= \!10^{-1}, \, 10^{-2}$ and $10^{-3}$.
    }
    \label{fig:gluon TMDPDF extraction 2}
\end{figure}

Fig.~\ref{fig:gluon TMDPDF extraction 1} compares the gluon TMD PDF obtained for a fixed value of $x\!=\!10^{-2}$ and three representative scales, namely $Q\!=\!3$, $12$ and $125.1~\mathrm{GeV}$.
In both approaches, the expected TMD evolution is clearly observed: as the hard scale increases, the distribution becomes broader in transverse momentum while its peak is progressively suppressed.
The two fitting procedures lead to very similar central predictions over the entire $k_{\sT}$ range, indicating that both methods provide a consistent description of the gluon TMD.
The main difference is found in the uncertainty band at low $Q$, where the fit constrained in $b_{\sT}$ space yields a significantly larger uncertainty, particularly in the small-$k_{\sT}$ region.
This behaviour is consistent with the larger uncertainties obtained for the non-perturbative parameters in this fit, as reflected by the values of $A$ and $B$.
By contrast, imposing the constraint directly in $q_{\sT}$ space results in a more localised parameter determination and consequently in narrower uncertainty bands, while preserving the same qualitative evolution pattern.

Fig.~\ref{fig:gluon TMDPDF extraction 2} shows the gluon TMD PDF at a fixed hard scale, $Q\!=\!12~\mathrm{GeV}$, for three representative momentum fractions, $x\!=\!10^{-1}$, $10^{-2}$, and $10^{-3}$.
As expected, the overall normalisation increases as $x$ decreases, reflecting the enhancement of the gluon density in the small-$x$ region.
The shape of the distributions is otherwise largely preserved, with all curves exhibiting a smooth decrease as a function of $k_{\sT}$.
The two fitting strategies again produce compatible central predictions throughout the considered transverse-momentum range.
Like in Fig.~\ref{fig:gluon TMDPDF extraction 1}, the largest differences are observed in the uncertainty bands at low transverse momentum, where the fit constrained in $b_{\sT}$ space leads to a broader uncertainty envelope, particularly at smaller values of $x$.
In contrast, the $q_{\sT}$-space constraint provides a more stable determination of the non-perturbative parameters, resulting in a noticeable reduction of the propagated uncertainty while leaving the central behaviour essentially unchanged.

While this work was being completed, another extraction of the gluon TMD $f_1^g$, based on Higgs production data, came out~\cite{Anedda:2026cox}.
The $S_{\rm NP}$ parametrisation used in that work is equivalent to ours for $n \!=\! 2$, upon identifying $A\! =\! 2 \, g_2$ and $B \!=\! g_1$, where $g_1$ and $g_2$ follow the notation of that paper.
However, they considered only $g_1$ as a free parameter, while $g_2$ is fixed (after Casimir scaling) to the value extracted in Ref.~\cite{Avkhadiev:2025wps}.
We also note that their extraction was performed using $b_{\sT,\rm max} \!=\! 1.5\,\rm GeV^{-1}$, while our analysis is performed with $b_{\sT,\rm max} \!=\! 0.5\,\rm GeV^{-1}$.
We find compatible results for $A \!=\! 0.36 \pm 0.20 \, \rm GeV^2$ and $2 \, g_2\! =\! 0.334 \pm 0.030 \, \rm GeV^2$, whereas the values of $B \!= \!1.04 \pm 0.68 \, \rm GeV^2$ and $g_1 \!=\! 14.4 \pm 5.1 \, \rm GeV^2$ are incompatible.
In particular, their extracted value of $g_1$ (or, equivalently, $B$) yields a steeper $S_{\rm NP}$, leading to a stronger suppression at large $b_{\sT}$, and consequently, a broader distribution in transverse-momentum space.
Future fits will help to clarify this discrepancy.

For completeness, in Figs.~\ref{fig:h1perpg extraction 1} and \ref{fig:h1perpg extraction 2}, the NNLL $k_{\sT}$-distributions of $x \, h_1^{\perp g}$ are shown using the fit parameters $A$ and $B$ from both $n\!=\!2$ and $n\!=\!10$ fit models. We stress that this is an assumption as our fit is not sensitive to
 $h_1^{\perp g}$ as we have just discussed in the previous subsection.
Several characteristic features emerge from the extracted distributions. First, the TMD exhibits the expected suppression in the small-$k_{\sT}$ region, vanishing as $k_{\sT} \!\to\! 0$, before developing a broad maximum at intermediate transverse momenta.
This behaviour reflects the intrinsic tensor structure of the linearly polarised gluon distribution, which requires a non-trivial transverse momentum to generate gluon polarisation inside an unpolarised hadron.
The evolution with the hard scale is also clearly visible.
As $Q$ increases, the peak of the distribution decreases while simultaneously shifting towards larger values of $k_{\sT}$. This behaviour is a direct consequence of TMD evolution, which progressively transfers probability from the low-$k_{\sT}$ region towards larger transverse momenta through Sudakov broadening. At the largest scale considered, $Q=125.1 \, \rm GeV$ , the distribution (multiplied by a factor of ten for readability) becomes considerably broader than at low scales, although its overall magnitude is significantly reduced.
Moreover, a comparison with the corresponding $f_1^g$, displayed in Figs.~\ref{fig:gluon TMDPDF extraction 1} and \ref{fig:gluon TMDPDF extraction 2}, highlights the strikingly different behaviour of the two gluon distributions in $k_{\sT}$ space,
and the effects associated with the linearly polarised gluons are found to be consistently sub-leading with respect to those generated by $f_1^g$.

\begin{figure}[htbp]
    \centering
    \subfloat[]{
        \includegraphics[width=\linewidth, trim=0.5cm 0.10cm 2.3cm 1.5cm, clip]{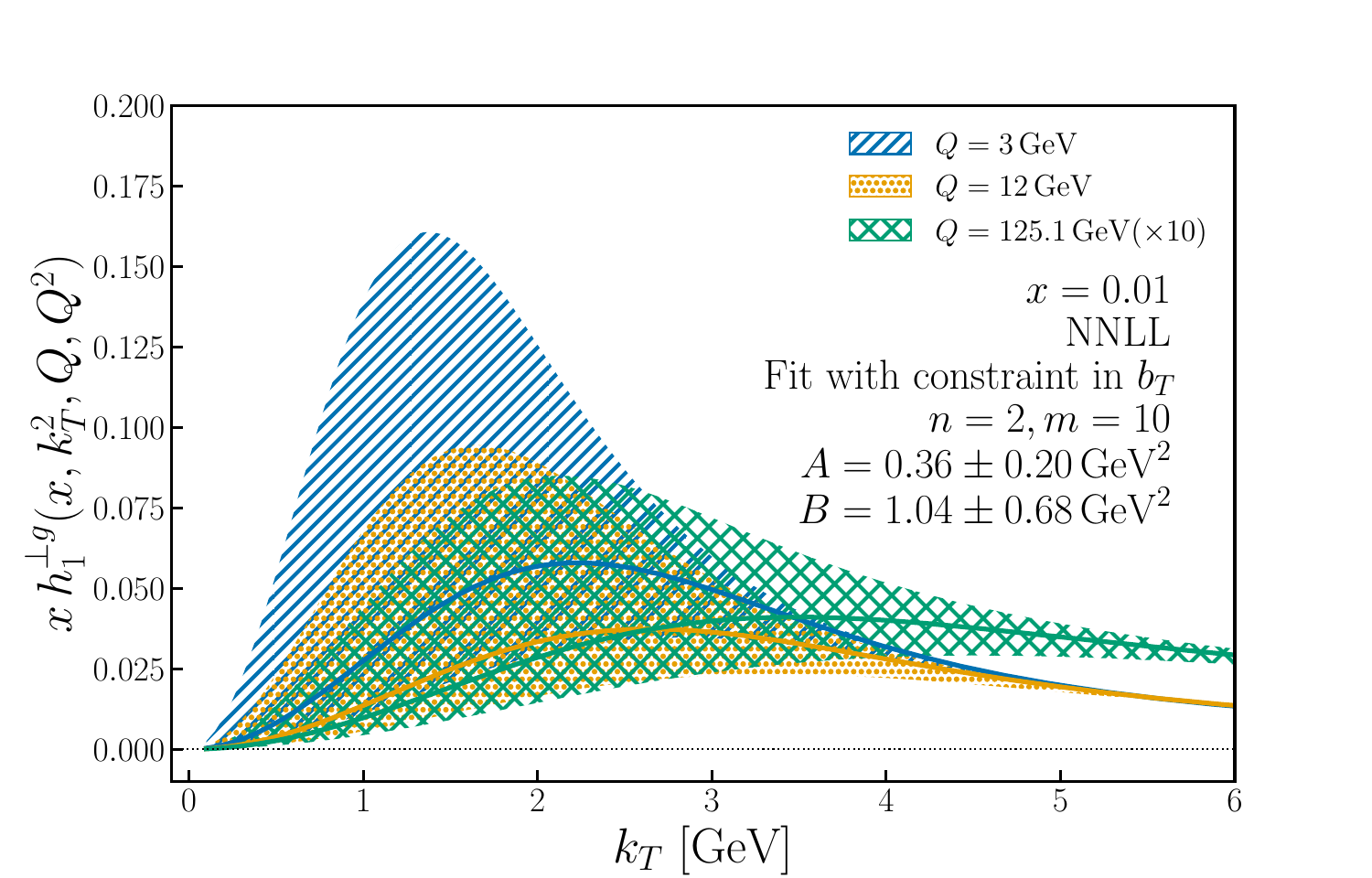}}\\
    \subfloat[]{
        \includegraphics[width=\linewidth, trim=0.5cm 0.10cm 2.3cm 1.5cm, clip]{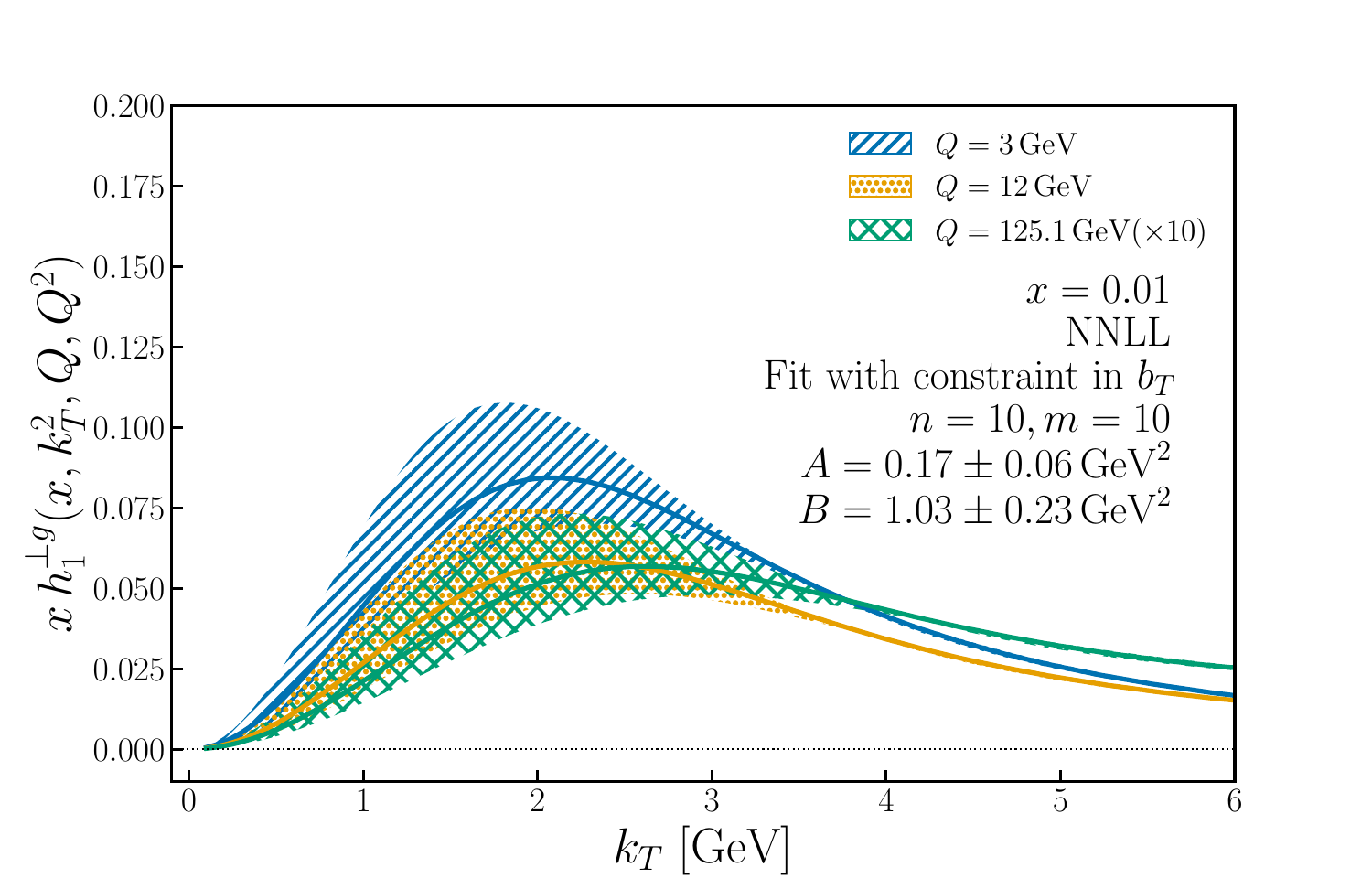}}
    \caption{
    Transverse-momentum distribution of the linearly polarised gluon TMD PDF, $x h_1^{\perp g}$,  using our extracted $S_{\rm NP}$ for $f_1^g$ for $x = 10^{-2}$ and $Q = 3, \, 12$ and $125.1$~GeV.
    The calculation is performed at NNLL with MSHT20 for $b_{\sT,\max} = 0.5 \, \rm GeV^{-1}$ and using the fit results of Sec.~\ref{sec:Minimising chi2 with regularisation and n2} (a) and Sec.~\ref{sec:Minimising chi2 with regularisation and n10} (b).
    The bands are obtained by error propagation of $A$ and $B$.
    }
    \label{fig:h1perpg extraction 1}
\end{figure}

\begin{figure}[htbp]
    \centering
    \subfloat[]{
        \includegraphics[width=\linewidth,trim=0.5cm 0.10cm 2.3cm 1.5cm, clip]{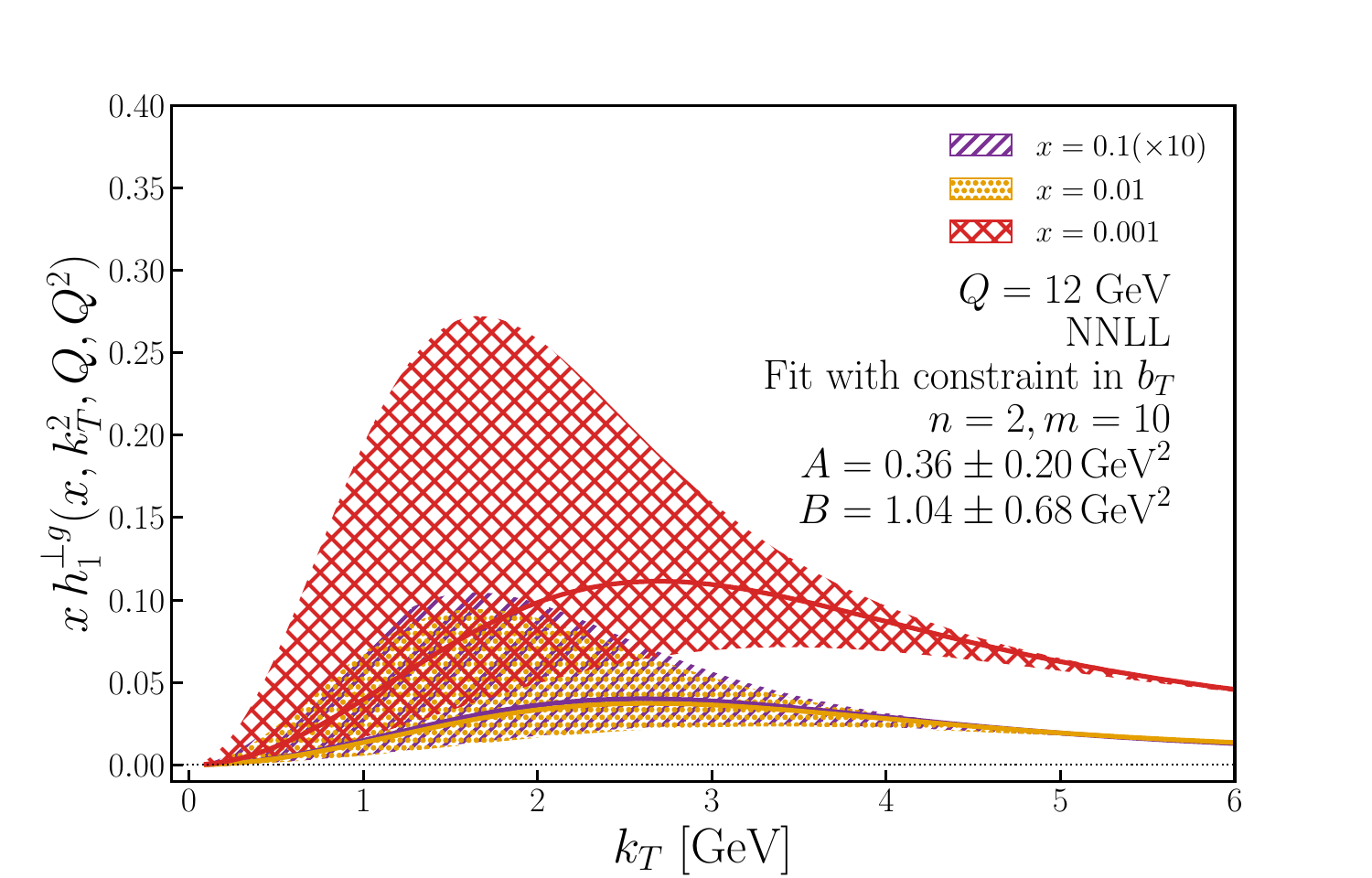}}\\
    \subfloat[]{
        \includegraphics[width=\linewidth,trim=0.5cm 0.10cm 2.3cm 1.5cm, clip]{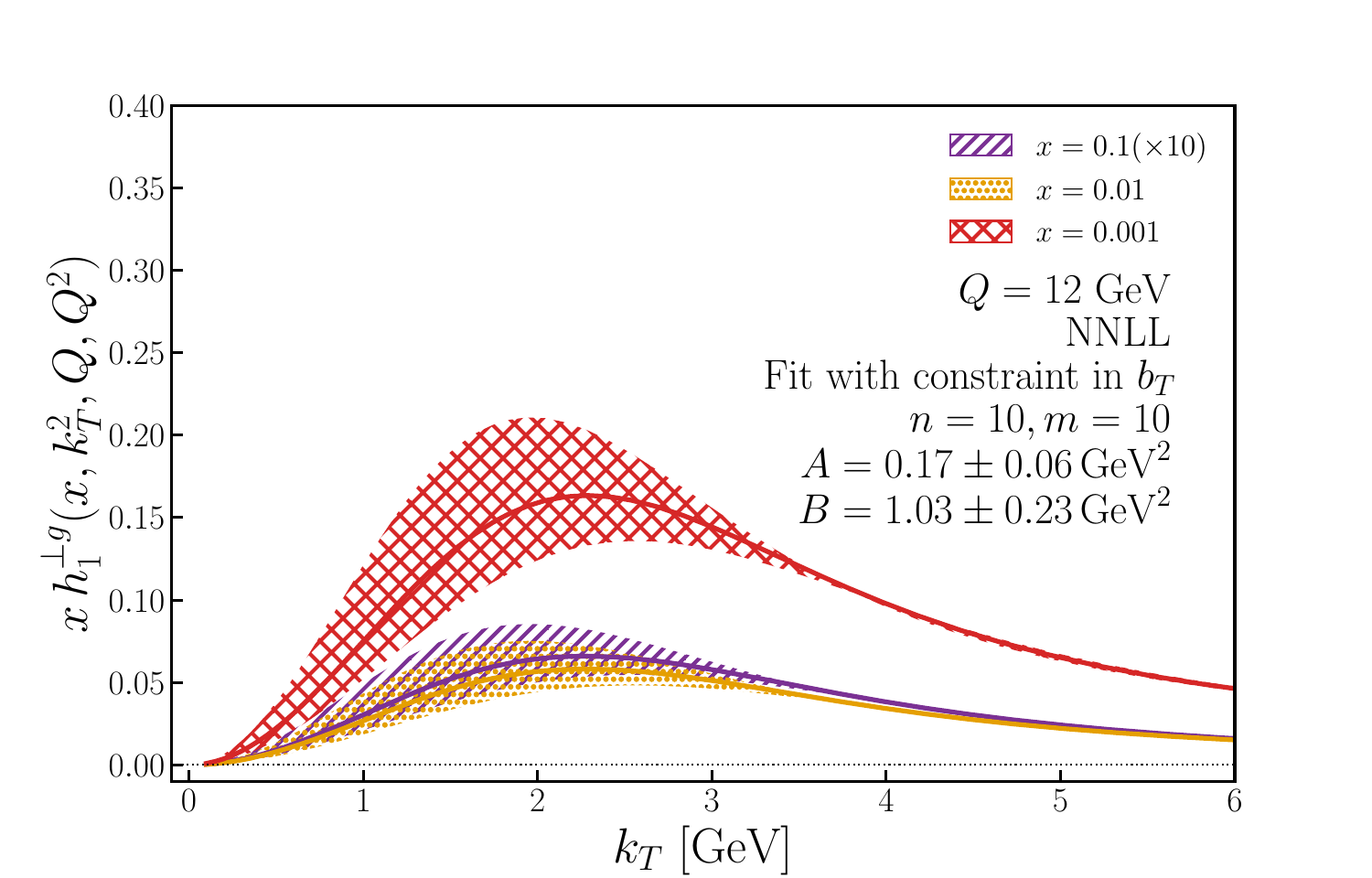}}
    \caption{
     Same as \cf{fig:h1perpg extraction 1} for $Q = 12$~GeV and $x = 10^{-1}, \, 10^{-2}$ and $10^{-3}$.
    }
    \label{fig:h1perpg extraction 2}
\end{figure}

\subsection{Our Collins-Soper kernel fit}

\begin{figure}[hbt!]
    \centering
\subfloat[]{
        \includegraphics[width=\linewidth]{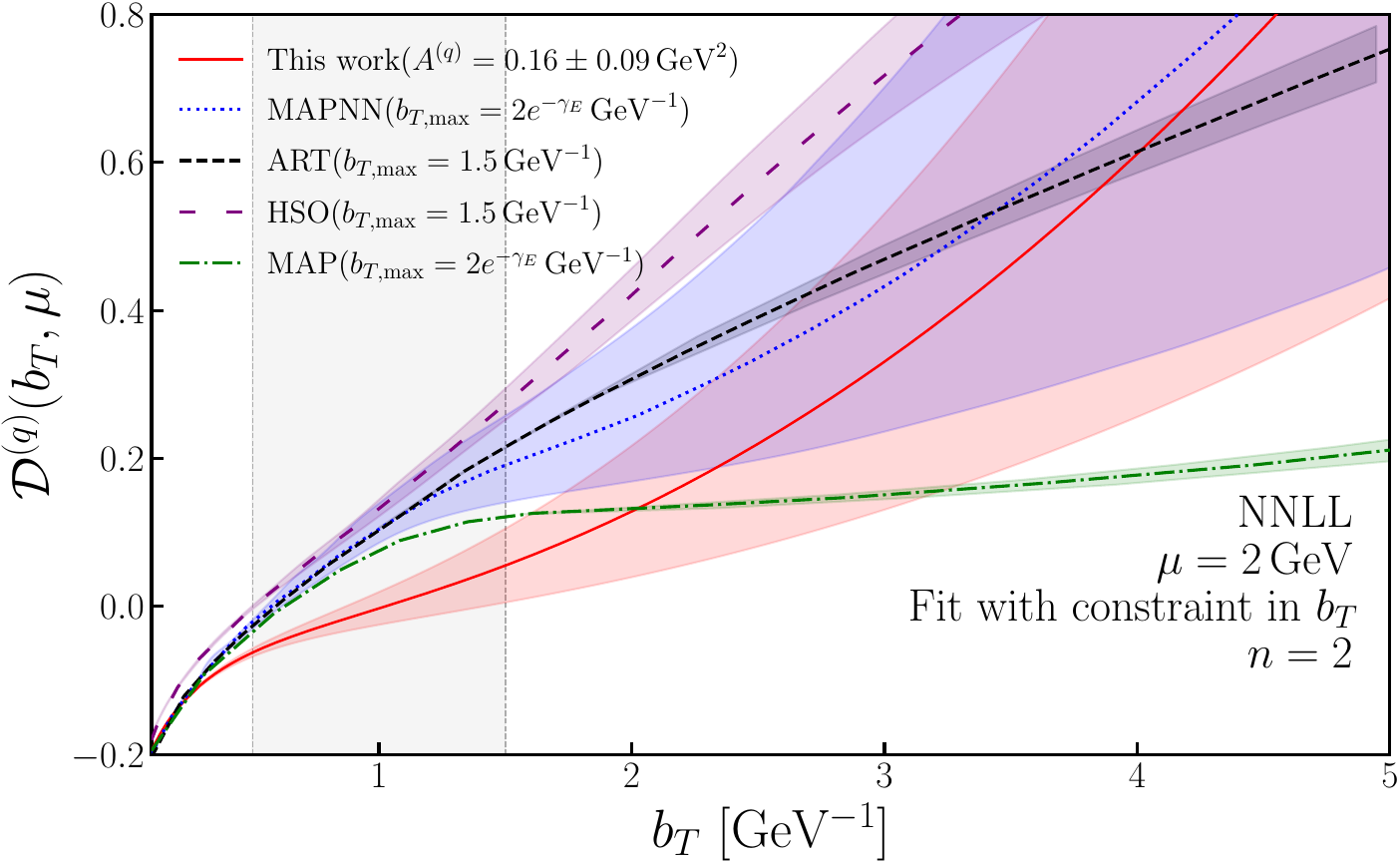}}\\
   \subfloat[]{
        \includegraphics[width=\linewidth]{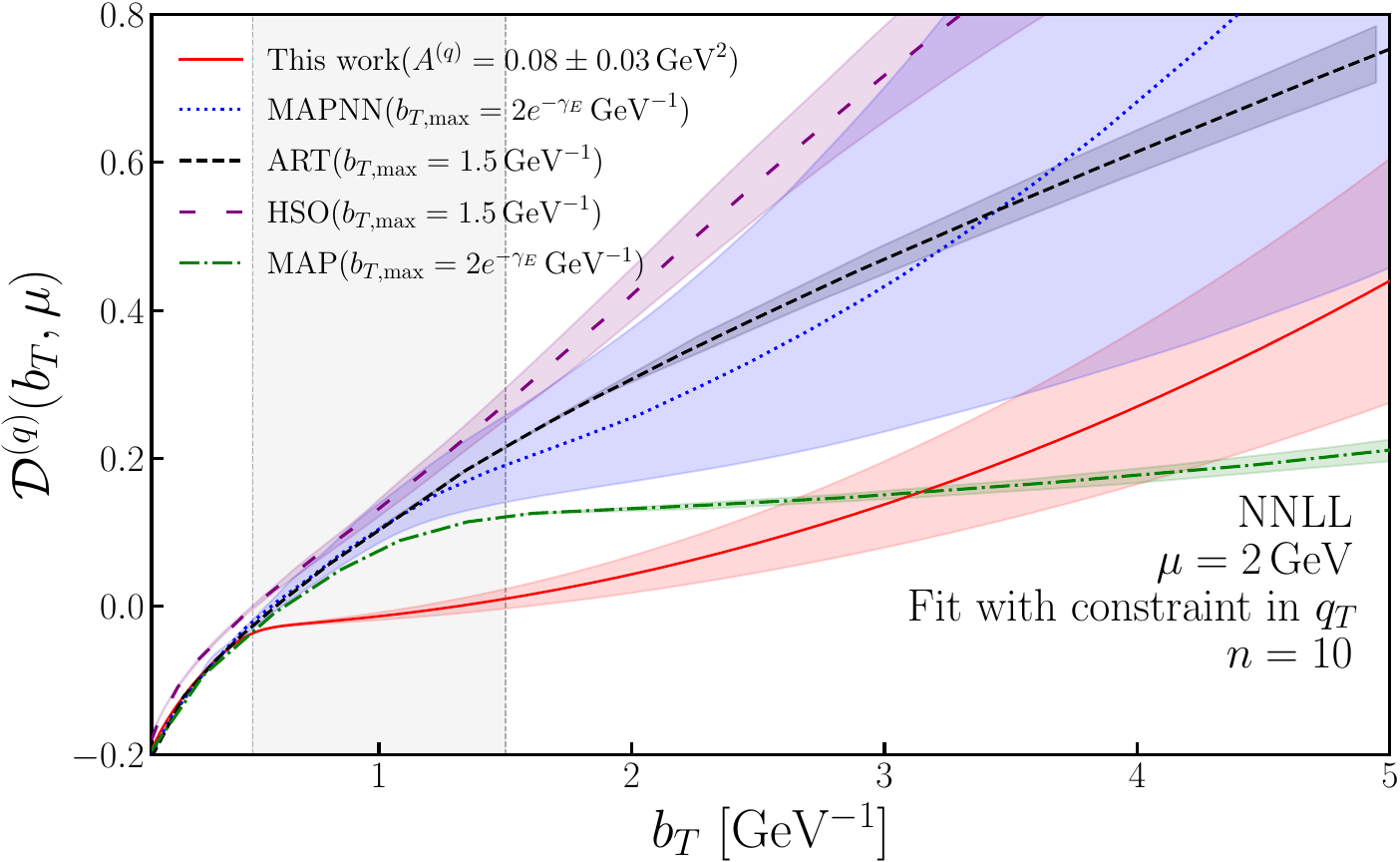}}
    \caption{
    $b_{\sT}$ distribution of the quark Collins-Soper kernel at NNLL (red solid line) obtained by Casimir scaling ($C_F/C_A$) the parameter $A$ extracted in Sec.~\ref{sec:Minimising chi2 with regularisation and n2} (a) and Sec.~\ref{sec:Minimising chi2 with regularisation and n10} (b) for $b_{\sT,\max} \!=\! 0.5 \, \rm GeV^{-1}$, together with previous extractions, namely MAPNN~\cite{Bacchetta:2025ara}, ART~\cite{Moos:2025sal}, 
HSO~\cite{Aslan:2024nqg} and
MAP~\cite{Bacchetta:2024qre}. The grey band indicates the range $b_{\sT,\mathrm{max}}$ used in the different extractions, where the comparison might not be accurate.}
    \label{fig:CS kernel extraction}
\end{figure}

In this section, we present the Collins-Soper kernel, defined in Eq.~(\ref{eq:CS kernel}), with
\begin{equation}
    \mathcal{D}_{\rm NP} (b_{\sT}) = \frac{A}{4}  b_{\sT}^{\dagger 2} (b_{\sT}) \; .
\end{equation}
The calculations are performed at NNLL using the values of $A$ obtained in Secs.~\ref{sec:Minimising chi2 with regularisation and n2} and \ref{sec:Minimising chi2 with regularisation and n10} for $Q_{\rm NP} = 1 \, \rm GeV$ and $b_{\sT,\max} = 0.5 \, \rm GeV^{-1}$.

To compare our results with extractions from previous works of quark TMD fitting, we rely on the Casimir scaling, according to which the quark Collins-Soper kernel is related to the gluon one by
\begin{equation}
    \mathcal{D}^{(q)} (b_{\sT}, \mu) = \frac{C_F}{C_A} \mathcal{D}^{(g)} (b_{\sT}, \mu) \; .
\end{equation}
This is exact for the perturbative expression up to four loops in perturbative QCD~\cite{vonManteuffel:2020vjv,Moult:2022xzt,Duhr:2022yyp}, and we extend it to the non-perturbative part:
\begin{equation}
\mathcal{D}_{\rm NP}^{(q)} (b_{\sT}) = \frac{A^{(q)}}{4} b_{\sT}^{\dagger 2} = \frac{C_F}{C_A} \frac{A}{4} b_{\sT}^{\dagger 2} \; ,
\end{equation}
where $A^{(q)} = 0.16 \pm 0.09 \, \rm GeV^2$ and $A^{(q)} = 0.08 \pm 0.03 \, \rm GeV^2$ come from the fit results of $A$ in Secs.~\ref{sec:Minimising chi2 with regularisation and n2} and Sec.~\ref{sec:Minimising chi2 with regularisation and n10}, respectively.
We note that these values do not seem to be compatible with the one found in Ref.~\cite{Landry:2002ix}, which also used $b_{T,\rm{max}}=0.5~\rm{GeV}^{-1}$.

Fig.~\ref{fig:CS kernel extraction} compares the corresponding $\mathcal{D}^{(q)}$ extracted with the two fitting strategies to several recent phenomenological determinations, namely MAPNN~\cite{Bacchetta:2025ara}, ART~\cite{Moos:2025sal}, 
HSO~\cite{Aslan:2024nqg} and
MAP~\cite{Bacchetta:2024qre}.
In both panels, the extracted kernel exhibits the expected perturbative behaviour at small $b_{\sT}$, where all determinations are in close agreement, reflecting the perturbative nature of this region.
Differences become more pronounced at larger values of $b_{\sT}$, where non-perturbative effects dominate.
The fit constrained in $b_{\sT}$-space (upper panel) leads to a steeper increase of the kernel together with a substantially larger uncertainty band, which reflects the weaker determination of the non-perturbative parameters obtained with this approach.
In contrast, imposing the constraint directly in $q_{\sT}$ space (lower panel) results in a flatter large-$b_{\sT}$ behaviour and significantly reduced uncertainties, yielding a prediction that lies within the range spanned by previous extractions.

The apparent lack of linear behaviour of our extraction at large $b_{\sT}$, compared with the other extractions, is due to the values of $A$ and $b_{\sT, \max}$.
In the latter, smaller values of $A$ are typically obtained, together with the use of larger values of $b_{\sT,\max}$.
As a result, the quadratic behaviour of the Collins-Soper kernel at large $b_{\sT}$ becomes visible only at significantly larger values of $b_{\sT}$ than in the gluon case.

Overall, the comparison shows that our extraction follows the same behaviour as previous determinations over the perturbative region, as expected, while the differences are confined to larger values of $b_{\sT}$.
This highlights that the non-perturbative large-$b_{\sT}$ behaviour of the Collins-Soper kernel remains only weakly constrained by current phenomenology, and that extending the Casimir scaling into the non-perturbative regime should be regarded with caution.

\subsection{Comparison at different perturbative accuracies}

Let us now investigate the importance of incorporating higher-order perturbative corrections into the theoretical predictions in order to achieve a better description of the experimental data.

To do so, we  have performed additional fits at NLL\footnote{We have used the same PDF set as in the NNLL calculation, i.e., NLO MSHT20.} using the procedures described in Secs.~\ref{sec:Minimising chi2 with regularisation and n2} and \ref{sec:Minimising chi2 with regularisation and n10}, and compared the corresponding results with those obtained at NNLL.

\begin{table}[hbt!]
\centering
\begin{tabular}{l l c c c}
\toprule
Fit method & Acc. & $A\,(\mathrm{GeV}^2)$ & $B\,(\mathrm{GeV}^2)$ & $\chi^2/\mathrm{dof}$ \\
\midrule
$b_{\sT}$-constrain{ed}  & NNLL & $0.36 \pm 0.20$ & $1.04 \pm 0.68$ & 1.18 \\
& NLL  & $0.21 \pm 0.15$ & $0.46 \pm 0.42$ & 0.50\\
$q_{\sT}$-constrain{ed}  & NNLL & $0.17 \pm 0.06$ & $1.03 \pm 0.23$ & 1.03 \\
& NLL  & $0.16 \pm 0.06$ & $1.00 \pm 0.27$ & 1.04 \\
\midrule
Conventional & NNLL & $0.23 \pm 0.07$ & -- & 0.30 \\
& NLL  & $ 0.20 \pm 0.06$ & -- & 0.35 \\
\bottomrule
\end{tabular}
\caption{Best-fit values of the non-perturbative parameters $A$ and $B$, together with the corresponding $\chi^2/\mathrm{dof}$, obtained using the different fitting methods at NLL and NNLL accuracy.}
\label{tab:fits at NLL and NNLL}
\end{table}

\begin{figure}[hbtp]
    \centering
    \subfloat[]{
        \includegraphics[width=0.96\linewidth]{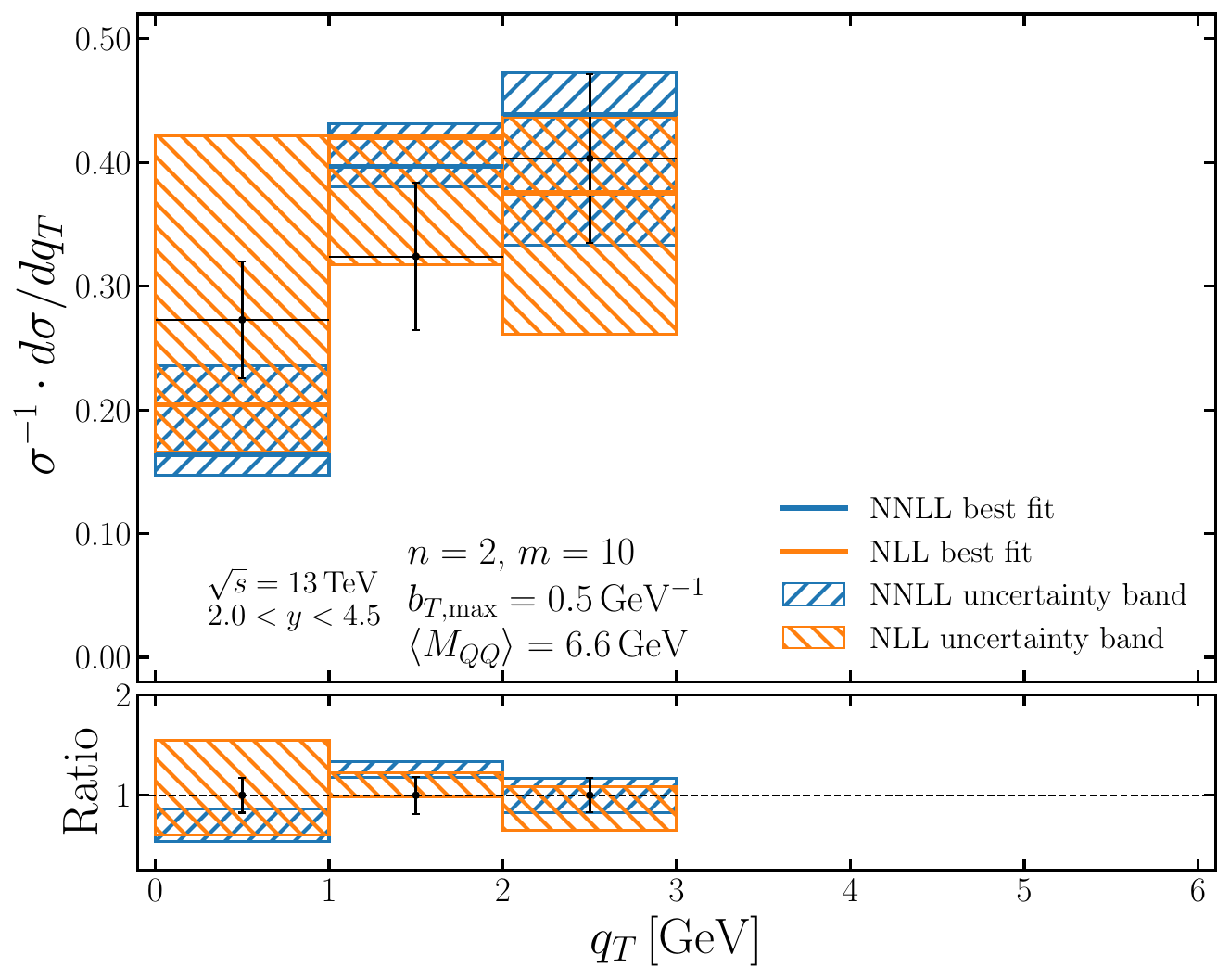}}\\
    \subfloat[]{
        \includegraphics[width=0.96\linewidth]{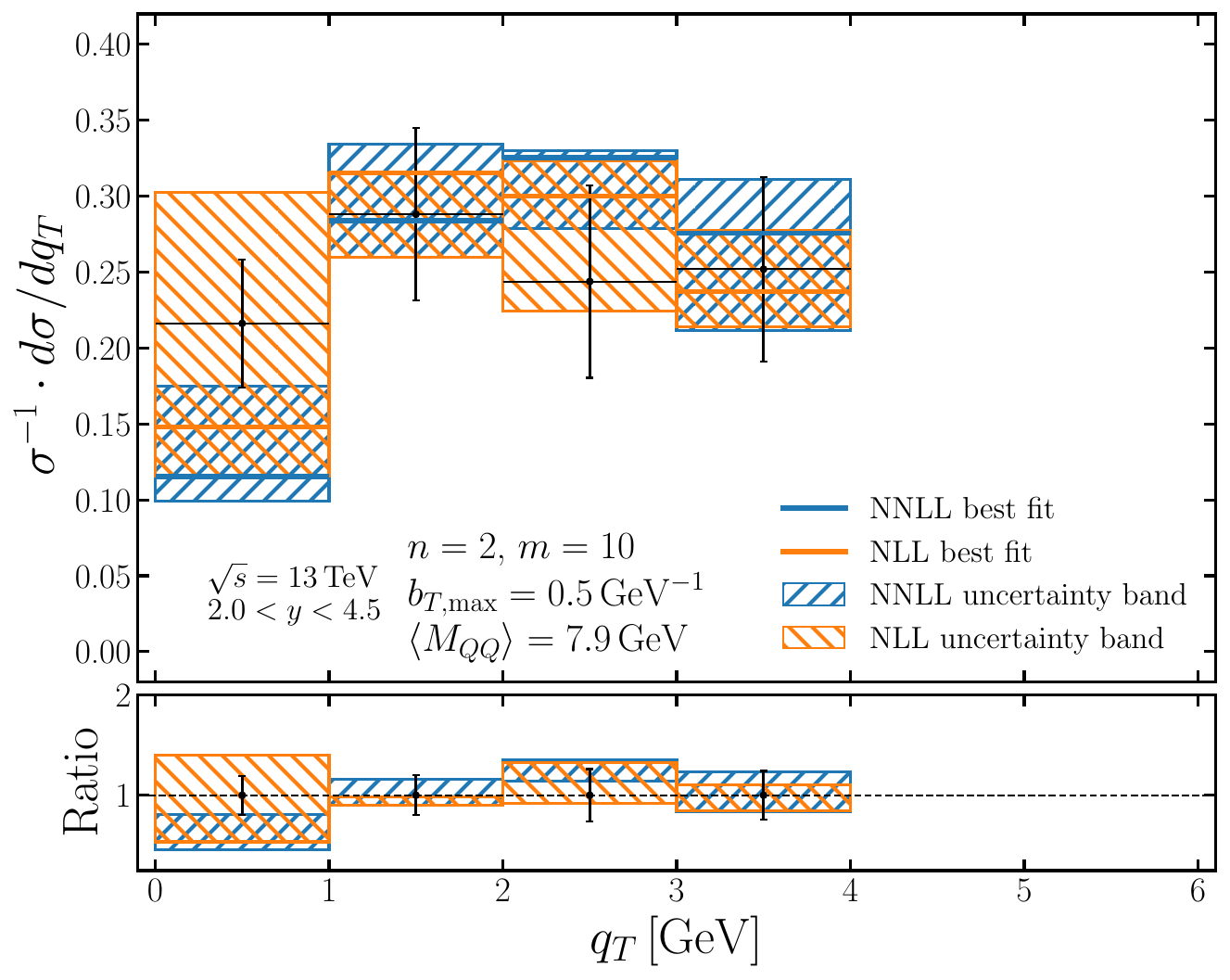}}\\
    \subfloat[]{
        \includegraphics[width=0.96\linewidth]{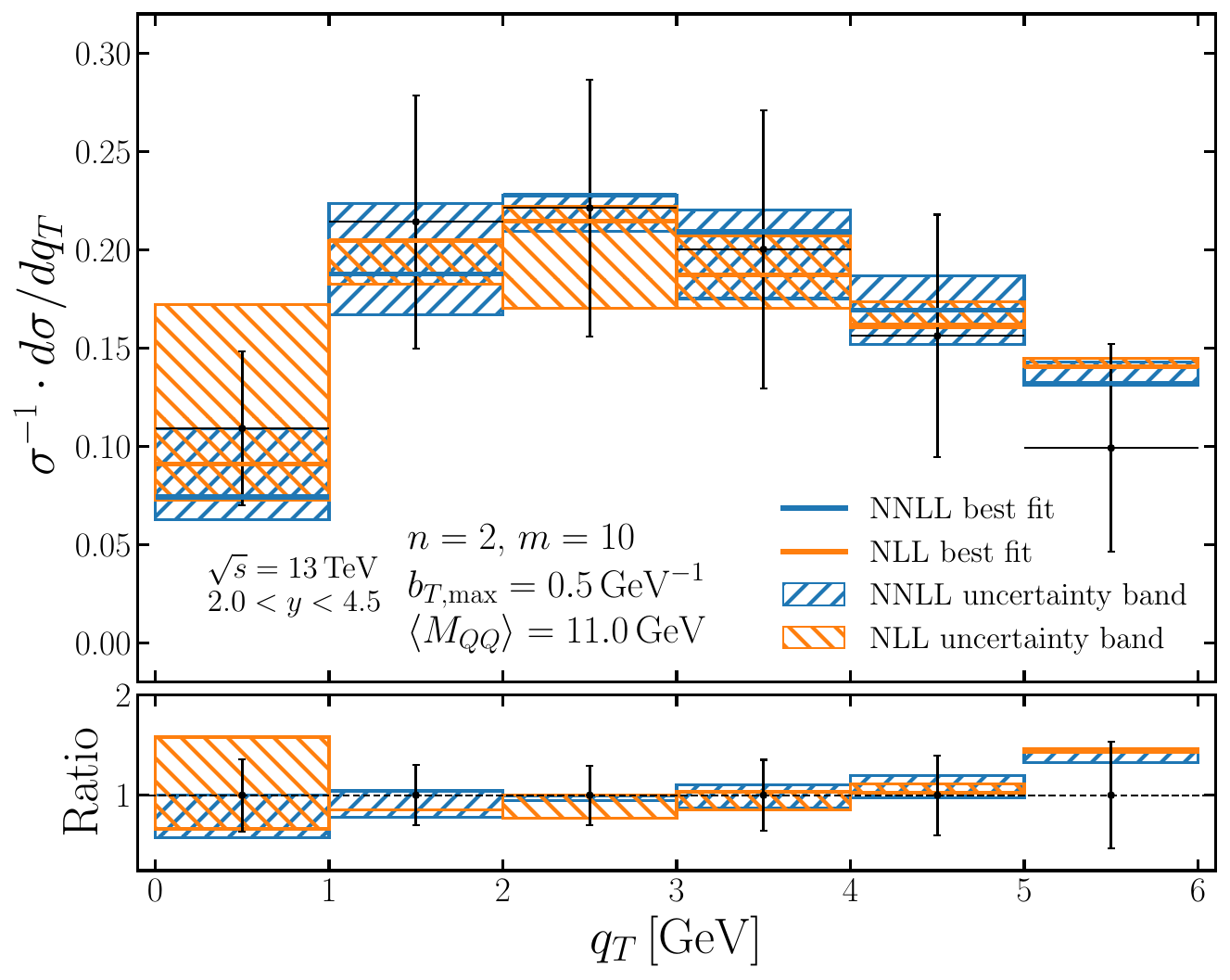}}
    \caption{
    Fit with theory constraints in $b_{\sT}$ space.
    Upper panels: normalised $q_{\sT}$ distribution of the theoretical and experimental differential cross section for the three bins of $\langle M_{QQ} \rangle$ at NLL and NNLL.
    The uncertainty bands correspond to the error propagation of the fit parameters.
    Lower panels: ratio between experimental data and theoretical results.
    }
    \label{fig:comparison pert acc n2}
\end{figure}

\begin{figure}[hbtp]
    \centering
    \subfloat[]{
        \includegraphics[width=0.98\linewidth]{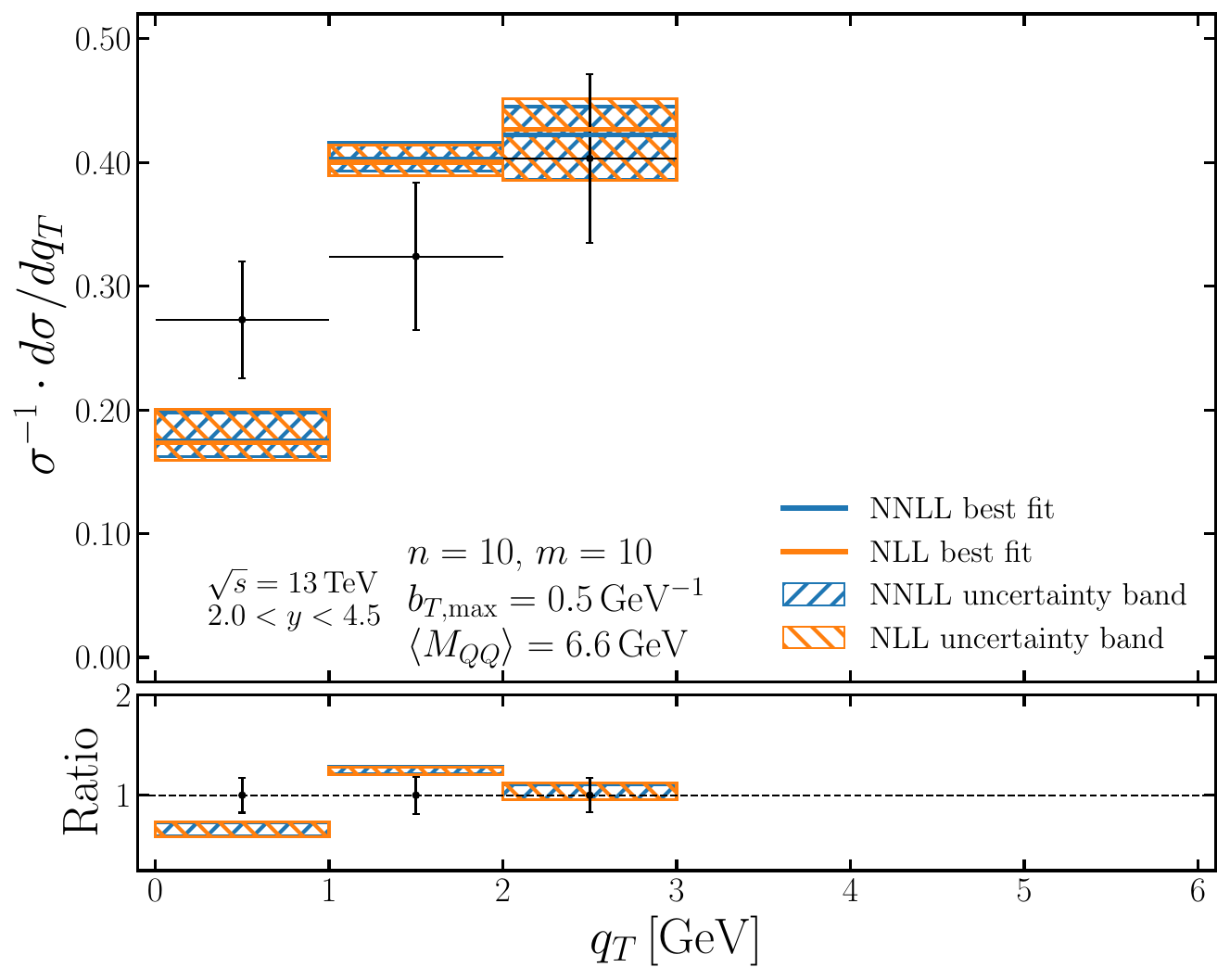}}\\
    \subfloat[]{
        \includegraphics[width=0.98\linewidth]{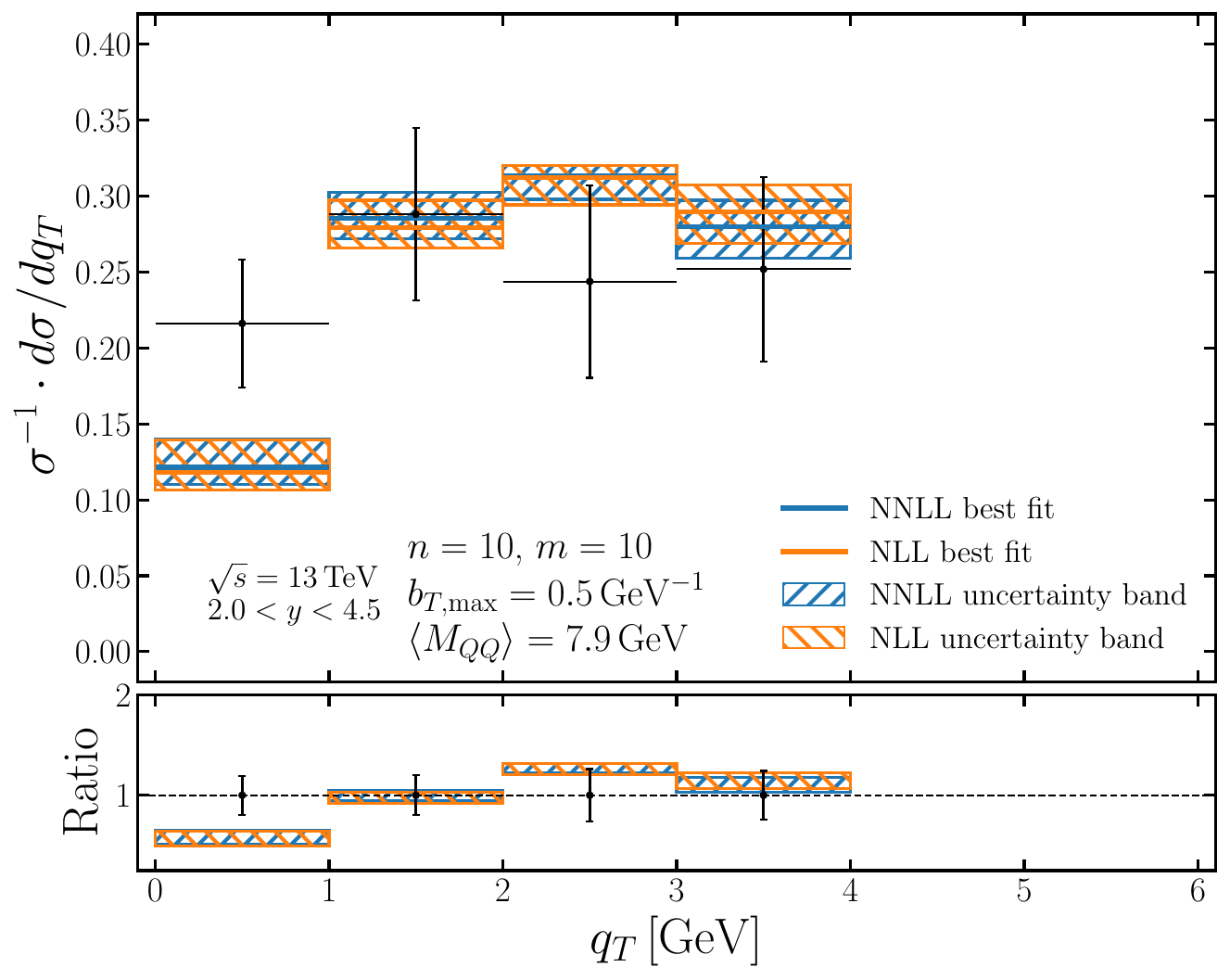}}\\
    \subfloat[]{
        \includegraphics[width=0.98\linewidth]{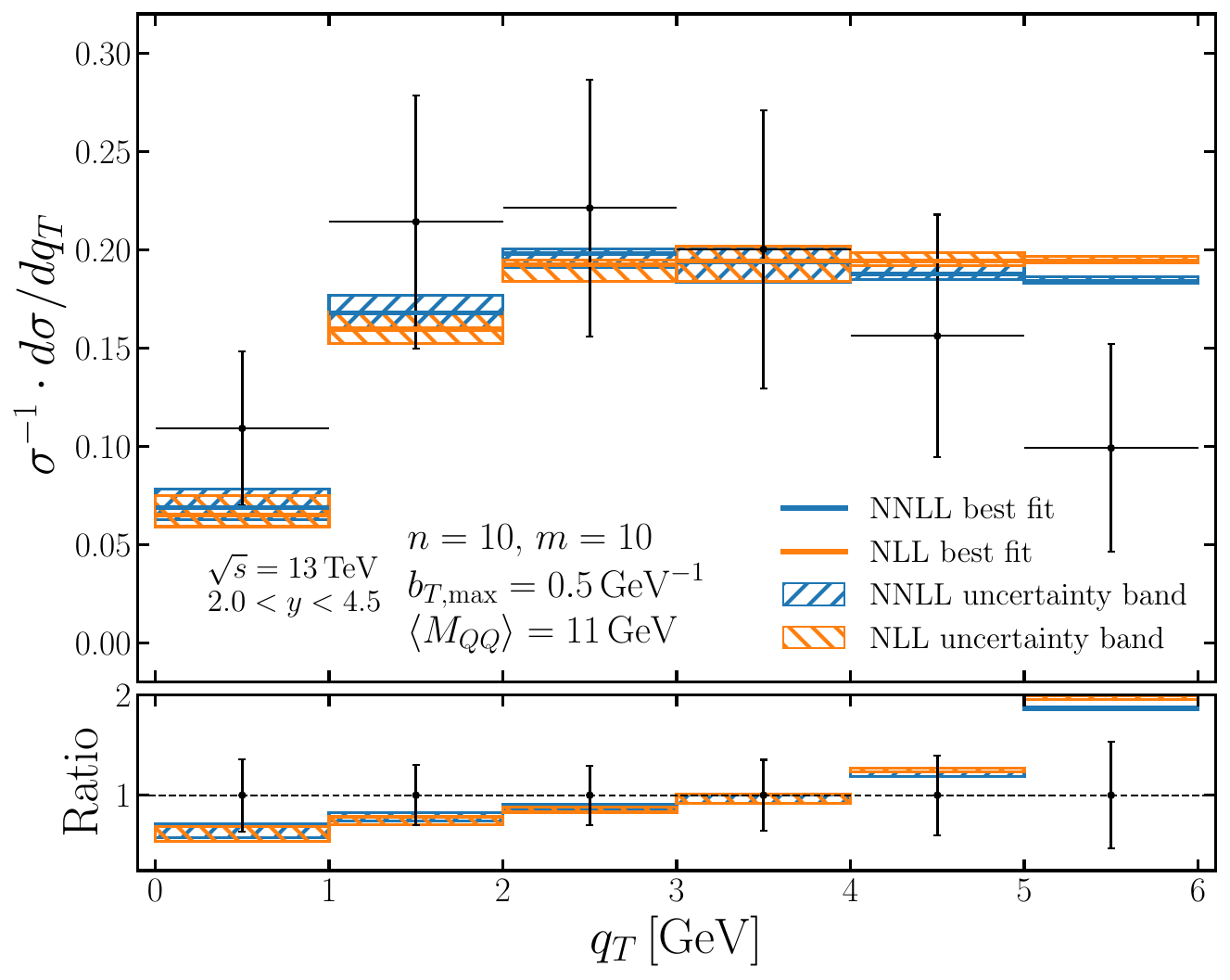}}\\
    \caption{
    Same as Fig.~\ref{fig:comparison pert acc n2} for our fit with theory constraints in $q_{\sT}$ space.
}
    \label{fig:comparison pert acc n10}
\end{figure}

They are summarised in Table~\ref{tab:fits at NLL and NNLL}, and compared with the results of Sec.~\ref{sec:classic_fit}.
The extracted non-perturbative parameters remain remarkably stable when the perturbative accuracy is reduced from NNLL to NLL.
For both fitting strategies, the central values of $A$ and $B$ are compatible within uncertainties, indicating that the determination of the non-perturbative contribution is robust against the perturbative order considered.
Nevertheless, the quality of the fit deteriorates slightly at NLL, as reflected by the $\chi^2/\mathrm{dof}$ values.

To provide a more direct comparison between the two resummation accuracies, the corresponding theoretical cross sections are confronted with the experimental data in Figs.~\ref{fig:comparison pert acc n2} and \ref{fig:comparison pert acc n10}.
As can be seen in Fig.~\ref{fig:comparison pert acc n2}, the agreement between the theoretical predictions and the experimental data improves significantly when the perturbative accuracy is increased.
By contrast, in Fig.~\ref{fig:comparison pert acc n10} both perturbative orders provide an almost identical description of the data as expected from the results in Table~\ref{tab:fits at NLL and NNLL}.

We find that the $D_{\rm pert}$ penalty introduced in the fit of Sec.~\ref{sec:Minimising chi2 with regularisation and n2} (the $b_{\sT}$-space constrained fit) has an almost negligible impact on the fitting procedure at NLL.
This is clearly illustrated in Table~\ref{tab:fits at NLL and NNLL}.
The $\chi^2/\rm dof$ decreases substantially when going from NNLL to NLL (approaching the result of the fit without constraint), while the relative uncertainties of both parameters, $A$ and $B$, increase significantly, indicating that they are only weakly constrained.
This behaviour is also evident in Fig.~\ref{fig:comparison pert acc n2}.
In general, the NLL uncertainty bands (orange) are systematically broader than the NNLL ones (blue) in all $q_{\sT}$ bins, and the effect is especially pronounced in the first one.
This is because, within the present setup, NLL accuracy is insufficient to constrain the parameter space as effectively as the NNLL fit, allowing parameter regions that give rise to unrealistic cross-section predictions.

This behaviour is fully consistent with the $b_{\sT}$-space constrained fitting strategy.
Since the perturbative uncertainty band is considerably broader at NLL than at NNLL, the allowed parameter space is correspondingly enlarged.
As a result, the penalty has a reduced discriminating power and cannot efficiently suppress parameter configurations that give rise to this kind of predictions.

\section{Conclusions}\label{conclusions}
In this work, we have presented the first extraction of the unpolarised gluon TMD PDF $f_1^g$ from $\sqrt{s} = 13 \, \rm TeV$ LHCb measurements of $J/\psi$-pair production.
The analysis is performed within the TMD factorisation framework at NNLL with the cut $q_T/\langle M_{QQ} \rangle < 0.5$.
Since the contribution of the linearly-polarised-gluon TMD $h_1^{\perp g}$ is negligible compared with that of $f_1^g$, it has not been included in the present analysis.

The conventional $\chi^2$-minimisation procedure yields $\chi^2/\rm dof = 0.30$, and a non-perturbative parameter $A = 0.23 \pm 0.07 \, \rm GeV^2$.
However, the conventional approach with the usual $b_{\sT}^*$ prescription leads to an unrealistically large influence of the non-perturbative contribution on the perturbative inputs, beyond what is tolerable at NNLL accuracy, and even at NLL accuracy. 
This is reflected in the shape of the theoretical cross section.
This behaviour highlights the limitations of the conventional modelling of non-perturbative effects for gluon-sensitive processes at low scales.

To address this issue, we have proposed two novel and independent fitting strategies.
Both approaches restrict the allowed parameter space, making it possible to determine simultaneously the two non-perturbative parameters, $A$ and $B$, constraining simultaneously the non-perturbative parts of the gluon TMD PDF and Collins-Soper kernel.
The first strategy injects, in the $\chi^2$-minimisation, information on the perturbative uncertainty obtained via scale variations.
The second strategy consists in injecting constraints in the physical $\qt$ space with a modified $b_{\sT}^*$ prescription, minimising the influence of the non-perturbative effects in the perturbative domain below $b_{T,\rm max}$.
Despite relying on different principles, both approaches remarkably lead to compatible results.
The first yields $\chi^2/\rm dof \!=\! 1.18$, $A \!=\! 0.36 \pm 0.20 \, \rm GeV^2$ and $B\! =\! 1.04 \pm 0.68 \, \rm GeV^2$, while the second gives $\chi^2/\rm dof \!=\! 1.03$, $A = 0.17 \pm 0.06 \, \rm GeV^2$ and $B \!=\! 1.03 \pm 0.23 \, \rm GeV^2$.

As a consequence, we have achieved the first independent extraction of the non-perturbative contributions to the gluon TMDs and the Collins--Soper kernel. 
The lack of sensitivity of the data to the rapidity $y$ of the produced $J/\psi$-pair however prevents us from constraining the $x$ dependence of the former. 
The Collins--Soper kernel is compared, after Casimir scaling, with previous extractions of quark TMDs.
Such a scaling seems to be compatible with the more recent fits, in line also with the recent work \cite{Avkhadiev:2026xyf}, while not with the one of Ref.~\cite{Landry:2002ix}, which used $b_{T,\rm{max}}=0.5~\rm{GeV}^{-1}$.
The validity of such scaling thus is not confirmed and calls for further investigation.

We have also redone the fits for different PDF sets and have found that results are consistent only if the PDF uncertainties of each set are properly accounted for. As such, we discourage using only central eigensets in future gluon TMD fits.

We have also performed fits at NLL to compare to our NNLL fits. Such a comparison is particularly pertinent for our first fit accounting for perturbative uncertainties. 
We have observed that the agreement between theory and data improves as the perturbative resummation accuracy is increased. 

The two alternative fitting strategies introduced in this work are not unique, and there is room for further refinement.
For instance, the penalty associated with the perturbative uncertainty could be defined using a larger number of scale variation points.
Likewise, the curvature penalty, implemented through the constraint in $q_T$ space, could be extended to different centre-of-mass energies, thereby mimicking multiple experimental conditions and probing a broader range of $x$.

Furthermore, extending this study to other processes and to more precise data will allow one to assess the extent to which these considerations remain necessary.

As we have discussed, future quarkonium-pair measurements, free of DPS contributions, will be extremely beneficial. One can identify four fronts: (i) extending the reach in $Q$; (ii) probing the dependence in $x$; (iii) testing the factorisation in the presence of quarkonia and (iv) pinning down the linearly polarised gluon TMD PDF. 

Regarding (i), ATLAS~\cite{ATLAS:2016ydt} and CMS~\cite{CMS:2014cmt} have already demonstrated that they can reach larger invariant masses. What is needed is double-differential data in $\qt\equiv P_{T,\psi\psi}$ and $Q\equiv M_{\psi \psi}$. 

Regarding (ii), data at different colliding energies from the LHC in the fixed-target modes would be ideal. Data triple differential in $\qt$, $Q$, and $y$ would also be welcome in the collider mode. 

Regarding (iii), $\qt-Q$ double differential di-$\Upsilon$ data extending the existing sample of CMS
would allow one to pin down possible $v^2$-suppressed phenomena, in addition to provide data at higher scales.

Regarding (iv), we are particularly excited by the non-zero (negative) value for $\langle\cos 4\phi_{\CS}\rangle$ with a significance of about 1.6~$\sigma$ recently obtained by LHCb~\cite{LHCb:2023ybt}, as it provides a very first hint of a non-zero $h_1^{\perp g}$. 
More data differential in $\cos \theta_{CS}$ should be able not only to observe the effect of linearly polarised gluons in unpolarised protons, but to start to test if their distribution can be described with the same framework as the unpolarised gluon distribution which we have fit in the present work. 

\section*{Acknowledgements}
We would like to thank T. Rabemananjara for essential inputs on the fitting methods.
We also acknowledge useful discussions with L. An, V.~Bertone, C.~Flore and C.~Pisano.

SFR acknowledges financial support from the fellowship
PIF22/219 of the EHU.
PT acknowledges support from the Research Foundation-Flanders (FWO) through fellowship no. 1233422N.

This project has also received funding from the Agence Nationale de la Recherche (ANR) via the grants ANR-20-CE31-0015 (``PrecisOnium''), and via the IDEX Paris-Saclay ``Investissements d'Avenir" (ANR-11-IDEX-0003-01) through the GLUODYNAMICS project funded by the ``P2IO LabEx (ANR-10-LABX-0038)". 
This work was also partly supported by the French CNRS via the IN2P3 projects ``GLUE@NLO" and ``QCDFactorisation@NLO".
It was also supported by the Spanish Ministerio de Ciencia,
Innovación y Universidades (MCIN/AEI/10.13039/501100011033) through grant No. PID2025-174203NB-I00, by the Basque Government through the grant IT1977-26, as well as by the European Union COST action No. 24159 (SHARP).

\appendix

\section{Complementary material to Sec.~\ref{Limitations of existing approaches}}
\label{appendix:Complementary material to sec}

\subsection{Choice of $b_{\sT,\max}$}
\label{appendix:selection of bmax}

In this section we discuss whether the kinematic region covered by the experimental data is genuinely sensitive to the non-perturbative dynamics of the process and, consequently, whether it allows one to isolate the non-perturbative contribution to the gluon TMD distribution.
We conclude that a low value of $b_{\sT,\max}$ is preferred.

This question is governed by the relative importance of the perturbative contribution in the prediction of the observable at a given collision energy $\sqrt{s}$ and hard scale $Q$.
Since perturbative and non-perturbative effects are naturally defined in impact-parameter $b_{\sT}$ space, whereas the observable is measured in transverse-momentum $q_{\sT}$ space, it is essential to analyse the Fourier transform in Eq.~(\ref{eq:Cff}) that connects these two representations.
If the integrand of the Fourier transform is dominated by the small-$b_{\sT}$ region, the contribution from non-perturbative physics is expected to be negligible.
Conversely, if the large-$b_{\sT}$ region provides a non-negligible contribution, the process becomes particularly suitable for probing the non-perturbative contribution of the gluon TMD.

\begin{figure}[htbp]
    \centering
    \includegraphics[width=\linewidth]{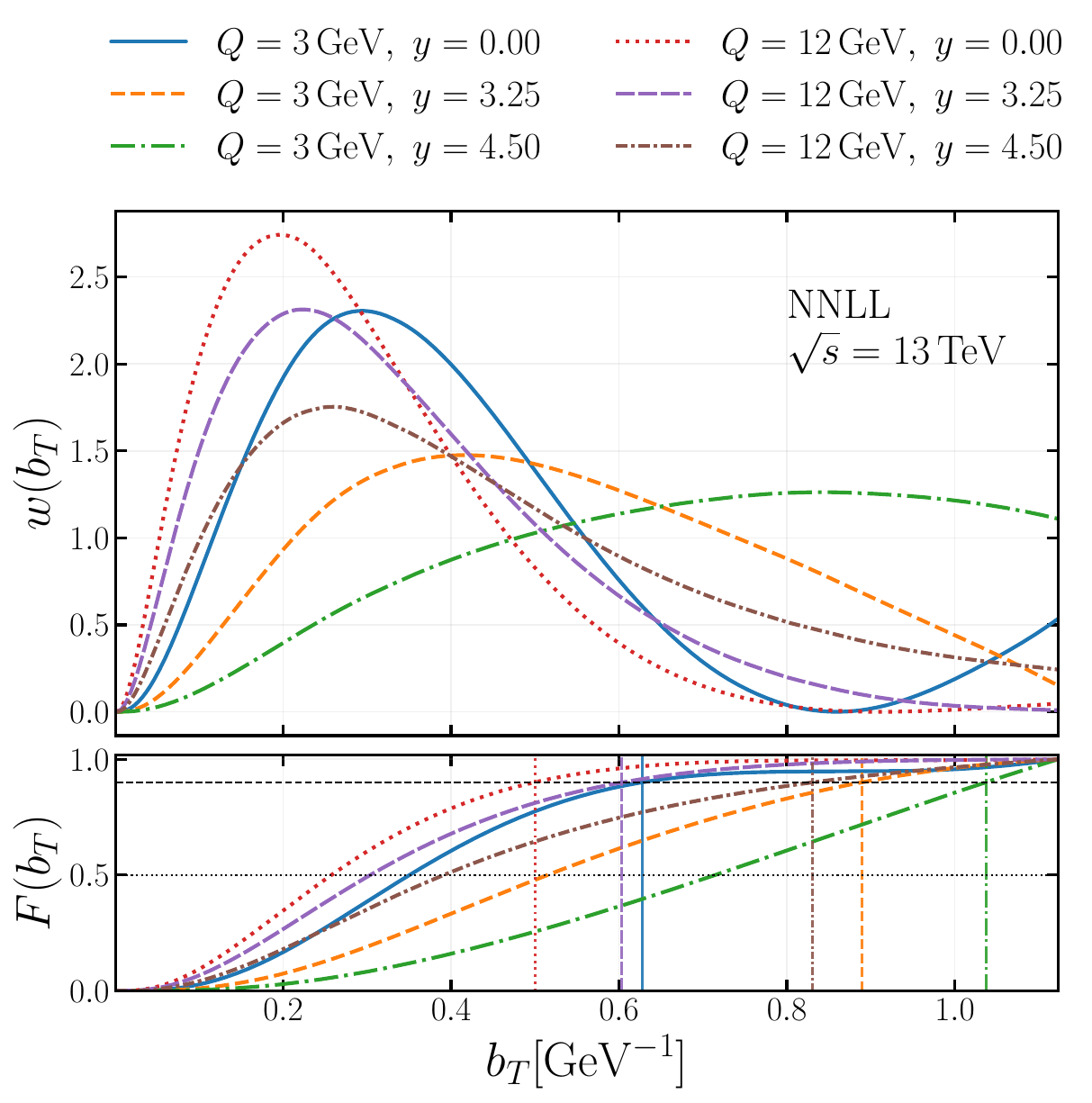}
    \caption{The upper panel shows the normalised distribution, while the lower panel displays the cumulative distribution of the perturbative part of the integrand of the convolution of two unpolarized gluon TMDs evaluated in $b_{\sT}$ space at $q_{\sT}Q \!=\!0$.
    That is, the integrand of Eq.~(\ref{eq:Cff}) evaluated at $b_{\sT}$ and $\mu_b$, with the non-perturbative contribution set to zero ($S_{\mathrm{NP}}\!=\!0$) and at $q_{\sT} \!=\! 0$.
    The vertical lines show the $90\%$ thresholds for the six distributions.
    The dashed and pointed horizontal lines display the $90\%$ and $50\%$ thresholds from the cumulative distribution.
    }
    \label{fig:Wper_norm_cum}
\end{figure}

In what follows, we use the perturbative calculation part from Eq.~(\ref{eq:Cff}) evaluated as a function of $b_{\sT}$ at NNLL, i.e., the object defined in Eq.~(\ref{eq:Wpert})
multiply by  ($b_{\sT}/2\pi$)
at NNLL.
Note that we evaluate the integrand of the convolution at $q_{\sT} \!=\! 0$, for which $J_0(q_{\sT} b_{\sT})=1$.
However, for $q_{\sT}\!>\!0$, the Bessel function provides an additional suppression of the large-$b_{\sT}$ region of the integration.
This suppression becomes stronger as $q_{\sT}$ increases, so it is sufficient to restrict the analysis to the $q_{\sT} \!=\! 0$ case.

To quantify where the perturbative information is predominantly concentrated, in Fig.~\ref{fig:Wper_norm_cum} (upper), we consider  the following normalised distribution:
\begin{equation}
    w(b_{\sT}) \equiv \frac{b_{\sT} W^{\rm pert}(b_{\sT})}{\int_{b_{\rm low}}^{b_{\rm high}} \mathrm{d}b_{\sT}\, b_{\sT} \, W^{\rm pert}(b_{\sT})} \; ,
\end{equation}
evaluated in $b_{\sT} \in [b_{\rm low}, b_{\rm high}]Q \!=\! [10^{-3},b_0/\mu_0] \, \text{GeV}^{-1}$ with $\mu_0$ being the minimum scale of the PDF set considered in the calculation; in particular, $\mu_0 \!=\!1 \, \text{GeV}$ for MSHT20 at NLO~\cite{Bailey:2020ooq}.
We perform the analysis for the same Q values considered in the analysis of Sec.~\ref{Limitations of existing approaches}, i.e., $Q \!=\! 3$ and 12~GeV, and three representative scenarios for the rapidity.
In this representation, the area under the curve is normalised to unity, allowing a direct comparison of the relative importance of different regions in $b_{\sT}$-space.

In addition, in the lower panel of Fig.~\ref{fig:Wper_norm_cum}, we show the following cumulative distribution
\begin{equation}
    F(b_{\sT})  \equiv \int_{b_{\rm low}}^{b_{\sT}} w(t) \, \mathrm{d}t \; .
\end{equation}
This distribution provides a quantitative measure of the integrated support of the underlying differential distribution, allowing  to determine how the total normalisation is progressively saturated as the integration variable increases.
The relative contribution from different regions of $b_{\sT}$ can be directly assessed, making it possible to identify the domain where the dominant perturbative contributions are concentrated.

In the lower panel, the dashed horizontal line denotes the $90\%$ saturation level of the cumulative integral, while the corresponding vertical lines define the values of $b_{\sT}$ at which this threshold is reached for each of the curves.
These intersections therefore delimit the effective support region containing the bulk of the perturbative information.
The most relevant feature emerging from this analysis is that the $90\%$ saturation thresholds (vertical lines) are systematically located below approximately $b_{\sT} \simeq 1.0~\mathrm{GeV}^{-1}$.
This behaviour indicates that the dominant contribution to the observable originates from the small-$b_{\sT}$ region, where the perturbative treatment remains theoretically reliable and under good control.

Once these curves are multiplied by $\exp[-S_{\rm NP}]$ and evaluated within the $b_{\sT}^*$-prescription, the contribution of the non-perturbative parameters in $S_{\rm NP}$ to the $q_{\sT}$ shape of the observable becomes negligible for $b_{\sT,\max} \gtrsim 1.0~\mathrm{GeV}^{-1}$.
In this configuration, the contribution arising from the large-$b_{\sT}$ becomes comparatively suppressed.
As a result, the corresponding theoretical predictions exhibit a significantly reduced sensitivity to the specific modelling of the non-perturbative sector, thereby improving the perturbative stability and robustness of the phenomenological description.

Furthermore, we find that the value of $y$ plays a relevant role in addressing the question raised in this section.
As $y$ increases, the width of the Gaussian distribution becomes larger, causing a greater fraction of the information encoded in it to extend into the non-perturbative large-$b_{\sT}$ region.
On the other hand, $Q$ produces the opposite effect: as its value decreases, the width of the distribution increases and the peak shifts toward larger $b_{\sT}$ values.
This trend can be observed from the intersection of the curves in the lower panel with the horizontal dashed line, i.e., $F(b_{\sT})Q \!=\!0.5$.
This finding is in agreement with the conclusions of Ref.~\cite{Grewal:2020hoc}, which indicate that the predictive power of the theoretical description for unpolarised TMD reaches its maximum in the large-$Q$ region with one of the gluons probed at low $x$ and the other at high $x$.

Based on these results, we conclude that for $\eta_{c}$ ($Q \!=\! 3 \, \rm GeV$) production, the distribution is broader and centred at larger $b_{\sT}$ values, extending significantly into the non-perturbative region.
As a result, the observable exhibits a strong sensitivity to the large-$b_{\sT}$ physics.
The case considered in this work, namely di-$J/\psi$ production ($Q \!=\! 12 \, \rm GeV$), still constitutes a suitable process for extracting information on non-perturbative effects.
Nevertheless, the numerical value of $b_{\sT,\max}$ must be chosen carefully, since contributions from the region $b_{\sT} > 1 \, \text{GeV}^{-1}$ may become effectively irrelevant for determining the shape of the distribution in $q_{\sT}$ space.

Finally, although it is not shown here, the corresponding distribution for Higgs production ($Q \!=\! 125.1\, \rm GeV$) is sharply peaked and narrowly centred around $b_{\sT} \simeq 0.1\,\text{GeV}^{-1}$ reaching the $90\%$ threshold at $b_{\sT} \simeq 0.3\,\text{GeV}^{-1}$.
Consequently, it is expected that this observable  is basically insensitive to the values of the non-perturbative parameters in $S_{\rm NP}$, in line with the conclusions of Ref.~\cite{Echevarria:2015uaa}.

\begin{figure*}[htbp]
    \centering

\includegraphics[width=0.95\textwidth]{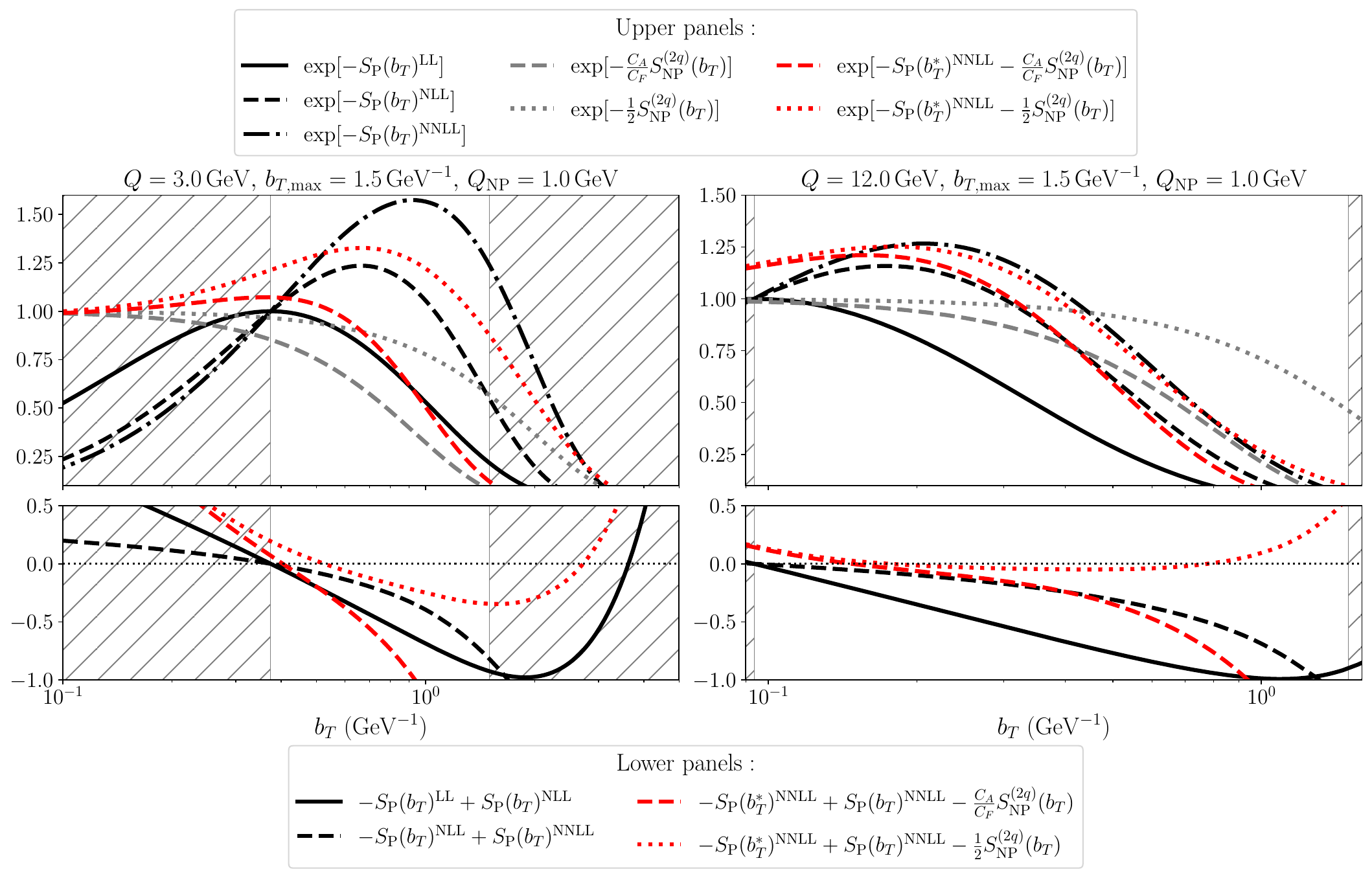}
    \caption{
    Upper panel:    perturbative LL, NLL and NNLL Sudakov factors (3 black curves), 
    non-perturbative Sudakov factors using $S_{\text{NP}}^{(g)}\!=\!(1/2, C_A/C_F)S_{\text{NP}}^{(q)}$  for $b_{\sT,\text{max}}=1.5\text{ GeV}^{-1}$ (see Eq.~\eqref{eq:twoquarkSudakov}) (2 gray curves) and their combination at NNLL (2 red curves) for $Q \!=\!(3,12)~\text{ GeV}$ (left, right).
    Lower panel: 
    differences between the perturbative Sudakov factors at successive logarithmic orders
    %we show the LL, NLL and NNLL Sudakov factors 
    evaluated at $b_{\sT}$ (black), compared with the difference between the NNLL Sudakov evaluated at $b_{\sT}$ and the corresponding Sudakov at $b_{\sT}^*$ including the just defined $S_{\text{NP}}$ (red).
    The white areas represent the region $b_{\sT} \in [b_0/Q, b_{\sT,\text{max}}]$.} 
    \label{fig:sudakov_combined2}
\end{figure*}

\subsection{Analysis at NLL and for $b_{\sT,\max}\!=\!1.5 \, \rm GeV^{-1}$}
\label{appendix:Complementary discussion on NLL and bmax15}

\begin{figure*}[ht]
    \centering

\includegraphics[width=0.95\textwidth]{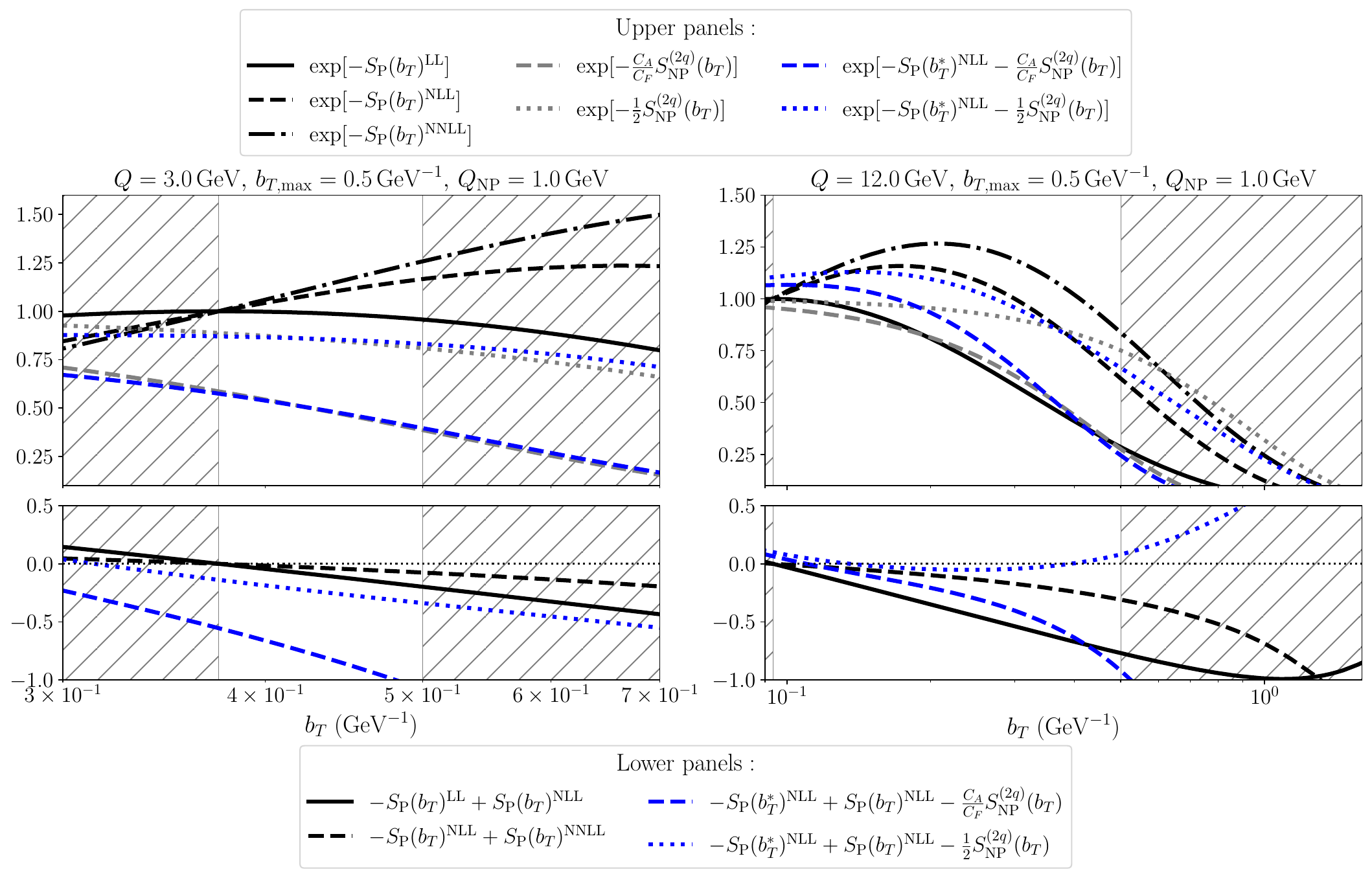}
    \caption{Upper panel:    perturbative LL, NLL and NNLL Sudakov factors (3 black curves), 
non-perturbative Sudakov factors using $S_{\text{NP}}^{(g)}\!=\!(1/2, C_A/C_F)S_{\text{NP}}^{(q)}$  for $b_{\sT,\text{max}}=0.5\text{ GeV}^{-1}$ (see Eq.~\eqref{eq:twoquarkSudakov}) (2 gray curves) and their combination at NLL (2 blue curves) for $Q \!=\!(3,12)~\text{ GeV}$ (left, right)
    Lower panel: 
    differences between the perturbative Sudakov factors at successive logarithmic orders
    %we show the LL, NLL and NNLL Sudakov factors 
    evaluated at $b_{\sT}$ (black), compared with the difference between the NLL Sudakov evaluated at $b_{\sT}$ and the corresponding Sudakov at $b_{\sT}^*$ including the just defined $S_{\text{NP}}$ (blue).
    The white areas represent the region $b_{\sT} \in [b_0/Q, b_{\sT,\text{max}}]$.} 
    \label{fig:sudakov_combined3}
\end{figure*}

\begin{figure*}[ht]
    \centering

\includegraphics[width=0.95\textwidth]{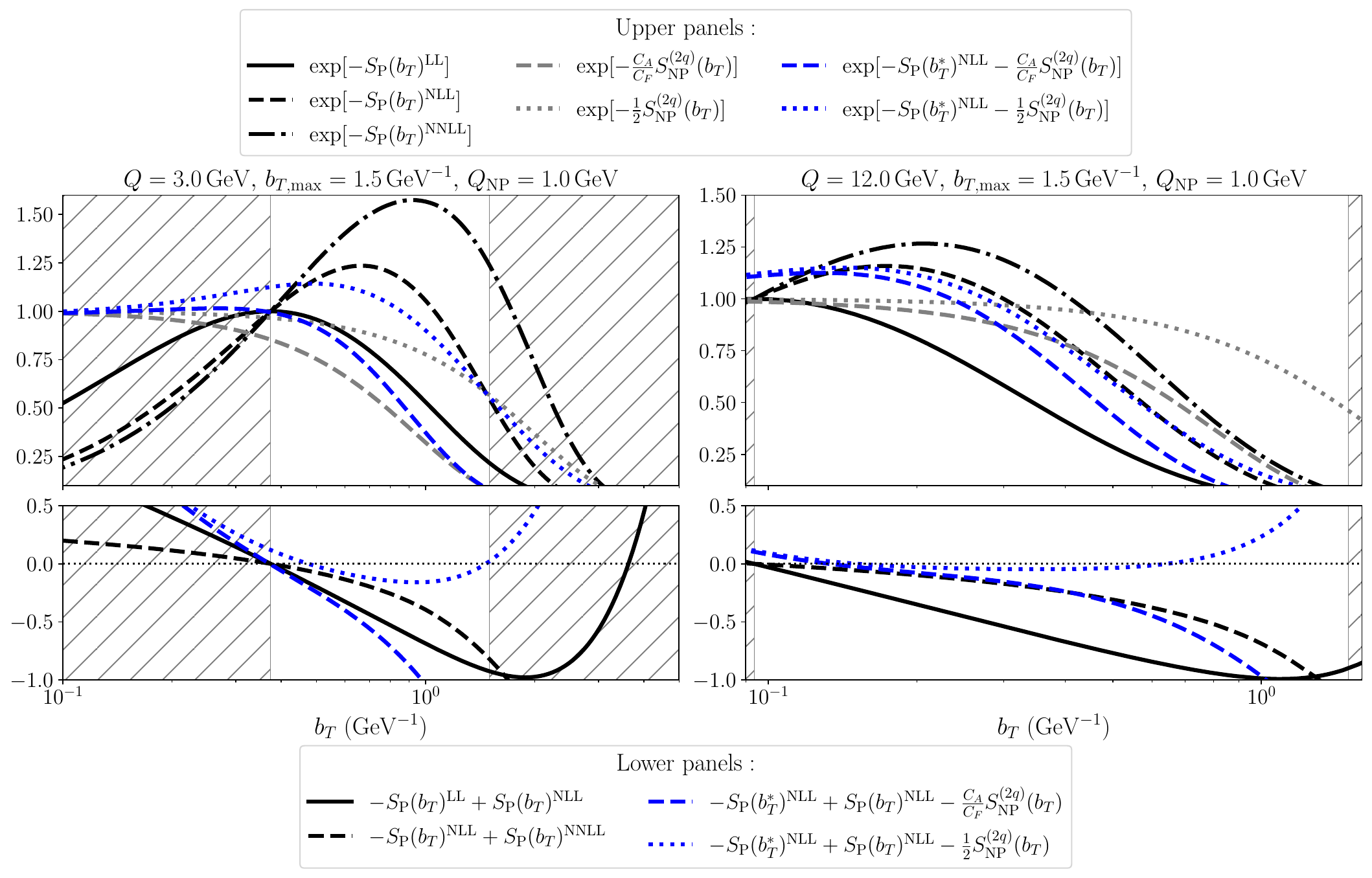}
    \caption{Upper panel:    perturbative LL, NLL and NNLL Sudakov factors (3 black curves), 
    non-perturbative Sudakov factors using $S_{\text{NP}}^{(g)}\!=\!(1/2, C_A/C_F)S_{\text{NP}}^{(q)}$  for $b_{\sT,\text{max}}=1.5\text{ GeV}^{-1}$ (see Eq.~\eqref{eq:twoquarkSudakov}) (2 gray curves) and their combination at NLL (2 blue curves) for $Q \!=\!(3,12)~\text{ GeV}$ (left, right)
        Lower panel: 
    differences between the perturbative Sudakov factors at successive logarithmic orders
    %we show the LL, NLL and NNLL Sudakov factors 
    evaluated at $b_{\sT}$ (black), compared with the difference between the NLL Sudakov evaluated at $b_{\sT}$ and the corresponding Sudakov at $b_{\sT}^*$ including the just defined $S_{\text{NP}}$ (blue).
    The white areas represent the region $b_{\sT} \in [b_0/Q, b_{\sT,\text{max}}]$.} 
    \label{fig:sudakov_combined4}
\end{figure*}

In this appendix, we present an analysis analogous to that of Sec.~\ref{Systematic effects induced by the conventional modelling of non-perturbative effects}, considering instead the case $b_{\sT,\text{max}} \!=\! 0.5\, \text{GeV}^{-1}$ at NLL, and the case $b_{\sT,\text{max}}\! =\! 1.5\, \text{GeV}^{-1}$ at NLL and NNLL.

In Ref.~\cite{Konychev:2005iy}, for $b_{\sT,\max}\!=\!1.5 \, \text{GeV}^{-1}$, the authors consider the
following non-perturbative model for one quark TMD PDF:
\begin{equation}
\label{eq:one quark 2}
    S^{(q)}_{\text{NP}}(Q,x) = \frac{a_2}{2} \ln \left( \frac{Q}{3.2\text{GeV}} \right) +  
    \frac{a_1}{2} + a_3 \ln (10 x) ,
\end{equation}
with $a_1\! =\! 0.201 \pm 0.011 \, \text{GeV}^2$, $a_2\! =\! 0.184 \pm 0.018 \, \text{GeV}^2$ and $a_3 \!=\! -0.026 \pm 0.007\, \text{GeV}^2$.
According to Eqs.~(\ref{eq:SNPsimple}) and (\ref{eq:B(x)}), we consider here the following non-perturbative parameters $A \!=\! 0.184\,\text{GeV}^2$, $b_1 \!=\! -0.026 \,\text{GeV}^2$ and $b_2\!=\!-0.007$.

Fig.~\ref{fig:sudakov_combined2} displays the case of $b_{\sT,\text{max}}\! =\! 1.5 \, \text{GeV}^{-1}$ at NNLL.
For $Q \!=\! 3\,$GeV, in comparison with the case of $b_{\sT,\text{max}} \!=\! 0.5 \, \text{GeV}^{-1}$ (Fig.~\ref{fig:sudakov_combined}), we observe here that the peak of the $S_\text{P}$ distributions (black curves) are now in the white area.
Furthermore, as can be seen in the upper panel, the differences between $S_\text{p}(b_{\sT}^*)+S_{\text{NP}}$ (red curves) and $S_\text{P}(b_{\sT})$ (black curves) are significantly larger.
Particularly, as quantified in the lower panel, for $(C_A/C_F) S_{\text{NP}}$ (red dashed curve), the absolute difference exceeds the value of one at $b_{\sT}\!=\!b_{\sT,\max}$.
As for $b_{\sT,\max} \!=\! 0.5 \, \text{GeV}^{-1}$, this deviation is larger than the black dashed curve, probing that  the conventional model (with $n \!=\! 2$) produces an excessively large deviation in the perturbative region for some non-perturbative parameters.
For $Q \!=\! 12\,$GeV, the lower panel shows that the difference at $b_{\sT}\!=\!b_{\sT,\max}$ between NLL and NNLL calculations is slightly larger than that between LL and NLL calculations, indicating that $\alpha_s(\mu_b)$ begins to lose its perturbative behaviour.

For completeness, we also show the corresponding NLL plots in Fig.~\ref{fig:sudakov_combined3} ($b_{\sT,\max} \!=\! 0.5 \, \text{GeV}^{-1}$) and Fig.~\ref{fig:sudakov_combined4} ($b_{\sT,\max}\!=\!1.5 \, \text{GeV}^{-1}$).
The conclusions are similar to those obtained at NNLL; the conventional approach ($n \!=\! 2$) is not suitable for some non-perturbative parameters if one aims to avoid significant modifications of $S_\text{P}(b_{\sT})$ within the perturbative region (white area).

\subsection{Further discussion on $b_{\sT}^*(n)$}
\label{appendix:Further discussion on bT*(n)}

\begin{figure}[htbp]
    \centering
    \includegraphics[width=\linewidth]{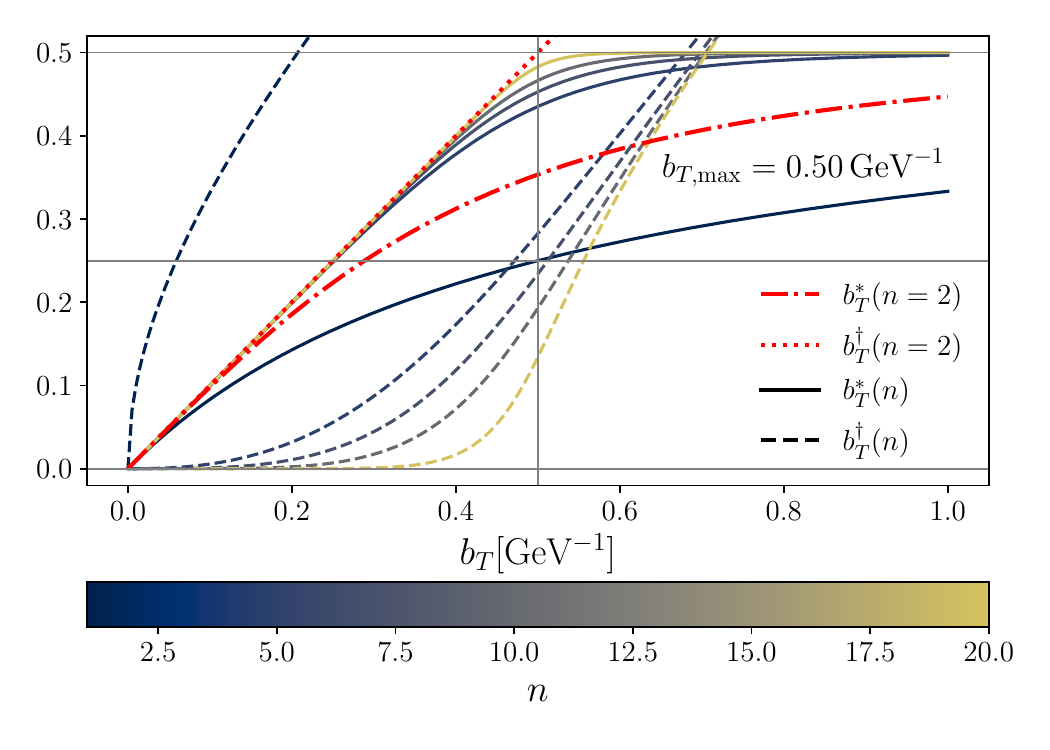}
    \caption{$b_{\sT}$-distributions of $b_{\sT}^*$ (solid lines) and $b_{\sT}^\dagger$ (dashed lines) for $n\!=\![1,20]$ at $b_{\sT,\max} \!=\! 0.5 \, \rm GeV^{-1}$.
    }
    \label{fig:n-scan of bTstar and bTdagger}
\end{figure}

Fig.~\ref{fig:n-scan of bTstar and bTdagger} compares $b_{\sT}^*(n)$ (solid lines) and $b_{\sT}^\dagger(n)$ (dashed lines) as a function of $b_{\sT}$ in the range $n\!=\![1,20]$ at $b_{\sT,\max} \!=\! 0.5 \, \rm GeV^{-1}$.
The corresponding distributions to the conventional approach (i.e., $n \!=\! 2$) are highlighted in a red dash-dot line for $b_{\sT}^*(n\!=\!2)$ and in a red dotted line for $b_{\sT}^\dagger(n\!=\!2)$.

It is clearly seen that $b_{\sT}^*$ defines an intermediate region centred at $b_{\sT}\!=\!b_{\sT,\max}$, where the freezing transition is smoothly interpolated between the small- and large-$b_{\sT}$ regimes ($b_{\sT}\leq b_{\sT,\max}$ and $b_{\sT}>b_{\sT,\max}$, respectively).
As $n$ increases, this transition region becomes progressively narrower.
Particularly, for $n \!=\! 2$, $b_{\sT}^* \simeq 0.35 \, \rm GeV^{-1}$ at $b_{\sT,\max}$ and it reaches $b_{\sT}^*\!=\!b_{\sT,\max}$ in $b_{\sT} > 1 \, \rm GeV^{-1}$.
As a consequence, defining $b_{\sT,\max}$ as the separation point between these two regions becomes meaningless.
By contrary, for $n \!=\! 20$, the transition region becomes very narrow with $b_{\sT}^* \simeq b_{\sT,\max}$ at $b_{\sT,\max}$.
Moreover, we observe a significant change in the $b_{\sT}^*$ distribution when going from $n \!=\! 2$ to $n\!=\!5$, after which it stabilizes for $n > 7$.
As for $b_{\sT}^\dagger$, its distribution in the region $b_{\sT} < b_{\sT,\max}$ decreases as $n$ increases. Therefore, if the goal is to minimise the contribution of $S_{\rm NP}$ in this region, larger values of $n$ are preferred.
Naturally, this behaviour extends to any value of $b_{\sT,\max}$.

\subsection{Deviating from the perturbative calculation}
\label{appendix:Deviation from the perturbative part contribution}

\begin{figure}[htbp]
    \centering
    \includegraphics[width=\linewidth]{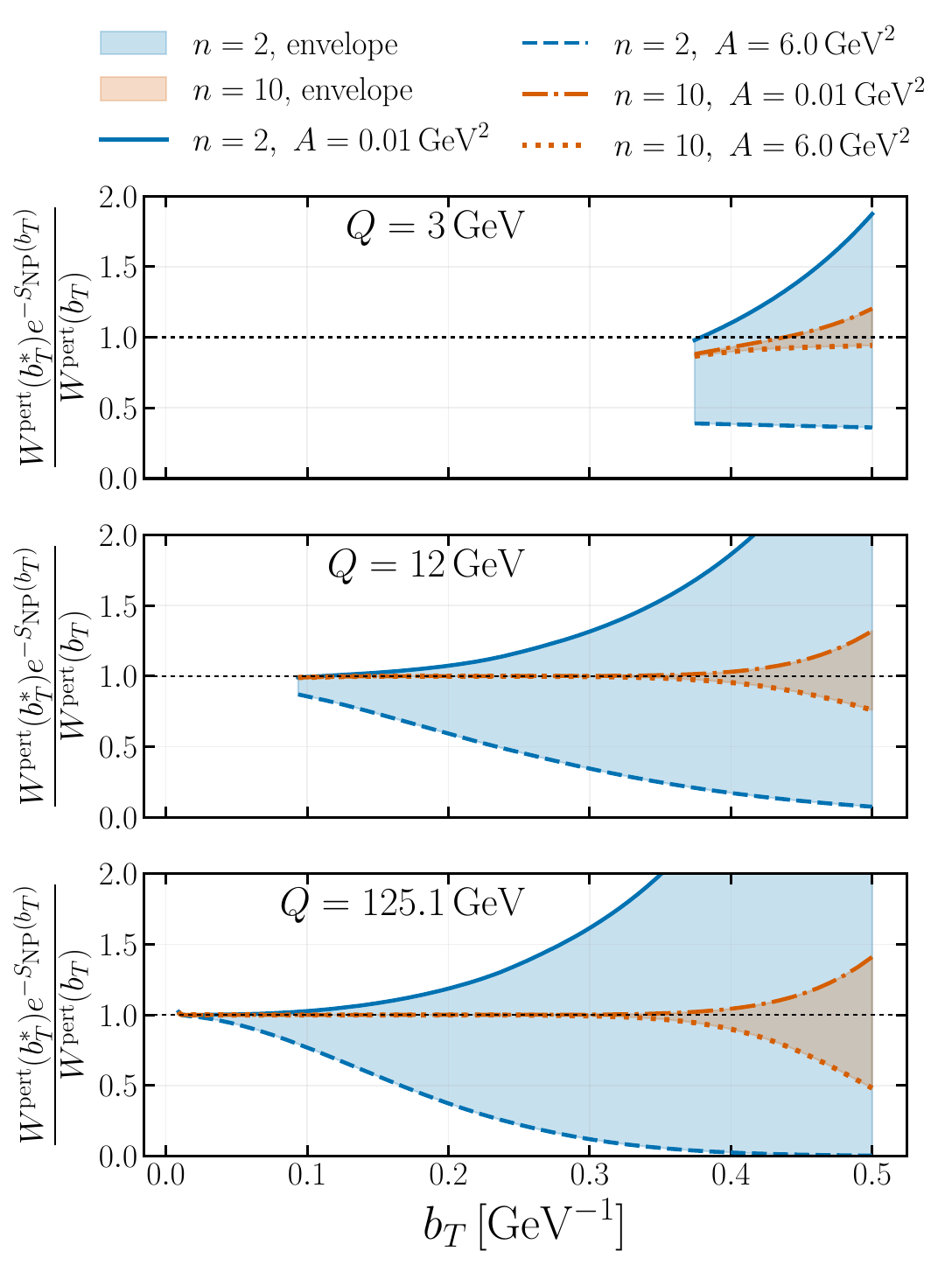}
    \caption{
    $b_{\sT}$ distribution of the ratio defined in Eq.~(\ref{eq:ratio deviation}) for $n \!=\! 2$ and $n \!=\! 10$, for $Q \!=\! 3, \, 12$ and $125.1$~GeV, for $A \!=\![0.01, 6] \, \rm GeV^2$ and for $y\!=\!0$ at $\sqrt{s}\!=\!13 \, \rm GeV$.
    The ratios are calculated at NNLL in the region $b_{\sT} \in [b_0/Q, b_{\sT,\text{max}}]$ for $b_{\sT,\max} \!=\! 0.5 \, \rm GeV^{-1}$.
    }
    \label{fig:deviation from pert part}
\end{figure}

In Sec.~\ref{sec:Beyond the conventional bstar prescription}, we propose an alternative approach to the conventional methods for describing the non-perturbative effects (already discussed in Sec.~\ref{Systematic effects induced by the conventional modelling of non-perturbative effects}).
In this appendix, we analyse in detail the effect of the alternative proposed method by comparing with the conventional ones in impact-parameter space.

The motivation for defining the Eq.~(\ref{EqNonPertS}) is to obtain a complete description of the cross section over the entire $b_{\sT}$ range for processes whose cross section is proportional to the convolution of two gluon TMD PDFs.
Note that Eq.~(\ref{EqNonPertS}) is, in fact, a direct consequence of attempting to describe the three-dimensional distribution of gluons inside the proton while acknowledging that only the small-$b_{\sT}$ region can be treated perturbatively.
Therefore, in Eq.~(\ref{EqNonPertS}), the left-hand side is expected to coincide with the right-hand side within the perturbative region, whose extent is determined by the prescription used to define it, in this case the $b_{\sT}^*$ prescription.
It is possible that the effect induced by evaluating $W^{\rm pert}$ at $b_{\sT}^*$ generates a deviation with respect to $W^{\rm pert}(b_{\sT})$ that is compensated by the contribution from $\exp[-S_{\mathrm{NP}}]$. However, the fundamental limitation of conventional approaches is that they provide no control over such a compensation mechanism.

To examine this behaviour, we evaluate the ratio
\begin{equation}
\label{eq:ratio deviation}
    \left. \frac{W^{\rm pert}(b_{\sT}^*) e^{-S_{\text{NP}}(b_{\sT})}}{W^{\rm pert}(b_{\sT})} \right|_{b_{\sT} = [b_0/Q,b_{\sT,\text{max}}]}
\end{equation}
for three representative cases of $Q$.
For the $b_{\sT}^*$ and the $S_{\text{NP}}$, we use the ones defined in Sec.~\ref{sec:Beyond the conventional bstar prescription} with $b_{\sT,\text{max}} \!=\! 0.5 \, \text{GeV}^{-1}$, $B \!=\!0$ and $Q_{\text{NP}}\!=\!1\, \text{GeV}$ for the range $A \!=\![0.01, 6]\, \text{GeV}^2$.
We again stress that $n \!=\! 2$ represent the conventional approaches.
The calculation is performed at NNLL with MSHT20~\cite{Bailey:2020ooq} at NLO as the PDF set using its $\alpha_s$ evaluation.

The results are shown in the Fig.~\ref{fig:deviation from pert part}.
The bands represent the envelopes with respect to the value of $A$ for the case with $n \!=\! 2$ (blue) and $n \!=\! 10$ (orange).
We also show the considered extreme values of $A$ for each scenario, which coincide with the boundaries of the bands.
For all three values of $Q$ and both choices of $n$, the ratio reaches its maximum at $b_{\sT}\!=\!b_{\sT,\text{max}}$ for the range $A$, as expected.
More importantly, increasing the $n$ from 2 to 10 leads to a substantial reduction of the ratio.
While for $n \!=\! 2$ the deviation can reach $100\%$ for extreme values of $A$ before $b_{\sT,\text{max}}$, for $n \!=\! 10$ it remains below approximately $40\%$ even at the endpoint $b_{\sT}\!=\!b_{\sT,\text{max}}$.
Moreover, we observe that the deviation from unity appears at smaller values of $b_{\sT}$ as $Q$ decreases.
For $Q \!=\! 3 \, \text{GeV}$, a non-zero deviation is already visible at $b_{\sT}\!=\!b_0/Q$, even in the case $n \!=\! 10$.
This can be traced back to the fact that, for lower values of $Q$, the perturbative region extends closer to the peak of $W(b_{\sT})$.
Consequently, the $b_{\sT}^*$ prescription modifies a region where the distribution varies more rapidly, leading to a larger deviation from the purely perturbative calculation.
Finally, we have verified that a similar behaviour reported in this appendix is obtained for different values of $b_{\sT,\text{max}}$.

Therefore, the analysis presented in this appendix further supports the importance of the  implementation of the $b_{\sT}^*$ prescription combined with the non-perturbative Sudakov $S_{\text{NP}}$,
discussed in Sec.~\ref{sec:Beyond the conventional bstar prescription}.
If one were to perform the fit using the conventional choice $ n \!=\! 2$, the extracted non-perturbative parameters could lead to a value of the ratio in Eq.~(\ref{eq:ratio deviation}) that is excessively large, implying a significant modification of the perturbative region.
This highlights the importance of controlling the interplay between the perturbative and non-perturbative components when extracting non-perturbative parameters from data.

\section{Complementary material to Sec.~\ref{sec:classic_fit}}

\subsection{Oscillations in $q_{\sT}$-space from broad large-$b_{\sT}$ non-perturbative models}
\label{appendix:oscillations in qT}

\begin{figure}[t]
    \centering
    \includegraphics[width=\linewidth]{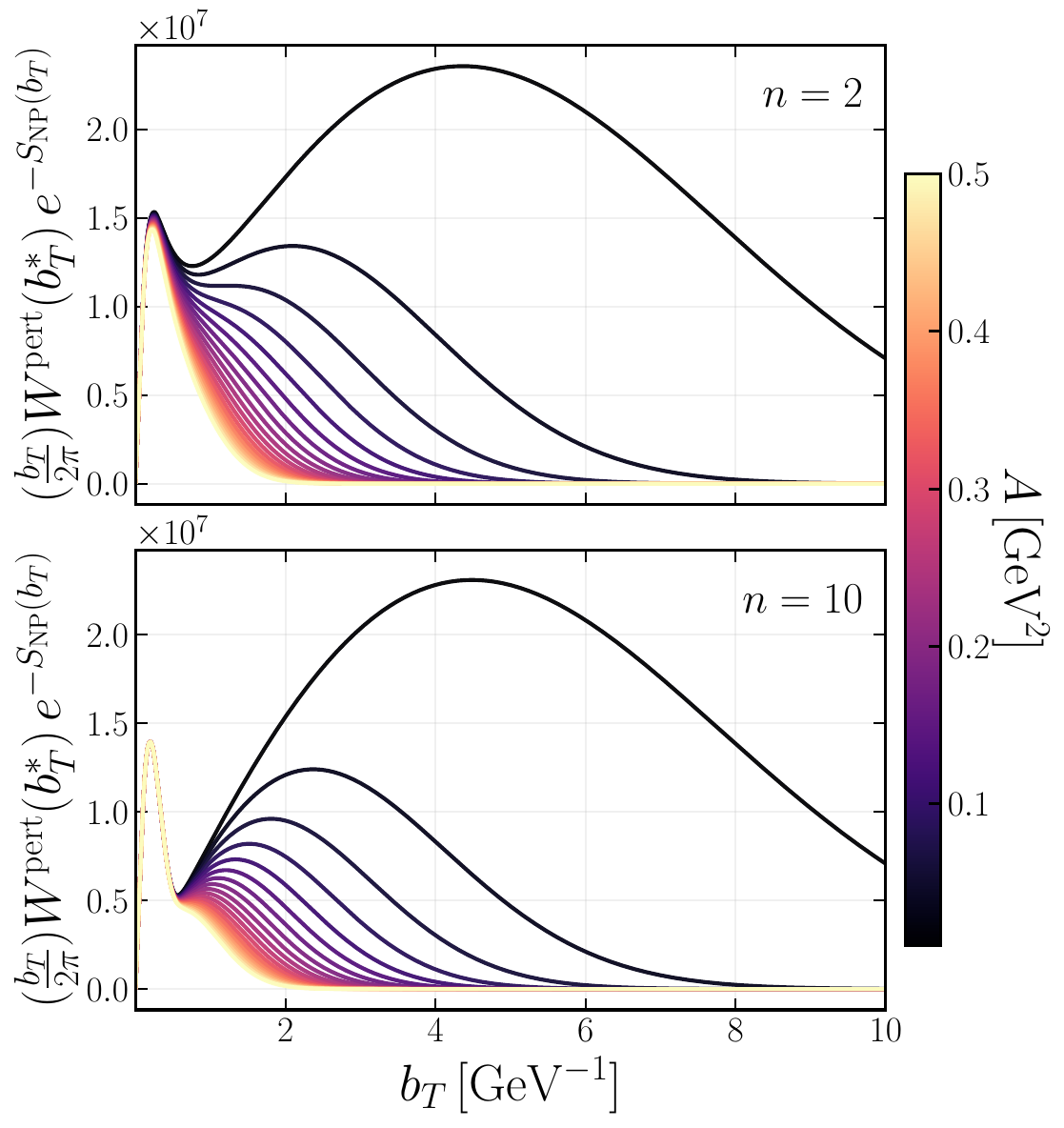}
    \caption{
    $b_{\sT}$ distribution of $(b_{\sT}/2\pi)W^{\rm pert} e^{-S_{\text{NP}}}$ obtained at NNLL accuracy including freezing effects for different values of the non-perturbative parameter $A$ in $S_{\text{NP}}$.
    The colour scale indicates the value of $A$ in the range $A=[0.01,0.5]\,\rm GeV^2$. 
    The distributions are shown for $Q\!=\!12~\mathrm{GeV}$, $n\!=\!2$ (top), $n \!=\! 10$ (bottom), $b_{\sT,\max}\!=\!0.5~\mathrm{GeV}^{-1}$ and for $y\!=\!0$ at $\sqrt{s} \!=\! 13 \, \rm TeV$.
    }
    \label{fig:bumps in bt}
\end{figure}

As discussed in the following analysis, since $W^{\rm pert}$ is centred at small values of $b_{\sT}$ ($\lesssim 1.0\,\text{GeV}^{-1}$) and exhibit a very narrow width, gluon TMD phenomenology presents limitations in the definition of the non-perturbative region.

Within the large freedom allowed in the functional form used to describe the non-perturbative large-$b_{\sT}$ region, the resulting behaviour is generally expected to provide a sufficiently strong suppression as $b_{\sT} \!\to\! \infty$, while Gaussian-like profiles in the intermediate-$b_{\sT}$ regime are often favoured by phenomenological studies~\cite{Collins:1984kg}. The strength of this suppression is governed by the non-perturbative parameters entering the model. Since these qualitative features are largely independent of the particular parametrisation adopted, the conclusions of the following analysis are expected to be of general validity, despite the fact that we explicitly employ the model defined in Eq.~(\ref{eq:SNPsimple}).

In Fig.~\ref{fig:bumps in bt}, we show the integrand of the convolution of two unpolarised gluon TMD PDFs defined in Eq.~(\ref{eq:Cff}) at $q_{\sT} \!=\! 0$, using the $S_{\text{NP}}$ model introduced in Eq.~(\ref{eq:SNPsimple}) with $Q_{\text{NP}}= 1.0\, \text{GeV}$, $B \!=\!0$ and for $A \!=\![0.01,0.5]\,\text{GeV}^2$.
For the kinematics, we choose $y\!=\!0$ and $Q \!=\! 12\, \text{GeV}$ at $\sqrt{s} \!=\! 13 \, \rm TeV$, since this configuration corresponds to the most sharply peaked distribution for the considered configurations in Fig.~\ref{fig:Wper_norm_cum}, and therefore provides the clearest setup for the purpose of this appendix.
In addition, the calculation is performed using the prescription of Eq.~(\ref{eq:modified b*}) for $m\!=\!10$, considering $n \!=\! 2$ in the upper panel and $n \!=\! 10$ in the lower panel, with $b_{\sT,\text{max}} \!=\! 0.5\, \text{GeV}^{-1}$.
The most distinctive feature observed in both panels is the abrupt change in curvature that develops in the intermediate-$b_{\sT}$ region as the values of $A$ decreases.
This effect is particularly evident for the $n \!=\! 10$ case, where the suppression becomes highly localized around $b_{\sT}\!=\!b_{\sT,\text{max}}$.
As a result, the integrand develops a pronounced shoulder-like structure before rapidly decreasing at larger $b_{\sT}$.
This occurs because, as $A$ decreases, the $S_{\text{NP}}$ model becomes increasingly flatter in the intermediate-$b_{\sT}$ region while decreasing significantly at larger $b_{\sT}$ values, which can lead to the appearance of a second peak that is, in some cases (e.g., for $A \!=\!0.01\,\text{GeV}^2$), dominant.

\begin{figure}[t]
    \centering
    \includegraphics[width=\linewidth]{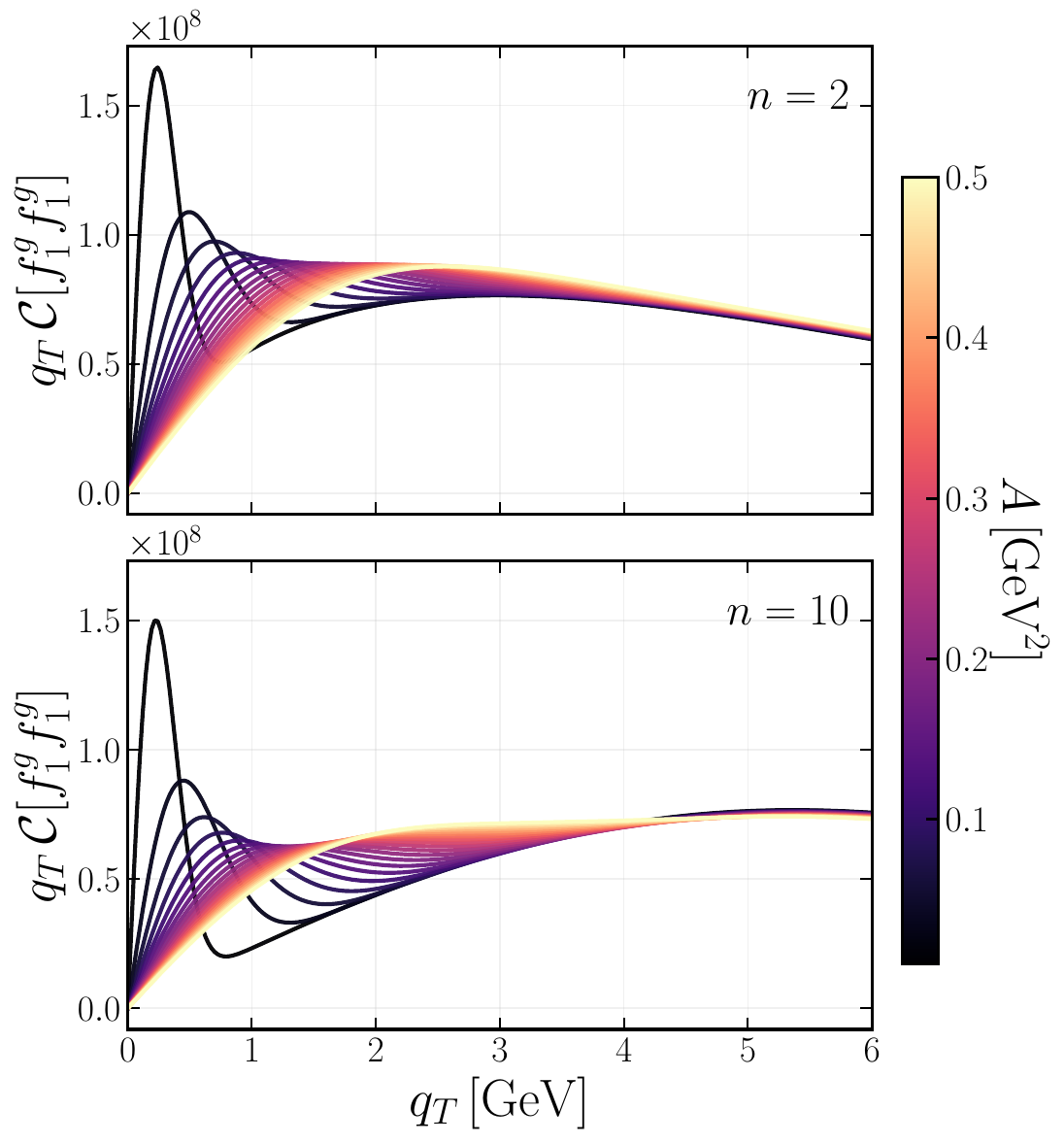}
    \caption{
    $q_{\sT}$ distribution of $q_{\sT} \cdot \mathcal{C}[f_1^g f_1^g]$ obtained at NNLL accuracy including freezing effects for different values of the non-perturbative parameter $A$ in $S_{\text{NP}}$.
    The colour scale indicates the value of $A$ in the range $A=[0.01,0.5]\,\rm GeV^2$. 
    The distributions are shown for $Q\!=\!12~\mathrm{GeV}$, $n\!=\!2$ (top), $n\!=\!10$ (bottom), $b_{\sT,\max}\!=\!0.5~\mathrm{GeV}^{-1}$ and for $y\!=\!0$ at $\sqrt{s} \!=\! 13 \, \rm TeV$.  
    }
    \label{fig:bumps in qT}
\end{figure}

The distributions observed for $A \!\lesssim\! 0.5\,\text{GeV}^2$ develop a pronounced tail toward large $b_{\sT}$, indicating that the Fourier transform receives sizeable contributions from regions where the kernel $J_0(b_{\sT} q_{\sT})$ undergoes multiple oscillations.
Consequently, stronger interference effects among different $b_{\sT}$ domains are expected to emerge in $q_{\sT}$-space, potentially producing oscillatory patterns or rapidly varying curvatures in the resulting $q_{\sT}$ spectrum, as shown in Fig.~\ref{fig:bumps in qT}.

This is precisely what is observed, as can be seen in Sec.~\ref{sec:classic_fit}.
Since the physical observable is proportional to Eq.~(\ref{eq:Cff}), these rapidly varying features are not expected.
In this sense, the appearance of strong oscillatory patterns can be interpreted as an indication that the corresponding non-perturbative configuration does not provide sufficient suppression of long-distance contributions, leading to a loss of phenomenological stability of the TMD description.

Remarkably, imposing the condition $S_{\text{NP}}^{(g)} \!>\! (1/2) S_{\text{NP}}^{(q)}$, as discussed in Sec.~\ref{Limitations of existing approaches}, and using the results of the quark fit of Ref.~\cite{Landry:2002ix}, leads to a lower bound of $A \!\gtrsim\!  0.46, \, 0.51, \, 0.54 \, \text{GeV}^{2}$ for $Q \!=\! 11, \, 7.9, \, 6.6 \, \rm GeV$.
This constraint completely excludes the parameter region analysed above.
Consequently, the physically motivated hierarchy between gluon and quark non-perturbative contributions could naturally prevent the TMD instability presented here.
As discussed in the main text, this feature is particularly relevant for the fit procedure.

\subsection{Comparison between $b_{\sT,\max} \!=\! 0.5 \; \rm{and}\; 1.5 \, \rm GeV^{-1}$ in $q_{\sT}$ space}
\label{sec:Comparison with LHCb data at N$^2$LL}

\begin{figure*}[t]
\centering
\includegraphics[width=\textwidth]{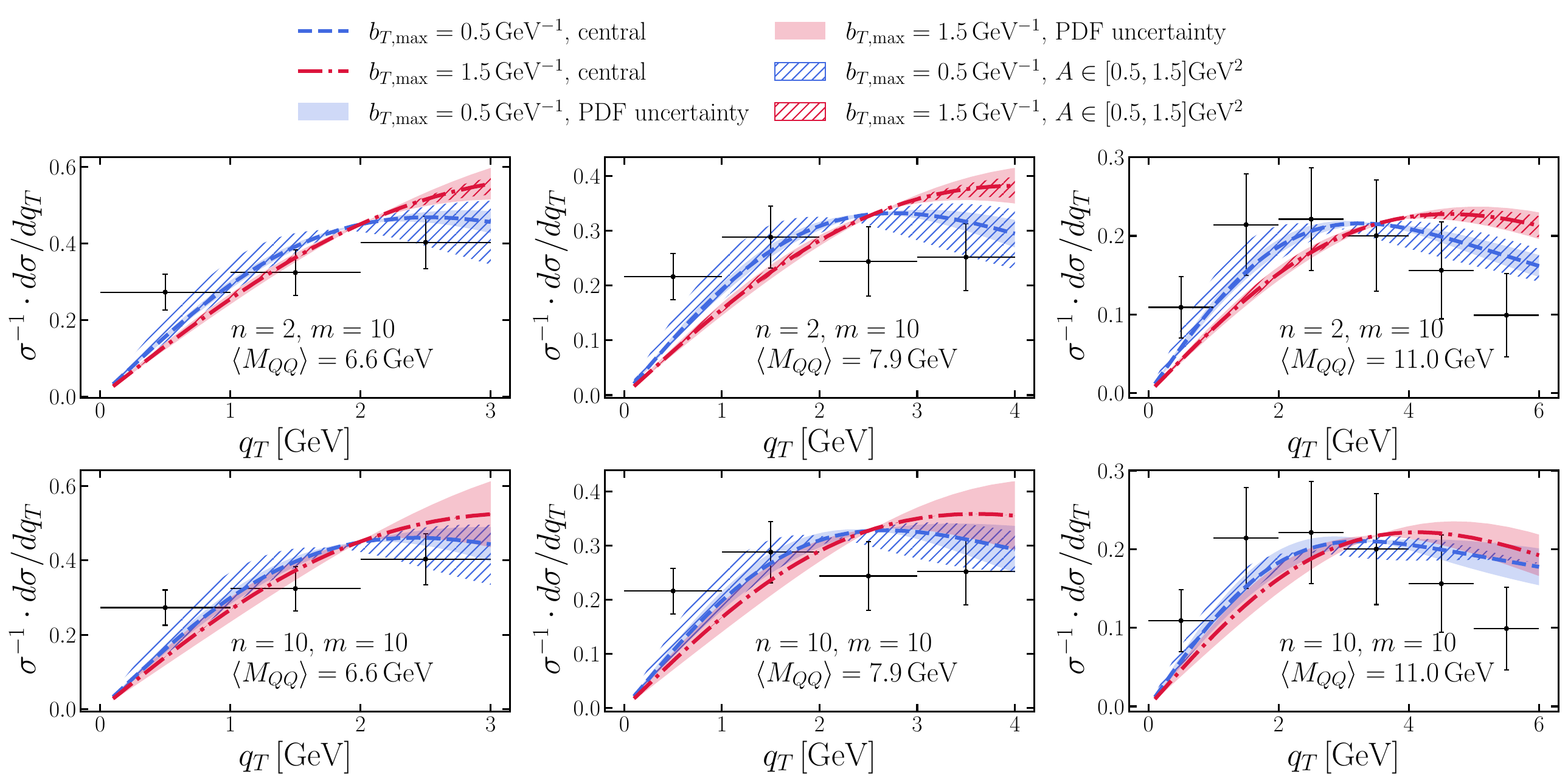}
\caption{
$q_{\sT}$ distributions of the experimental normalised differential cross section for three values of the $J/\psi$-pair invariant-mass, namely $\langle M_{QQ} \rangle \!=\! 6.6, 7.9,11\,$GeV at $\sqrt{s} \!= \!13 \, \rm TeV$ LHCb data~\cite{LHCb:2023ybt}.
We compare with the theoretical normalised cross section at NNLL accuracy for $b_{\sT,\text{max}} \!=\! 0.5\,$ and $1.5 \,$GeV$^{-1}$, and for two different scenarios, each shown in a separate row.
In the first row, we consider $n\!=\!10$ in Eq.~(\ref{eq:modified b*}) with $S_{\text{NP}}$ defined in Eq.~(\ref{eq:SNPsimple}) for $B \!=\! 0$ and $Q_{\text{NP}} \!= \!1\,$GeV.
In the second row, we use $n\!=\!2$.
The solid bands represent the PDF uncertainty of the NLO MSHT20 set.
The hatched bands represent the variation induced by varying the non-perturbative parameter $A$ within the range $A \!=\![0.5,1.5]\,$GeV$^2$.
}
\label{fig:comparison with LHCb}
\end{figure*}

In this section, we compare the normalised cross section using the conventional approach ($n \!=\! 2$) with the approach with $n \!=\! 10$ in $q_{\sT}$ space.
For $S_{\rm NP}$, we use Eq.~(\ref{eq:SNPsimple}) with $Q_{\rm NP} \!=\! 1\,$GeV, and for simplicity, we fix $B\!=\!0$ and perform a scan with respect to $A$.
Additionally, the corresponding predictions are compared with the latest LHCb data~\cite{LHCb:2023ybt} for $J/\psi$-pair production in the region $q_{\sT}\! < \!Q/2$ with $Q \!= \!\langle M_{QQ} \rangle\!= \!6.6, \, 7.9$ and 11~GeV, where  the TMD formalism is expected to be reliable.

We present the results in Fig.~\ref{fig:comparison with LHCb} for two choices of $b_{\sT,\max}$, namely $b_{\sT,\max} \!=\! 0.5 \, \rm GeV^{-1}$ (blue) and $b_{\sT,\max}\! = \!1.5 \, \rm GeV^{-1}$ (red).
The dashed and dot-dash lines represent the central predictions for each one (where $A \!=\! 1.0 \, \rm GeV^2$ was fixed).
The solid bands represent the PDF uncertainty, while the hatched bands show the variation induced by varying the non-perturbative parameter $A$ within the range $A \!=\! [0.5,1.5]\,$GeV$^2$.
This interval is motivated by the analysis presented in Appendix~\ref{appendix:oscillations in qT}, where we find that values $A \!\lesssim\! 0.5 \, \rm GeV^2$ produce oscillatory features in the predicted spectrum, which are not characteristic of a physical differential cross section.

In the first row of panels, we present the theoretical predictions for $n \!= \!10$.
Since we take $n\!=\!10$, such that the perturbative calculation in the $b_{\sT} \!<\! b_{\sT,\text{max}}$ is unaffected by the $S_{\text{NP}}$, the obtained shapes stem purely from the impact of the $b_{\sT}\! > \!b_{\sT,\text{max}}$ region, as described by the parameter $A$ through the functional form of $S_{\text{NP}}$.
We find that the $A$-band corresponding to $b_{\sT,\text{max}} \!= \!1.5\,$GeV$^{-1}$ is not visible, demonstrating that the observable is practically insensitive to contributions from the $b_{\sT}\! > \!b_{\sT,\text{max}}$ region.
By contrast, for $b_{\sT,\text{max}} \!= \!0.5\,$GeV$^{-1}$, varying $A$ leads to a noticeable distortion of the spectrum.
Moreover, this variation modifies the shape of the spectrum across the entire TMD region and, for the considered data sets, tends to improve the agreement with the experimental measurements.
Finally, the $A$-band exhibits only a partial overlap with the PDF uncertainty band (particularly for $Q \!=\!6.6\,$GeV), showing that the uncertainty associated with A is not fully covered by the PDF uncertainty.
In the second row, we present the theoretical predictions for $n \!= \!2$.
In this case, the $A$-band for $b_{\sT,\text{max}} \!=\! 1.5\,$GeV$^{-1}$ becomes visible.
However, this effect is driven by the freezing prescription.
In this scenario, besides the contribution from $b_{\sT} \!>\! b_{\sT,\text{max}}$ region, we also observe the impact of $S_{\rm NP}$ in the $b_{\sT} \!<\! b_{\sT,\text{max}}$ region (as well as that of $W^{\rm pert}$ in $b_{\sT} \!>\! b_{\sT,\text{max}}$).

Fig.~\ref{fig:comparison between models 2} quantifies the difference between $n\!=\!10$ and $n\!=\!2$ predictions.
This figure isolates the impact of freezing at $b_{\sT}\! =\! b_{\sT,\text{max}}$ as we are evaluating both models with the same $S_{\text{NP}}$.
We plot 20 curves in the interval $A \!=\! [0.5,1.5]\,$GeV$^2$ for both values of $b_{\sT,\text{max}}$.
For $\langle M_{QQ}\rangle\! =\! 6.6$ and $7.9\,$GeV, the averaged difference between the two freezing methods is larger for $b_{\sT,\text{max}}\! = \!1.5\,$GeV$^{-1}$.
The origin of this behaviour lies in $S_{\text{NP}}$.
For $n\!=\!10$, $S_{\text{NP}}$ remains very close to unity in $b_{\sT} \!<\! b_{\sT,\text{max}}$.
By contrast, for $n\!=\! 2$, $S_{\text{NP}}$ already deviates significantly from unity in this region.
As a consequence, for $b_{\sT,\text{max}} \!=\! 1.5\,$GeV$^{-1}$, the $n\!=\!2$ model introduces sizeable non-perturbative distortions over a broad intermediate-$b_{\sT}$ region.
On the other hand, for $b_{\sT,\text{max}} \!= \!0.5\,$GeV$^{-1}$, the effective region over which the two prescriptions differ significantly is considerably reduced.

Overall we conclude that we correctly describe the data.
At NNLL, it prefers a low value of $b_{\sT,\text{max}}$ like $0.5 \, \rm GeV^{-1}$.
Moreover, non-perturbative effects are significant in obtaining the $q_{\sT}$-distribution shape consistent with the experimental data, showing sensitivity to the non-perturbative parameters of Eq.~(\ref{eq:SNPsimple}) even accounting for the PDF uncertainty and disentangling the contribution from the effect of freezing at $b_{\sT}\!=\!b_{\sT,\text{max}}$.

So far, the discussion has been restricted up to NNLL accuracy.
Considering the calculation at higher orders could result in left-shifted peaks in the $q_{\sT}$-distributions of Fig.~\ref{fig:comparison with LHCb}, implying a significant improvement in the data description and reducing the the sensitivity of the observable to $A$ in $S_{\rm NP}$.
This is analysed in the Appendix~\ref{appendix:Suitability depending on the resummation accuracy}.
We conclude that although the shift in the peak position of the distribution is significant when going from LL to NLL, it becomes negligible from NLL to NNLL, and is expected to be even smaller beyond NNLL.

These findings provide compelling motivation for exploiting the LHCb data~\cite{LHCb:2023ybt} to achieve the first extraction of the large-$b_{\sT}$ behaviour of the unpolarised gluon TMD PDF at NNLL accuracy on $J/\psi$-pair production.

\begin{figure}[t]
    \centering
    \includegraphics[width=1\linewidth]{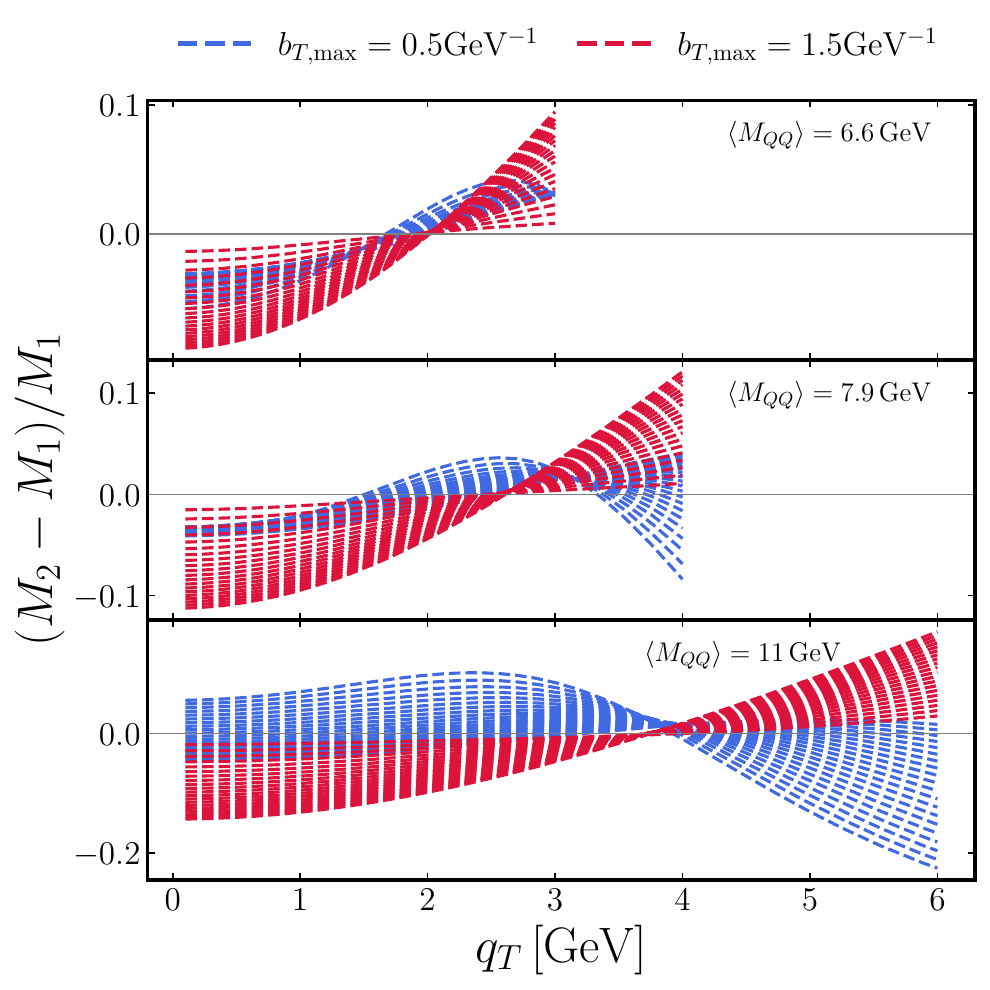}
    \caption{
    Comparison of the relative deviation between the two theoretical approaches in the second and third rows of Fig.~\ref{fig:comparison with LHCb} and for the two values $b_{\sT,\max}\!=\!0.5~\mathrm{GeV}^{-1}$ and $b_{\sT,\max}\!=\!1.5~\mathrm{GeV}^{-1}$.
    We denote the $n \!=\! 2$ case by $M_2$, and the case $n \!= \!10$ as $M_1$.
    The three panels correspond to different invariant-mass regions, characterized by $\langle M_{QQ}\rangle\!=\!6.6$, $7.9$, and $11~\mathrm{GeV}$.
    The curves correspond to 20 values of $A$ in the interval $A \!=\! [0.5,1.5]\,$GeV$^2$.
    }
    \label{fig:comparison between models 2}
\end{figure}

\section{Non-perturbative sensitivity across resummation accuracies}
\label{appendix:Suitability depending on the resummation accuracy}

\begin{figure}
    \centering
    \includegraphics[width=1\linewidth]{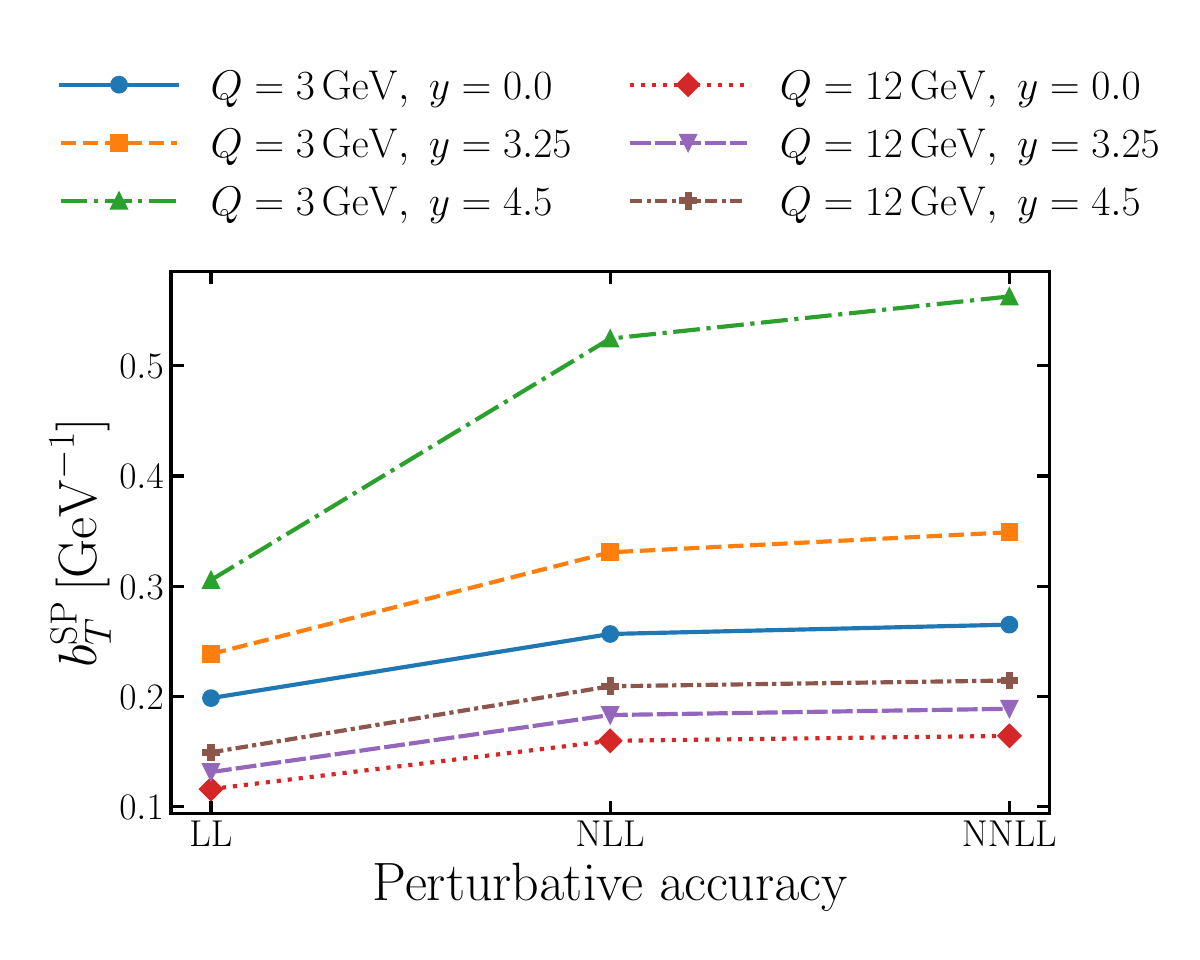}
    \caption{
    Dependence of the saddle-point parameter $b_{\sT}^{\mathrm{SP}}$ on the perturbative accuracy for different kinematic configurations $(Q,y)$. 
    The displayed values correspond to the solution of Eq.~(\ref{eq:saddle point approx}).
    }
    \label{fig:suitability from resum acc}
\end{figure}

\begin{table}[t]
    \centering
    \begin{tabular}{|c|c|c|c|}
        \hline
        $Q\,$[GeV] & $y$ & LL $\to$ NLL & NLL $\to$ NNLL \\
        \hline 
        \hline
        3 & 0 & 29$\%$ & 3$\%$ \\
        3 & 3.25 & 39$\%$ & 5$\%$\\
        3 & 4.50 & 72$\%$ & 7$\%$\\
        \hline
        12 & 0 & 38$\%$ &  3$\%$\\
        12 & 3.25 & 40$\%$ & 3$\%$\\
        12 & 4.50 & 40$\%$ & 3$\%$ \\
        \hline
    \end{tabular}
    \caption{Relative variations of the saddle-point parameter $b_{T}^{\mathrm{SP}}$ under successive increases of perturbative accuracy for the different kinematic configurations considered in the analysis.
    The quoted percentages quantify the relative shifts between LL and NLL, and NLL and NNLL predictions.}
    \label{tab:relative deviations}
\end{table}

In the following discussion, we investigate the effect of increasing the perturbative order in $W^{\rm pert}$ defined in Eq.~(\ref{eq:Wpert}).
Note that a shift of the peak of this distribution towards larger values translates into a shift of the peak of the observable in Sec.~\ref{sec:Comparison with LHCb data at N$^2$LL} towards smaller values in $q_{\sT}$ space.

Inspired by Refs.~\cite{Qiu:2000hf, Grewal:2020hoc}, to quantify how the integrate of Eq.~(\ref{eq:Cff}) is affected by the small-$b_{\sT}$ region, we use the saddle-point method.
At $q_{\sT}\! = \!0$ and for $S_{\rm NP} \!=\! 0$ in Eq.~(\ref{eq:Cff}), it is known that the integral is dominated by a saddle point at $b_{\sT} \!=\! b_{\sT}^{\rm SP}$, which is determined by,
\begin{equation}
\label{eq:saddle point approx}
    \left.\frac{\mathrm{d}}{\mathrm{d} b_{\sT}}
    \left[
    \ln \left(
    b^2_{\sT} W^{\rm pert}(b_{\sT})
    \right)
    \right]
    \right|_{b_{\sT} = b_{\sT}^{\rm SP}} = 0 \; .
\end{equation}

In Fig.~\ref{fig:suitability from resum acc}, we show the solution of the Eq.~(\ref{eq:saddle point approx}) as a function of the resummation accuracy.
The configurations of $(Q,y)$ correspond to $Q \!=\! 3$ and 12~GeV, and $y \!=\! 0, \, 3.25$ and 4.5 at $\sqrt{s} \!=\! 13 \, \rm TeV$.
Moreover, in Tab.~\ref{tab:relative deviations}, we compute the  relative shift of the saddle point when increasing the resummation procedure by one order.

The results exhibit a clear stabilisation pattern of the $b_{\sT}^{\rm SP}$ as the perturbative accuracy of $W^{\rm pert}$ is increased.
In particular, the relative variation between NLL and NNLL predictions is systematically smaller than the corresponding LL to NLL shift for all considered kinematic configurations, indicating a progressive convergence of the resummed perturbative series.
Furthermore, the relatively small NLL-to-NNLL deviations support the perturbative stability of the saddle-point region in the explored kinematics and reinforce the interpretation that the dominant contributions to the resummed TMD cross section remain largely controlled by perturbative dynamics.

Moreover, assuming perturbative convergence at higher orders of resummation accuracy beyond (N$^3$LL and so on), one expects the deviation of the $b_{\sT}^{\rm SP}$ with respect to the NNLL determination to remain below $3-7\%$, according to the results reported in Tab.~\ref{tab:relative deviations}.
We consider that such an effect has a negligible impact on the observable shown in Fig.~\ref{fig:comparison with LHCb}, thereby reinforcing the conclusions obtained in Sec.~\ref{sec:Comparison with LHCb data at N$^2$LL}.
In particular, the results of this appendix further support the interpretation of $J/\psi$-pair production as an optimal observable for extracting the non-perturbative information on the 3D gluon distribution inside the proton.

\section{On the theory constraint in \texorpdfstring{$\bt$}{bT} space}
\label{appendix: On the regularisation methods}

In Sec.~\ref{sec:Minimising chi2 with regularisation and n2}, to quantify the extent to which the non-perturbative approach defined by the $b_{\sT}^*$-prescription and $S_{\text{NP}}$ modifies the perturbative part and to regularise the $q_{\sT}$-oscillations found in Sec.~\ref{sec:Minimising chi2}, we construct a constraint based on the covariance matrix generated by perturbative scale variations.

For a fixed value of the hard scale $Q \!=\!\langle M_{QQ} \rangle$, we define the perturbative covariance matrix in Eq.~(\ref{eq:covmat}).
This  matrix characterizes the deformation patterns generated by perturbative scale variations and therefore defines the perturbative uncertainty manifold in the space of transverse-coordinate distributions.

To identify the statistically relevant deformation modes, we perform the spectral decomposition as follows
\begin{equation}
    C_{\rm th}=V \Lambda V^T \; ,
\end{equation}
where $V\!=\!(v_1,\ldots,v_n)$ contains the orthonormal eigenvectors and
\begin{equation}
    \Lambda={\rm diag}(\lambda_1,\ldots,\lambda_n),
    \qquad
    \lambda_1\ge\lambda_2\ge\cdots\ge\lambda_n \; ,
\end{equation}
with $\lambda$ the eigenvalues.
This decomposition is equivalent to a Principal Component Analysis (PCA), or Karhunen–Loève decomposition, and provides an optimal orthogonal basis ordered according to the perturbative variance carried by each deformation mode.
Large eigenvalues correspond to deformation patterns strongly supported by the perturbative uncertainty estimate, whereas small eigenvalues correspond to directions only weakly explored by the scale variations.

Given a model prediction $\tilde{f}(A)$, the deformation with respect to the perturbative central prediction is
\begin{equation}
\Delta =\tilde{f}(A) -T^{0} \; .
\end{equation}
Projecting this deformation onto the eigenbasis,
\begin{equation}
p_i =v_i^T \Delta ,
\end{equation}
the natural measure of its size relative to the perturbative uncertainty is,
\begin{equation}
\begin{aligned}
\label{eq:PCA of Dpert}
D_{\rm pert} & = \left( \tilde{f}_i(A) - T_i^{0} \right) (C_{th})^{-1}_{ij} \left( \tilde{f}_j(A) - T_j^{0} \right) \\
& =  \sum_i \frac{p_i^2}{\lambda_i} \; .
\end{aligned}
\end{equation}

In the present case, the covariance matrix is estimated from a finite set of scale variations.
Consequently, the smallest-eigenvalue modes are increasingly sensitive to the limited sampling of the perturbative uncertainty and may receive disproportionately large weights in the inverse covariance matrix.
Therefore, considering all eigenmodes in Eq.~(\ref{eq:PCA of Dpert}) yields a $D_{pert}(A)$ with a sharp well-defined minimum, implying that the minimisation of Eq.~(\ref{eq:minimisation chi2 + Dpert}) is dominated by $D_{pert}$.
To avoid this, it is necessary to exclude from Eq.~(\ref{eq:PCA of Dpert}) those modes that give rise to such large weights.
Rather than introducing a hard truncation of the spectrum, we employ a continuous spectral regularisation, so the penalty term is modified as,
\begin{equation}
    D_{\rm pert} \to D_{\rm pert} = \sum_i \frac{\lambda_i}{\lambda_i +\kappa \lambda_{\rm max}} \frac{p_i^2}{\lambda_i},
\end{equation}
where $\lambda_{\rm max} \equiv\lambda_1$ is the largest eigenvalue, and the regularisation parameter $\kappa$ is determined from the cumulative variance,
\begin{equation}
    R = \frac{\sum_{i=1}^{n_c}\lambda_i}{\sum_j\lambda_j} \; ,
\end{equation}
where $n_c$ corresponds to the point where the $R \!=\! R_{n_c}$ of the perturbative variance has been accumulated with $R_{n_c}$ being the minimum $R$ such as the fit is stable.
Therefore, $\kappa$ is defined as $\kappa\! \equiv\! \lambda_{n_c}/\lambda_{\rm max}$.
This method introduces a smooth transition between well-determined and weakly-determined eigenmodes without requiring an arbitrary spectral cut-off.
Modes responsible for less than $(1 \!-\! R_{n_c})\%$ of the total perturbative variance are gradually suppressed, while the dominant perturbative deformation patterns remain unaffected.\label{appendices}

\bibliography{Bibliography}
\bibliographystyle{ieeetr}

\end{document}